\documentclass[%
 reprint,
 amsmath,amssymb,
 aps,
 prl,
 nolongbibliography,
 superscriptaddress,
 floatfix,
 preprintnumbers
]{revtex4-2}

\usepackage{graphicx}
\usepackage{dcolumn}
\usepackage{bm}
\usepackage{siunitx}
\usepackage{xcolor}
\usepackage[left]{lineno}
\usepackage{hyperref}
\usepackage{float}
\usepackage{placeins}
\usepackage{bibunits}
\makeatletter
\AtBeginDocument{\auto@bib@empty}
\newcommand{\bibunitrevtexcontrols}{%
  \@startbibunitorrelax
  \immediate\write\@bibunitaux{\string\citation{REVTEX42Control}}%
  \immediate\write\@bibunitaux{\string\citation{apsrev42Control}}%
}
\makeatother

\newcommand{\CCQE}{\nu_\mu\text{CC}1p0\pi}
\newcommand{\papertitle}{Measurements of transverse kinematic imbalance in muon neutrino charged-current interactions on argon with a single proton and no pions in the final state with SBND}

\begin{document}

\preprint{FERMILAB-PUB-26-0759-PPD}

\defaultbibliographystyle{apsrev4-2}
\defaultbibliography{references}

\begin{bibunit}[apsrev4-2]

\title{\papertitle}

\newcommand{\ABC}{Universidade Federal do ABC, Santo Andre, SP 09210-580, Brazil}
\newcommand{\ANL}{Argonne National Laboratory, Lemont, IL 60439, USA}
\newcommand{\Alfenas}{Universidade Federal de Alfenas, Pocos de Caldas, MG 37715-400, Brazil}
\newcommand{\BNL}{Brookhaven National Laboratory, Upton, NY 11973, USA}
\newcommand{\Bern}{Universit{\"a}t Bern, Bern CH-3012, Switzerland}
\newcommand{\CIEMAT}{CIEMAT, Centro de Investigaciones Energeticas, Medioambientales y Tecnologicas, Madrid E-28040, Spain}
\newcommand{\CSU}{Colorado State University, Fort Collins, CO 80523, USA}
\newcommand{\CTI}{Center for Information Technology Renato Archer, Campinas, SP 13069-901, Brazil}
\newcommand{\Campinas}{Universidade Estadual de Campinas, Campinas, SP 13083-970, Brazil}
\newcommand{\Chicago}{University of Chicago, Chicago, IL 60637, USA}
\newcommand{\Columbia}{Columbia University, New York, NY 10027, USA}
\newcommand{\Edinburgh}{University of Edinburgh, Edinburgh EH9 3FD, United Kingdom}
\newcommand{\FNAL}{Fermi National Accelerator Laboratory, Batavia, IL 60510, USA}
\newcommand{\Florida}{University of Florida, Gainesville, FL 32611, USA}
\newcommand{\GeorgiaTech}{Georgia Institute of Technology, Atlanta, GA 30332, USA}
\newcommand{\Granada}{Universidad de Granada, Granada E-18071, Spain}
\newcommand{\Hanyang}{Hanyang University, Seoul 04763, Korea}
\newcommand{\Holyoke}{Mount Holyoke College, South Hadley, MA 01075, USA}
\newcommand{\IIS}{Indian Institute of Science, Centre for High Energy Physics, Bangalore 560012, India}
\newcommand{\IIT}{Illinois Institute of Technology, Chicago, IL 60616, USA}
\newcommand{\Imperial}{Imperial College London, London SW7 2AZ, United Kingdom}
\newcommand{\Kansas}{University of Kansas, Lawrence, KS 66045, USA}
\newcommand{\LANL}{Los Alamos National Laboratory, Los Alamos, NM 87545, USA}
\newcommand{\LSU}{Louisiana State University, Baton Rouge, LA 70803, USA}
\newcommand{\Lancaster}{Lancaster University, Lancaster LA1 4YW, United Kingdom}
\newcommand{\Liverpool}{University of Liverpool, Liverpool L69 7ZE, United Kingdom}
\newcommand{\Manchester}{University of Manchester, Manchester M13 9PL, United Kingdom}
\newcommand{\Michigan}{University of Michigan, Ann Arbor, MI 48109, USA}
\newcommand{\Minnesota}{University of Minnesota, Minneapolis, MN 55455, USA}
\newcommand{\NotreDame}{University of Notre Dame, Notre Dame, IN 46556, USA}
\newcommand{\Oxford}{University of Oxford, Oxford OX1 3RH, United Kingdom}
\newcommand{\Palermo}{Universit{\`a} degli Studi di Palermo, Palermo I-90123, Italy}
\newcommand{\Penn}{University of Pennsylvania, Philadelphia, PA 19104, USA}
\newcommand{\QueenMary}{Queen Mary University of London, London E1 4NS, United Kingdom}
\newcommand{\Rutgers}{Rutgers University, Piscataway, NJ 08854, USA}
\newcommand{\SaoJose}{Instituto Tecnol{\'o}gico de Aeron{\'a}utica S{\~a}o Jos{\'e} dos Campos, SP 12228-900, Brazil}
\newcommand{\Sheffield}{University of Sheffield, School of Mathematical and Physical Sciences, Sheffield S3 7RH, United Kingdom}
\newcommand{\Sussex}{University of Sussex, Brighton BN1 9RH, United Kingdom}
\newcommand{\Syracuse}{Syracuse University, Syracuse, NY 13244, USA}
\newcommand{\TAMU}{Texas A{\&}M University, College Station, TX 77843, USA}
\newcommand{\Tufts}{Tufts University, Medford, MA 02155, USA}
\newcommand{\UCSB}{University of California Santa Barbara, Santa Barbara, CA 93106, USA}
\newcommand{\UTA}{University of Texas at Arlington, Arlington, TX 76019, USA}
\newcommand{\VirginiaTech}{Virginia Tech, Center for Neutrino Physics, Blacksburg, VA 24060, USA}
\newcommand{\Warwick}{University of Warwick, Coventry CV4 7AL, United Kingdom}
\affiliation{\ABC}
\affiliation{\ANL}
\affiliation{\Alfenas}
\affiliation{\BNL}
\affiliation{\Bern}
\affiliation{\CIEMAT}
\affiliation{\CSU}
\affiliation{\CTI}
\affiliation{\Campinas}
\affiliation{\Chicago}
\affiliation{\Columbia}
\affiliation{\Edinburgh}
\affiliation{\FNAL}
\affiliation{\Florida}
\affiliation{\GeorgiaTech}
\affiliation{\Granada}
\affiliation{\Hanyang}
\affiliation{\Holyoke}
\affiliation{\IIS}
\affiliation{\IIT}
\affiliation{\Imperial}
\affiliation{\Kansas}
\affiliation{\LANL}
\affiliation{\LSU}
\affiliation{\Lancaster}
\affiliation{\Liverpool}
\affiliation{\Manchester}
\affiliation{\Michigan}
\affiliation{\Minnesota}
\affiliation{\NotreDame}
\affiliation{\Oxford}
\affiliation{\Palermo}
\affiliation{\Penn}
\affiliation{\QueenMary}
\affiliation{\Rutgers}
\affiliation{\SaoJose}
\affiliation{\Sheffield}
\affiliation{\Sussex}
\affiliation{\Syracuse}
\affiliation{\TAMU}
\affiliation{\Tufts}
\affiliation{\UCSB}
\affiliation{\UTA}
\affiliation{\VirginiaTech}
\affiliation{\Warwick}
\author{R.~Acciarri} \affiliation{\FNAL}
\author{L.~Aliaga-Soplin} \affiliation{\UTA}
\author{D.~Andrade Aldana} \affiliation{\IIT}
\author{C.~Andreopoulos} \affiliation{\Liverpool}
\author{A.~Antonakis} \affiliation{\UCSB}
\author{S.~Balasubramanian} \affiliation{\Holyoke}
\author{A.~Barnard} \affiliation{\Oxford}
\author{V.~Basque} \affiliation{\Edinburgh}
\author{J.~R.~Bateman} \affiliation{\Manchester}\affiliation{\Imperial}
\author{A.~Beever} \affiliation{\Sheffield}
% ...
\collaboration{The SBND Collaboration}
\date{\today}

\begin{abstract}
This letter presents the first neutrino-argon cross-section measurement from the Short-Baseline Near Detector (SBND) at the Fermi National Accelerator Laboratory. We report the nominal flux-averaged single-differential cross sections of muon neutrino charged-current interactions on argon with a single proton and no pions in the final state as functions of the transverse kinematic imbalance variables. Using a subset of Run~1 data corresponding to $8.8\times10^{19}$ protons-on-target, we select 12,804 signal candidates with a purity of 91\%. 
The extracted cross sections discriminate strongly among event-generator predictions, owing to the high statistics and the design of the analysis.
Among the compared models, none successfully describes all three variables, with the largest discrepancies in the transverse missing momentum variable. 
These results set new benchmarks for neutrino-argon interaction modeling, which is crucial for high-precision measurements at current and future accelerator neutrino experiments.
\end{abstract}

\maketitle

Accelerator neutrino experiments extract oscillation parameters from the rates and energy distributions of detected neutrino interactions. These spectra depend on the neutrino flux, the interaction cross sections, detector effects, and the analysis techniques, which makes their interpretation challenging. In the few-GeV regime, multiple interaction channels compete, and heavy nuclear targets like argon introduce nuclear effects that alter the kinematics of the final-state particles. Understanding these neutrino-nucleus interactions is therefore crucial for oscillation measurements and for the discovery of new physics~\cite{Abi:2020,Abi:2020epjc,Abi:2021epjc,NAGU2020114888,xsec_on_osc_1,xsec_on_osc_2,nustec_white_paper}. Uncertainties on the interaction models are significant sources of systematic uncertainties for long-baseline programs such as T2K and NOvA~\cite{Abe:2018pwo,Acero:2018prd,PhysRevLett.111.221802} and must be further constrained to meet the discovery goals of next-generation experiments such as the Deep Underground Neutrino Experiment (DUNE)~\cite{Abi:2020} and Hyper-Kamiokande (HK)~\cite{Hyper-Kamiokande}.

The $\nu_\mu$ charged-current quasielastic-like (CCQE-like) final state, representative of the CCQE scattering process, is frequently defined as the pionless final state with a single muon and a proton (``$\CCQE$"). This clear experimental signature enables high-purity event selection. In addition, measurements of the CCQE-like final-state kinematics serve as a sensitive probe of nuclear effects, such as the nuclear ground state and final-state interactions (FSI). Neutrino event generators that embed models for these processes rely on such measurements as benchmarks to evaluate their predictions of emergent final-state topologies~\cite{Andreopoulos:2015,Formaggio:2013kya}. On argon, $\CCQE$ cross sections have been measured by MicroBooNE~\cite{Abratenko:2020,TKI_DD_uboone,MicroBooNE:2023prd,MicroBooNE_GKI,MicroBooNE_CCQE_angle}, and measurements on more inclusive pionless final states have been reported by ICARUS~\cite{ICARUS:0pixsec} and MicroBooNE~\cite{uBooNE:0pi_0, uBooNE:0pi_1, uBooNE:0pi_2}. Measurements on other targets have been reported by MiniBooNE on mineral oil~\cite{Aguilar-Arevalo:2010}, by T2K on carbon and oxygen/water~\cite{Abe:2016prd,Abe:2020prd}, and by MINERvA on hydrocarbon~\cite{Lu:2018prl}.

The transverse kinematic imbalance (TKI) variables~\cite{PhysRevC.94.015503} quantify the momentum imbalance on the plane transverse to the neutrino direction (taken to be the beam axis). The variables are derived from the final-state particle kinematics.
The transverse missing momentum vector is defined as $\delta\vec{p}_T = \vec{p}_T^{\,\mu} + \vec{p}_T^{\,p}$, with magnitude $\delta p_T = |\delta\vec{p}_T|$, where $\vec{p}_T^{\,\mu}$ and $\vec{p}_T^{\,p}$ are the projections of the muon and proton momentum vectors on the transverse plane, respectively.
The shape of the $\delta p_T$ distribution reflects the nuclear ground state, or Fermi motion, combined with other nuclear effects.
The directional orientation of the imbalance is quantified by two angular variables. The variable $\delta\alpha_T$, defined as $\delta\alpha_T = \arccos\!\left[(-\vec{p}_T^{\,\mu}\cdot\delta\vec{p}_T)/(|\vec{p}_T^{\,\mu}|\,\delta p_T)\right]$, quantifies the angle between the transverse component of the muon momentum and the transverse missing momentum.
Due to the isotropic nature of Fermi motion, $\delta\alpha_T$ is expected to be uniformly distributed in the absence of additional nuclear effects. With FSI, the proton momentum is generally reduced and the $\delta\alpha_T$ distribution peaks toward 180$^\circ$~\cite{PhysRevC.94.015503}.
The azimuthal imbalance $\delta\phi_T$, defined as $\delta\phi_T = \arccos\!\left[(-\vec{p}_T^{\,\mu}\cdot\vec{p}_T^{\,p})/(|\vec{p}_T^{\,\mu}|\,|\vec{p}_T^{\,p}|)\right]$, provides complementary sensitivity to nuclear effects.
A simultaneous measurement of the TKI variables helps disentangle distinct nuclear effects.

This letter reports the first measurement of the flux-averaged differential cross section of $\CCQE$ interactions on argon as a function of TKI variables with SBND~\cite{Antonello:2015,SBNProgram,SBND_Detector} at a mean neutrino energy of $\langle E_\nu\rangle = \SI{0.75}{GeV}$. The signal is defined as a charged-current muon neutrino interaction producing exactly one muon with momentum between $0.22$ and $1.0$\,GeV/$c$,  exactly one proton with momentum between $0.3$ and $1.0$\,GeV/$c$ and no additional protons with momentum above $0.3$\,GeV/$c$, no charged pions with momentum above $70$\,MeV/$c$, no neutral pions or heavier mesons, and any number of neutrons. 
The bounds on the final-state particle momenta are set by the reconstruction efficiency, the candidate-event selection performance, and the detector acceptance. 

The SBND detector~\cite{SBND_Detector} is located 110\,m downstream from the target of the Booster Neutrino Beam (BNB)~\cite{miniboone_flux}, a predominantly muon-neutrino beam, at Fermilab.
SBND consists of a central liquid argon time projection chamber (LArTPC)~\cite{LArTPC} with dimensions $4\times4\times5$\,m$^3$, corresponding to a liquid argon active mass of $112$\,t.
LArTPCs operate by applying a uniform electric field across a high-purity liquid argon volume. When charged particles traverse the detector, they ionize argon atoms. 
The resulting ionization electrons are drifted towards an anode comprised of a stack of three wire planes with 3\,mm spacing between wires and between planes.
The wires in the two induction planes are angled at $\pm60^\circ$ relative to the vertical collection plane wires.
SBND consists of two drift regions, each $2$\,m long, separated by a central cathode, with a $500$\,V/cm drift field. 
The photon collection system~\cite{SBND:2024epjc} provides timing information from the scintillation light and aids the identification of neutrino interactions. 
The drift time and wire readout combined enable three-dimensional reconstruction of particle trajectories and calorimetric measurements of energy depositions.

Cuts on the detector data acquisition conditions and beam quality are applied to the data to ensure the high quality of the analyzed sample.
The results presented here are based on data from the beginning of SBND Run 1 corresponding to $8.8\times10^{19}$ protons on the BNB target (POT), after data quality cuts.

The simulation and reconstruction are performed within the \texttt{LArSoft} software framework~\cite{LArSoft}.
The neutrino beam flux is simulated with the BNB flux simulation~\cite{miniboone_flux}. 
Neutrino interactions are simulated with the \texttt{GENIE v3.4.0 AR23\_20i\_00\_000} neutrino event generator~\cite{GENIEAR23_0, GENIEAR23_1}. Cosmic-ray activity is simulated with the \texttt{CORSIKA v7.4003} generator~\cite{CORSIKA} and overlaid on the neutrino interaction simulation to account for the cosmic-ray contamination in neutrino interaction events. In addition, a separate simulation is generated for the cosmic-ray events that trigger the readout in the absence of a neutrino interaction.
Particles from interaction simulations are then propagated through \texttt{Geant4 v4\_10\_6}~\cite{GEANT4, G4_0, G4_1} with physics list \texttt{QGSP\_BERT\_HP}, which simulates the passage of particles through material such as liquid argon and detector material.
Finally, the detector response simulation converts the energy deposits simulated by \texttt{Geant4} into the charge detected by the wire planes, taking into account argon and detector properties such as electric field non-uniformities, electron-ion recombination~\cite{ArgoNeuT_recombination,ICARUSEnergyScale}, drifting electron diffusion, electric field response, and the electronics response.

The kinematics of detected neutrino interactions are extracted from charge deposits in the detector through the reconstruction of events. Charge deposits processed from wire waveforms~\cite{MicroBooNE_SP} are clustered and matched across planes with the \texttt{Pandora} pattern recognition toolkit~\cite{Acciarri:2018}, which forms particle objects that provide 3D spatial and calorimetric information, such as interaction vertices and the directions and lengths of particle trajectories.
Track momenta for particles that stop within the detector volume are obtained from stopping range via the continuous-slowing-down approximation. An alternate momentum estimate is obtained from multiple Coulomb scattering (MCS)~\cite{MCS} and compared to the range-based estimate to flag misreconstructed tracks.
Particle species identification uses calorimetric likelihood scores calculated by comparing the measured energy deposited per unit length ($dE/dx$) as a function of residual range with the Bethe-Bloch prediction~\cite{pdg} for different particle types.

The fiducial volume (FV) is selected to ensure uniform detector performance, and comprises the two drift regions separated by a gap around the cathode. The FV is defined by a \SI{50}{\cm} inset on the downstream face, \SI{10}{\cm} insets on all other detector faces and the cathode, and an additional exclusion in the upper-rear corner. Selected $\CCQE$ candidate events must be fully contained within a single drift region.

A sequential event selection procedure identifies the $\CCQE$ events. First, cosmic-ray backgrounds are rejected by discarding events flagged as cosmic-ray activity by \texttt{Pandora}, applying a cut on the neutrino-likeness boosted decision tree (BDT) score based on the spatial distribution of the event, and requiring the reconstructed neutrino interaction vertex to lie within the FV. Next, a two-track topology selection is applied, requiring exactly two reconstructed track-like particle objects with a common vertex, both fully-contained in the FV.
Lastly, muon and proton candidates are identified using calorimetry and kinematics (see the Supplemental Material~\cite{SM} for more details).
The muon candidate must exceed $50$\,cm in length to suppress pions, have range- and MCS-based momenta agree within 20\% to reject poorly reconstructed tracks, satisfy muon particle identification (PID) score cuts, and have reconstructed momentum within the signal definition.
The remaining track must pass a proton PID cut and satisfy the signal definition.

A total of 12{,}804 data events are selected as signal candidates. The containment requirement has an efficiency of 28\%, and the event selection has an efficiency of 34\% for fully contained signal events. 
The selected sample has a signal purity of 91\% and is 80\% CCQE, as predicted by the \texttt{GENIE} event generator.
The non-CCQE processes (meson exchange currents, resonance production, and shallow/deep inelastic scattering) that mimic the final-state topology of the signal channel can qualify as signal events. This occurs when additional particles fall below detection thresholds or when FSI (e.g., pion absorption) produces a muon-proton final state consistent with the signal definition.
Leading contributions to the background are muon-proton pairs outside the kinematic thresholds of the signal definition, multi-proton final states with obscured or overlapping tracks, and neutral-current events with pions misidentified as muons. 

\begin{figure*}[t]
\centering
\begin{tabular}{@{}ccc@{}}
\includegraphics[width=0.33\textwidth]{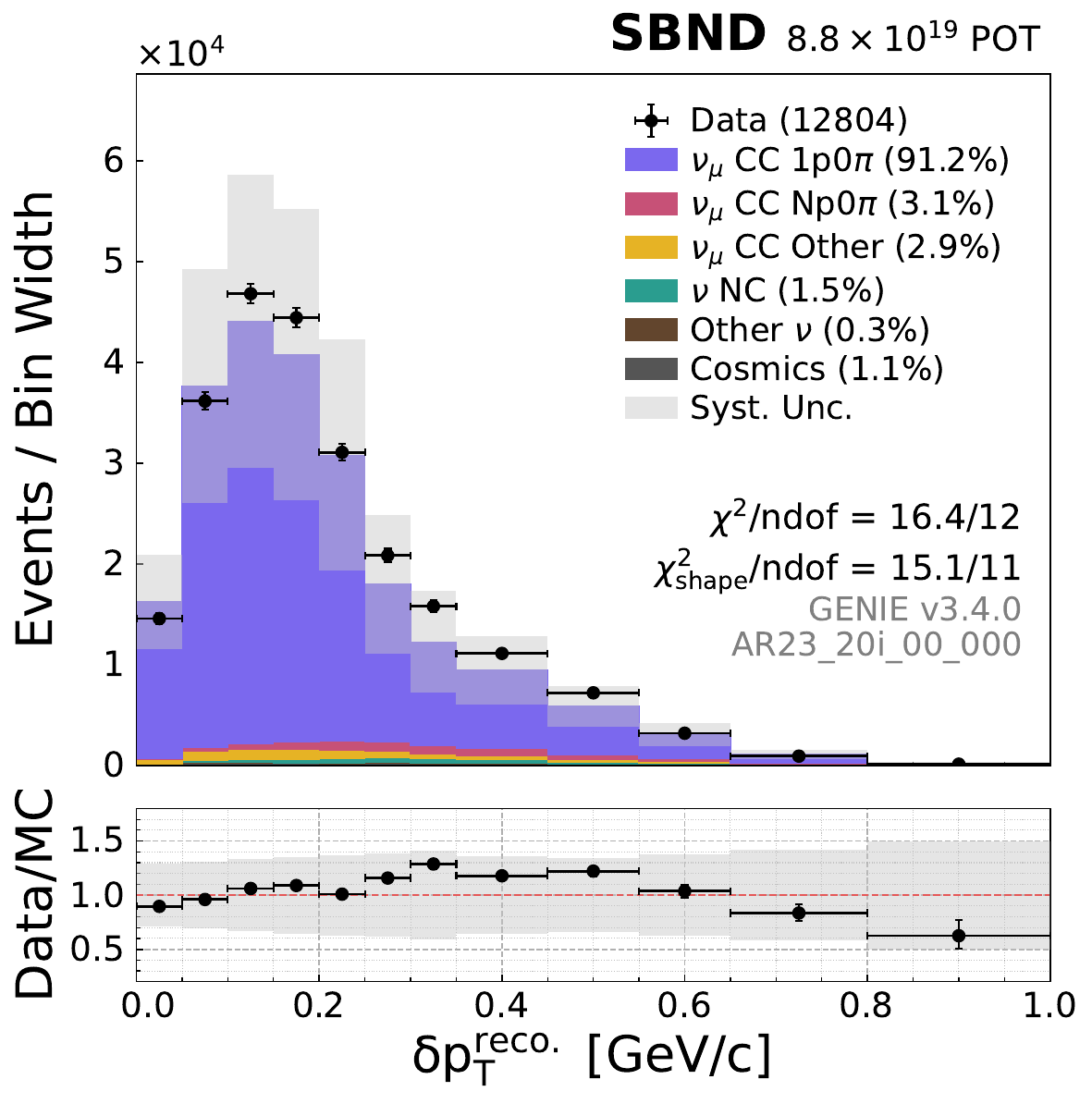} &
\includegraphics[width=0.324\textwidth]{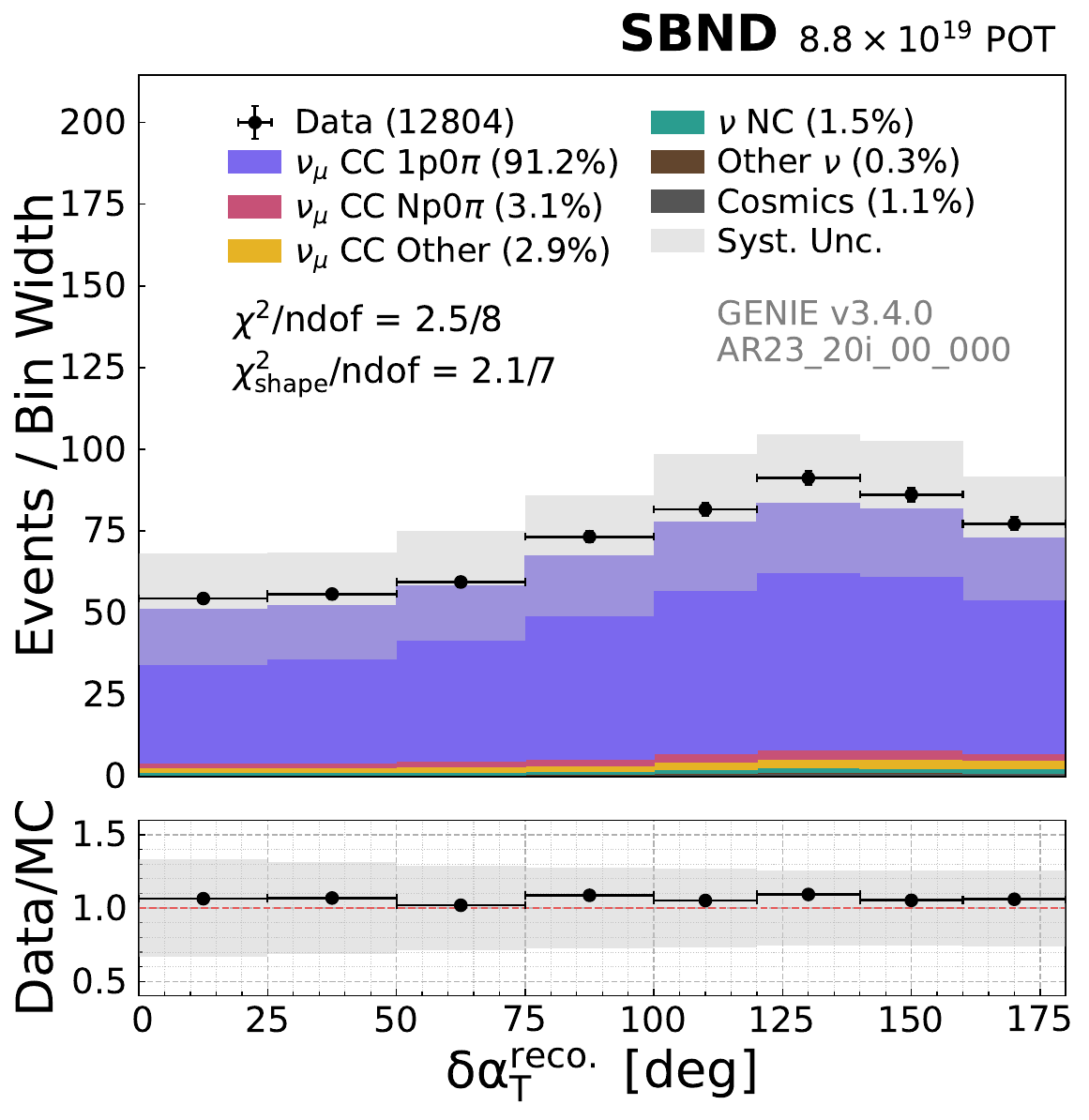} &
\includegraphics[width=0.325\textwidth]{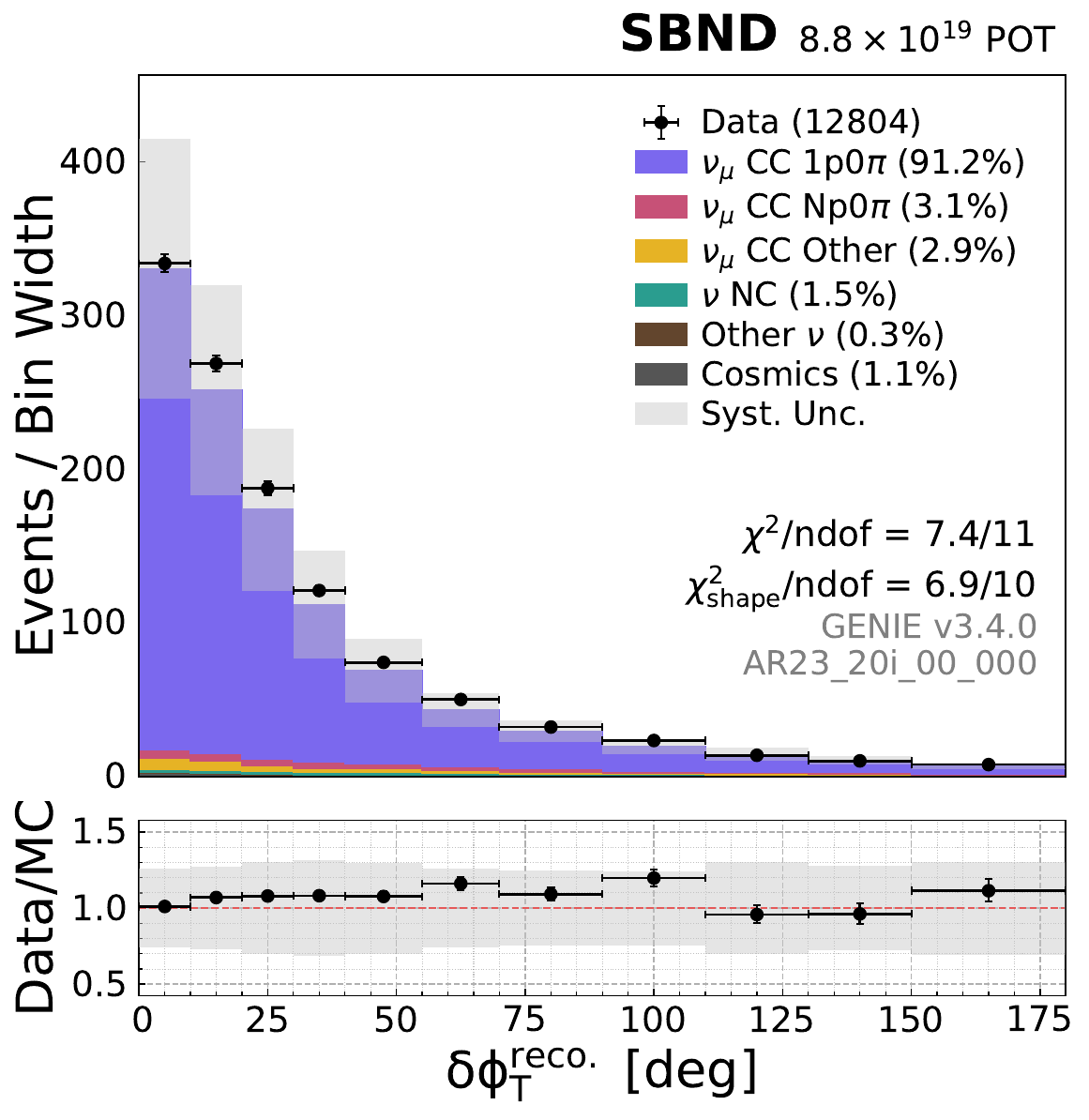} \\
(a) & (b) & (c) \\
\end{tabular}
\caption{Selected data and Monte Carlo (MC) simulation event rates as functions of (a)~$\delta p_T$, (b)~$\delta\alpha_T$, and (c)~$\delta\phi_T$.
Stacked histograms show the MC prediction by event category, and the shaded band shows the total systematic uncertainty.
The $\nu_\mu$ CC events are subdivided by the final-state topology. 
``Other~$\nu$'' includes $\bar{\nu}_\mu$ and $\nu_e$ events.
}
\label{fig:sel_tki}
\end{figure*}

Figure~\ref{fig:sel_tki} shows the selected data and Monte Carlo (MC) simulation event rates as functions of $\delta p_T$, $\delta\alpha_T$, and $\delta\phi_T$. 
The measurement bins are chosen for each variable to ensure that at least 70\% of simulated signal events are reconstructed in their true bin.
Data and MC show agreement in both normalization and shape within the evaluated uncertainties.
As shown, the TKI distributions differ among event topologies that arise from different interaction processes.

The systematic uncertainties in Fig.~\ref{fig:sel_tki} include contributions from the neutrino flux, the detector response, the neutrino-nucleus interaction model, secondary interactions, cosmic-ray background, the number of argon targets, the beam protons-on-target, and the statistical uncertainty of the MC sample.
Flux, interaction-model, and secondary-interaction uncertainties are evaluated by reweighting each simulated event to an ensemble of alternate scenarios.
The flux uncertainties account for beam instrumentation (horn current and skin effect), primary production of charged pions and of charged and neutral kaons, the inelastic and quasi-elastic cross sections of the primary protons, and the inelastic cross section of secondary pions.
The neutrino-nucleus interaction model uncertainties from \texttt{GENIE}~\cite{Andreopoulos:2015,GENIE_v3} cover the cross sections of all scattering channels as well as the FSI parameters.
The uncertainties on secondary interactions are evaluated using \texttt{Geant4Reweight v01\_20\_11}~\cite{G4Reweight} to reweight the simulated events by varying the probabilities of nucleon and pion absorption and charge exchange.
The detector response uncertainties are evaluated by varying the overall charge scale, the parameters of the electron-ion recombination model~\cite{ICARUSEnergyScale}, the electric field across the detector volume, and the charge response of the anode wires via a data-driven map~\cite{WireMod,WireMod_MicroBooNE}.
The cosmic-ray uncertainty is assessed by comparing beam-off data and simulation.
Normalization uncertainties of 2\% and 1\% are assigned to the beam protons-on-target and the number of argon targets, respectively.
Lastly, the MC statistical uncertainty is evaluated by bootstrapping, where each event is resampled using a Poisson distribution with a mean of 1.

The neutrino interaction model and its covariance are tested in this sample with a data-driven conditional constraint, following Ref.~\cite{Validation}.
The measured muon momentum and angle are used to update the predicted proton momentum and angle through the joint covariance of the two variables.
The constrained proton momentum and angle distributions remain consistent with the data (see the Supplemental Material~\cite{SM}).

We report the unfolded cross sections with Wiener-SVD unfolding~\cite{WienerSVD}, using a second-derivative regularization matrix and unit normalization of the smoothness constraint.
The flux-averaged differential cross sections in the reconstructed space are obtained by dividing the background-subtracted event count in each bin by the product of the integrated flux, the number of targets, and the bin width. 
Then, the unfolding procedure returns the unfolded differential cross sections and their covariance matrix in the regularized space, defined by the regularization smearing matrix $A_C$. The $A_C$ matrix is applied to the theory predictions for comparison to the reported results.

The systematic uncertainties are propagated to the unfolding procedure. 
The integrated contributions from each source are summarized in Table~\ref{tab:uncertainties}, and the uncertainties in each differential bin are provided in the Supplemental Material~\cite{SM}.
In particular, the neutrino-nucleus interaction model uncertainties on the signal channel are evaluated only for their impact on efficiency and bin migration. 
The dominant uncertainty arises from the neutrino flux, which primarily affects the precision on the normalization. 
However, the differential measurements provide sensitivity to the spectral shape of the cross section.

The unfolding procedure is tested on fake datasets in which the nominal \texttt{GENIE} prediction is replaced by smooth reweightings of interaction-mode rates and kinematic shapes, and by injected spectral bumps. In addition, fake data tests with the alternate generator \texttt{GiBUU} are used to assess the robustness to the model dependence of our MC, including acceptance and detector response effects.
In each case the unfolded result follows the injected spectrum with no statistically significant bias (see the Supplemental Material~\cite{SM}).

\begin{table}[]
\caption{Summary of uncertainties on the flux-averaged cross section.}
\label{tab:uncertainties}
\centering
\begin{ruledtabular}
\begin{tabular}{lc}
Source & Uncertainty [\%] \\
\hline
    Neutrino flux & 6.5 \\
    Detector & 3.7 \\
    Neutrino interaction model & 3.4 \\
    Secondary interactions & 1.4 \\
    Beam protons-on-target & 2.0 \\
    Number of Ar targets & 1.0 \\
    Cosmic-ray contamination & 0.4 \\
    MC statistics & 0.5 \\
\hline
    Total & 8.7 \\
\end{tabular}
\end{ruledtabular}
\end{table}

\begin{figure*}[t]
\centering
\begin{tabular}{@{}ccc@{}}
\includegraphics[width=0.33\textwidth]{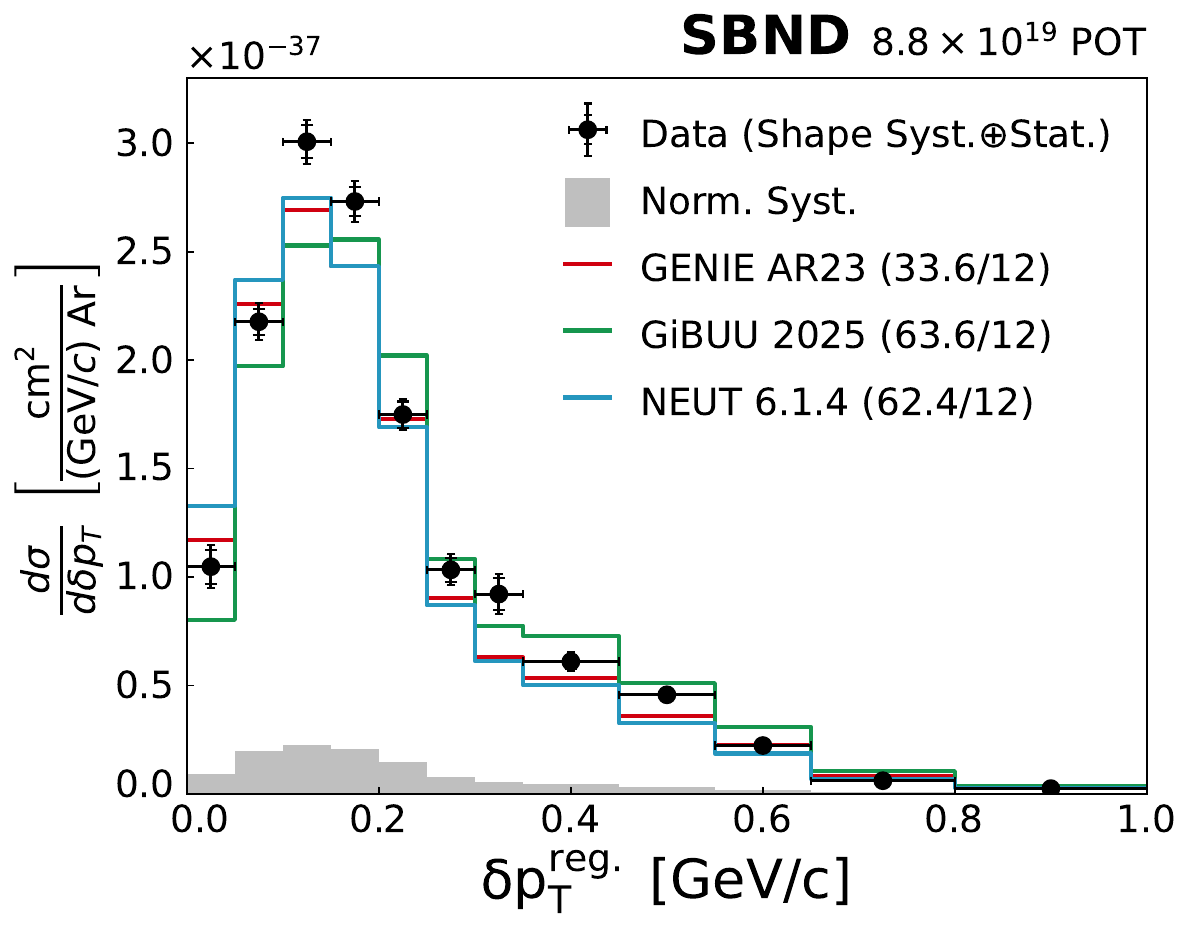} &
\includegraphics[width=0.315\textwidth]{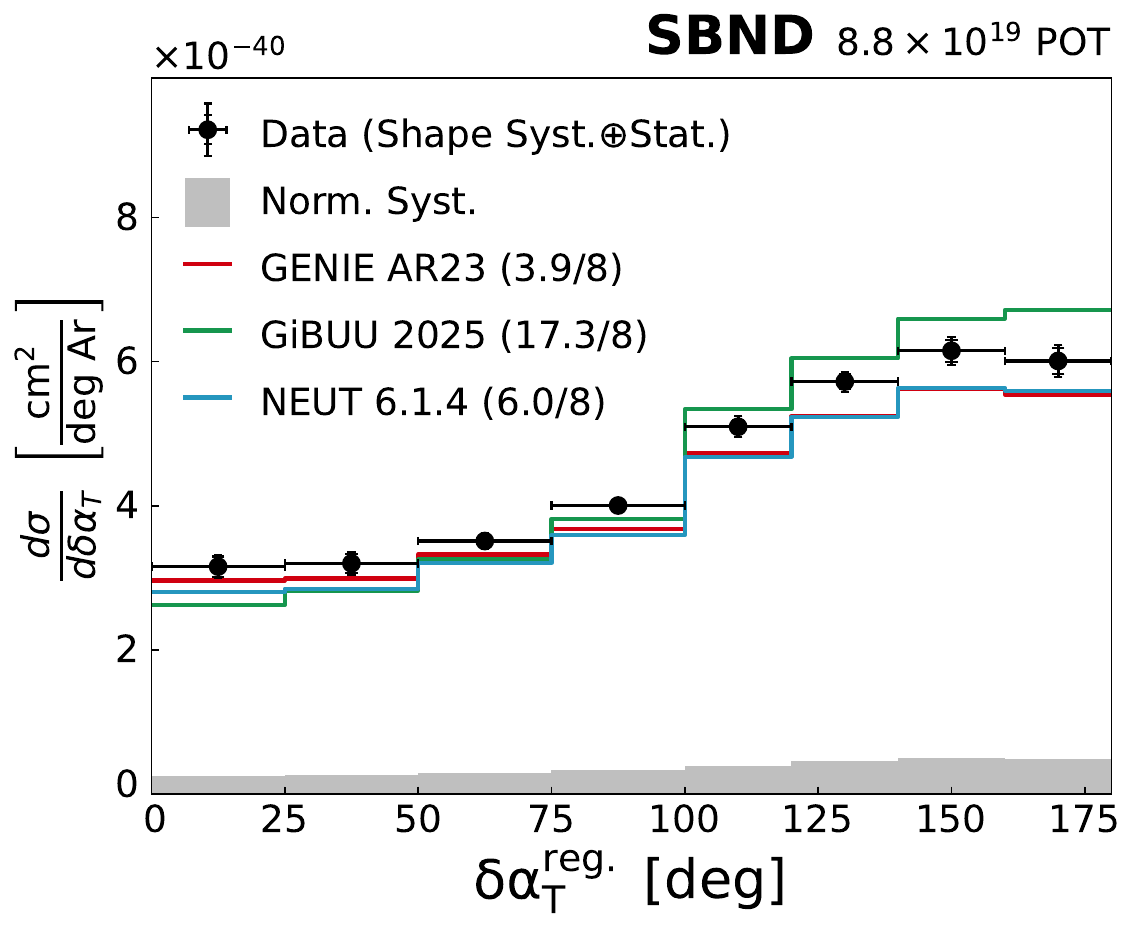} &
\includegraphics[width=0.327\textwidth]{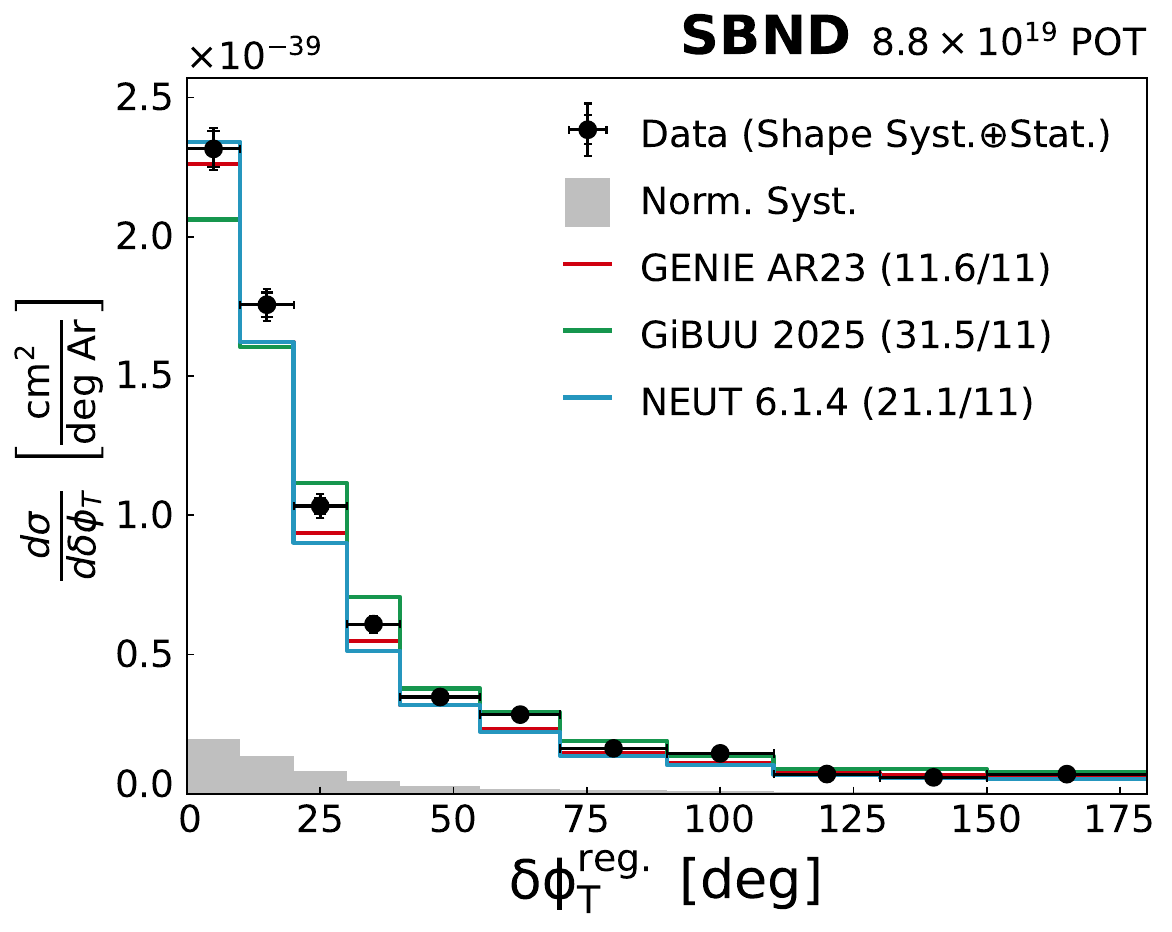} \\
(a) & (b) & (c) \\
\end{tabular}
\caption{
    Unfolded, flux-averaged $\CCQE$ differential cross sections as functions of (a)~$\delta p_T$, (b)~$\delta\alpha_T$, and (c)~$\delta\phi_T$, compared to generator predictions from \texttt{GENIE AR23\_20i\_00\_000}, \texttt{GiBUU 2025}~\cite{Mosel:2019vhx}, and \texttt{NEUT v6.1.4}~\cite{NEUT}. Inner error bars denote statistical uncertainties, while outer error bars represent the statistical and shape-only systematic uncertainties added in quadrature. The gray bands at the bottom indicate normalization uncertainties. Values of $\chi^2/\mathrm{ndof}$, evaluated using the full unfolded covariance matrix, are given in parentheses.
}
\label{fig:generators}
\end{figure*}

The unfolded differential cross section is shown in Fig.~\ref{fig:generators}, along with comparisons to predictions from various neutrino event generators. 
The spectral shape of the differential cross section shows distinct features that reflect nuclear effects.
The overall peaked shape of the $\delta p_T$ cross-section spectrum reflects the momentum distribution of the struck nucleon in the nuclear ground state. Nucleons that exist in correlated pairs within the nucleus~\cite{SRCStrength1,SRCStrength2,SRCStrength3,SRCStrength4}, as well as nuclear effects like FSI, contribute to the extended tail.
The rising slope in the $\delta\alpha_T$ spectrum is caused by outgoing protons that experience nuclear effects, while the falling edge in the $\delta\phi_T$ spectrum represents events driven out of muon-proton coplanarity. 

The high-purity selection and excellent detector performance provide the precision needed for stringent tests of the model predictions.
We compare our measurements to different event generator predictions using the \texttt{NUISANCE} framework~\cite{NUISANCE}. First, as shown in Fig.~\ref{fig:generators}, we compare to predictions from different event generators with different modeling philosophies: \texttt{GENIE AR23\_20i\_00\_000} (GENIE AR23) ~\cite{Andreopoulos:2015,GENIE_v3}, \texttt{GiBUU 2025} (GiBUU) ~\cite{Mosel:2019vhx}, and \texttt{NEUT v6.1.4} (NEUT) ~\cite{NEUT}.
For instance, GENIE and NEUT factorize the neutrino-nucleus interaction process into the nuclear ground state, a primary interaction, and subsequent final-state interactions, whereas GiBUU evolves the initial- and final-state dynamics coherently in a quantum-kinetic transport framework. 
We also compare to predictions of \texttt{GENIE} with alternate model choices: spectral function (SF) for the nuclear ground state~\cite{SF_argon,SF_Genie}, different axial form factor models~\cite{MINERvA_FormFactor, LQCD_FormFactor}, and CCQE models based on Hartree-Fock (HF) calculations~\cite{HF1,HF2,CRPA1,CRPA2}. These variations are further detailed in the Supplemental Material~\cite{SM}.

In $\delta p_T$, none of the models provides a satisfactory description of the measurement. While GENIE AR23 comes the closest in the comparison of Fig.~\ref{fig:generators}, it underpredicts the data in the peak and intermediate regions, and overpredicts in the high tail region. Some models show better predictions locally (i.e. GiBUU in the intermediate region), but do not enhance the overall agreement. 
Out of the \texttt{GENIE} variations, the HF-based models show the best agreement, as shown in Fig.~S25. 

Most generators describe the $\delta \alpha_T$ measurement reasonably well. However, the GiBUU model predicts a steeper slope than the measurement, as expected if nuclear effects distort the proton kinematics more strongly. Meanwhile, the HF models, observed to improve the $\delta p_T$ agreement, predict a slope that is too shallow.
The $\delta\phi_T$ measurement is also described reasonably well by most generators.
However, the GENIE AR23 axial form-factor variants, which also include the Nieves QE correction of Ref.~\cite{deltp_benchmark}, show significant discrepancies, as shown in Fig.~S23.

The GENIE AR23 model shows good agreement with data for the cross section in muon and proton kinematics (as shown in Fig.~S22), as well as the normalization, in contrast to previous reports showing more significant underestimation of the MicroBooNE data~\cite{deltp_benchmark}.
This suggests that the discrepancies in TKI modeling arise from modeling deficiencies of the underlying nuclear physics rather than a general mismodeling of the final-state kinematics.
Meanwhile, other generators show discrepancies for certain variables, such as GiBUU in muon momentum and both NEUT and GiBUU in the proton direction along the beam axis.

Our measurements exhibit some features consistent with previous TKI cross-section measurements~\cite{TKI_DD_uboone,deltp_benchmark}.
For instance, the $\delta p_T$ peak is underpredicted by the generators, and the preference among the generators is consistent with the measurement from MicroBooNE in Ref.~\cite{TKI_DD_uboone}, with GENIE favored over NEUT and GiBUU, albeit with different versions of each generator.
In addition, the GiBUU generator's prediction of a steeper $\delta\alpha_T$ slope than is observed in data was also noted in earlier observations~\cite{TKI_DD_uboone,T2K_DD_carbon_oxygen}.
On the other hand, while adopting lattice QCD form factors improved agreement with the MicroBooNE measurements~\cite{deltp_benchmark}, such adjustments do not provide the same level of agreement with this measurement.

In summary, we report SBND's first neutrino-argon cross-section measurement of the $\CCQE$ final state using 8.8$\times 10^{19}$ POT from the BNB. We report the flux-averaged single-differential cross sections in the TKI variables, extracted with Wiener-SVD unfolding from 12{,}804 candidates, predicted to have $91\%$ signal purity by our MC.
Compared to event generators, \texttt{GENIE AR23\_20i\_00\_000}, \texttt{NEUT v6.1.4}, and \texttt{GiBUU 2025} are all disfavored for $\delta p_T$, while \texttt{GENIE AR23\_20i\_00\_000} and \texttt{NEUT v6.1.4} remain broadly consistent with $\delta\alpha_T$ and $\delta\phi_T$.
Among model configuration variations of \texttt{GENIE}, SF and axial form-factor variants leave the $\delta p_T$ tension unresolved, and HF configurations improve $\delta p_T$ at the expense of $\delta\alpha_T$. These results demonstrate that the different TKI variables probe complementary aspects of the neutrino-argon interaction model that current generators do not yet fully capture.
With an anticipated dataset size corresponding to $10 \times 10^{20}$~POT, SBND will continue to deliver the high-statistics, multi-differential neutrino-argon scattering measurements needed to sharpen such details of neutrino-nucleus interaction modeling for the next generation of experiments.

\begin{acknowledgments}

This document was prepared by the SBND Collaboration using the resources of the Fermi National Accelerator Laboratory (Fermilab), a U.S. Department of Energy, Office of Science, Office of High Energy Physics HEP User Facility. Fermilab is managed by FermiForward Discovery Group, LLC, acting under Contract No. 89243024CSC000002. The SBND Collaboration acknowledges the generous support of the following organizations: the U.S. Department of Energy, Office of Science, Office of High Energy Physics; the U.S. National Science Foundation; Los Alamos National Laboratory for LDRD funding, the Science and Technology Facilities Council (STFC), part of United Kingdom Research and Innovation (UKRI), the UKRI Future Leaders Fellowship (grant number MR/V022407/1), and The Royal Society; the Swiss National Science Foundation; the Spanish Ministerio de Ciencia, Innovaci\'{o}n y Universidades (MICIU/ AEI/ 10.13039/ 501100011033) under grants No PRE2019-090468, CNS2022-136022, RYC2022-036471-I, PID2023-147949NB-C51 \& C53 and Comunidad
de Madrid (PEJ-2023-AI/COM-28399); the European Union’s Horizon 2020 research and innovation program under GA no 101004761 and the Marie Sklodowska-Curie grant agreements No 822185, 101081478, and 101003460; the S\~{a}o Paulo Research Foundation 1098 (FAPESP), the National Council of Scientific and Technological Development (CNPq) and Ministry of Science, Technology \& Innovations-MCTI of Brazil; the Minas Gerais research foundation (FAPEMIG), grants APQ-00544-23 and APQ-01249-24; the Anusandhan National Research Foundation (ANRF, India) under the Ramanujan Fellowship (Grant No. RJF/2025/000203). An award of computer time was provided by the ASCR Leadership Computing Challenge (ALCC) program. This research used resources of the Argonne Leadership Computing Facility, which is a U.S. Department of Energy Office of Science User Facility operated under contract DE-AC02-06CH11357.

\end{acknowledgments}

\bibunitrevtexcontrols
\putbib[mainNotes,references]

\end{bibunit}

\clearpage
\onecolumngrid
\clearpage
\setcounter{page}{1}
\setcounter{figure}{0}
\setcounter{table}{0}
\setcounter{equation}{0}
\setcounter{section}{0}
\setcounter{secnumdepth}{3}
\setcounter{dbltopnumber}{1}
\setcounter{topnumber}{6}
\setcounter{bottomnumber}{6}
\setcounter{totalnumber}{10}
\renewcommand{\thefigure}{S\arabic{figure}}
\renewcommand{\thetable}{S\arabic{table}}
\renewcommand{\theequation}{S\arabic{equation}}
\renewcommand{\thesection}{S\arabic{section}}
\renewcommand{\topfraction}{0.95}
\renewcommand{\bottomfraction}{0.95}
\renewcommand{\textfraction}{0.05}
\renewcommand{\floatpagefraction}{0.9}

\begin{bibunit}[apsrev4-2]

\title{Supplemental Material:\\ \papertitle}
\author{}
\date{\today}
\maketitle

\onecolumngrid

This Supplemental Material accompanies the Letter and provides additional details on the main results. The content includes the Booster Neutrino Beam flux spectrum over the analysis fiducial volume, the selection efficiency and response matrices, the systematic-uncertainty breakdowns and fractional covariance matrices in each differential bin, data-driven conditional-constraint validation tests of the systematic uncertainties, fake-data robustness tests of the Wiener-SVD unfolding, the regularization matrices $A_C$ and unfolded systematic covariances, and additional generator comparisons and model configurations beyond those shown in the Letter.

\section{Flux}
\label{sec:flux}

The Booster Neutrino Beam (BNB) flux varies across the SBND detector volume ($\sim$5\% variation in the total flux). We therefore compute a nominal flux by averaging the flux over the fiducial volume (FV) defined for this analysis. 
The FV consists of regions of the two TPC drift volumes, excluding a \SI{20}{cm} gap about the cathode, \SI{50}{cm} inset from the downstream face, \SI{10}{cm} insets from the other detector faces, and further restrictions in the upper-rear corner of the detector. These regions are excluded from the FV to ensure homogeneity of the detector response, rejecting regions with non-uniform detector response, such as electric field distortions.
We use the SBND detector coordinates in units of cm: $\hat{x}$ is perpendicular to the beam with the
cathode at $x=0$, $\hat{y}$ is vertically upward with $y=0$ at the vertical center of the
detector, and $\hat{z}$ is along the beam
(downstream) with $z=0$ at the upstream face of the active volume. In these coordinates, the FV consists of four rectangular boxes:
\begin{equation}
  \begin{cases}
  10 < |x| < 190,\ |y| < 190,\ 10 < z < 250, \\
  10 < |x| < 190,\ -190 < y < 100,\ 250 < z < 450.
  \end{cases}
\end{equation}

The volume-averaged flux is computed by ray-tracing. Each simulated neutrino in the flux prediction is treated as a straight ray with weight $w_i$, energy $E_i$, and direction $\hat{p}_i$. For each ray, we compute the path length $L_i$ through the FV defined above. The volume-averaged flux in energy bin $b$ is then computed as:
\begin{equation}
  \phi_b
  =
  \frac{1}{V\,\mathrm{POT}}
  \sum_{i:\,E_i\in b}
  w_i\,L_i\,,
\end{equation}
where $V$ is the volume of the FV, $\mathrm{POT}$ is the total number of protons on target, and the sum is over all simulated neutrinos in the flux prediction with energy in bin $b$. 

\begin{figure}[H]
\centering
\includegraphics[width=0.45\textwidth]{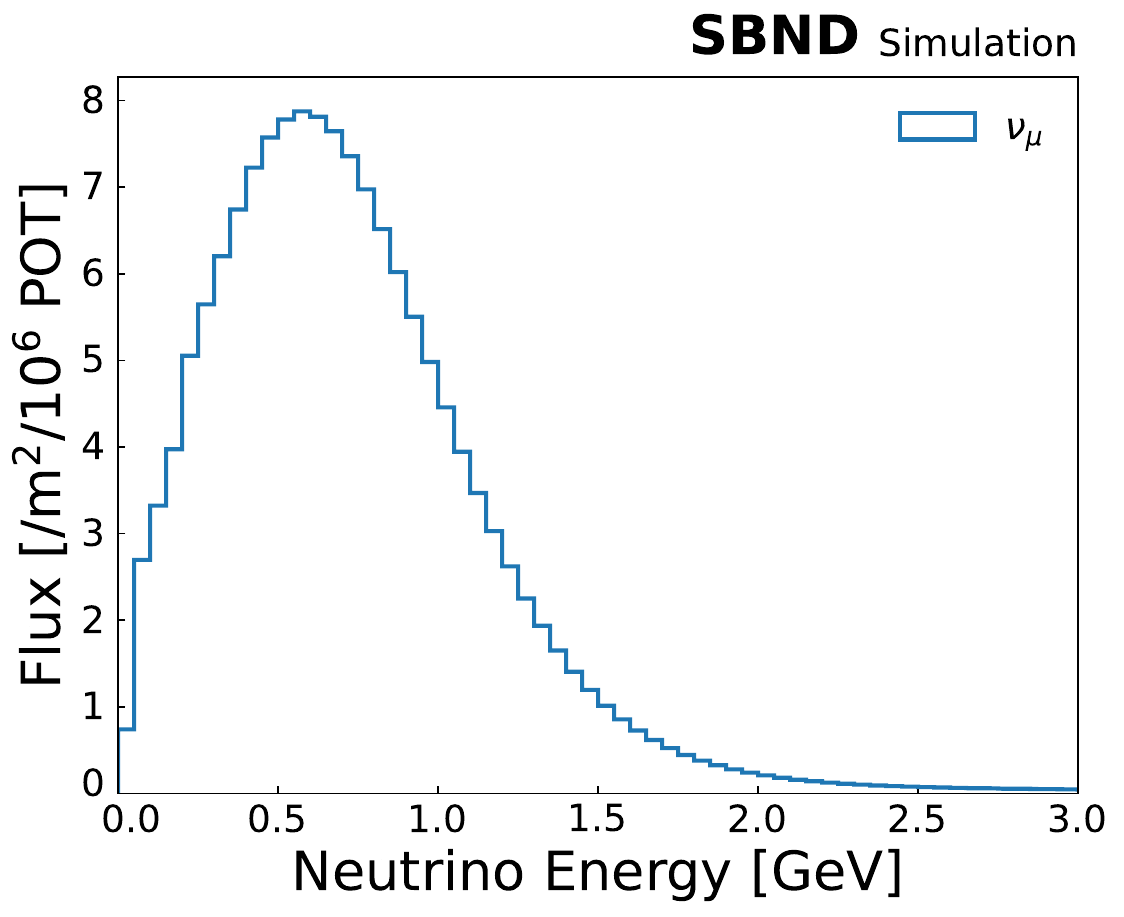}
\caption{BNB $\nu_\mu$ flux spectrum averaged over the analysis fiducial volume.}
\label{fig:flux}
\end{figure}

The flux is binned in \SI{50}{MeV} intervals, covering the range from 0 to \SI{10}{GeV}. The resulting $\nu_{\mu}$ flux spectrum, shown in Fig.~\ref{fig:flux}, is included in the data release and reported in units of neutrinos per \SI{}{m^2} per $10^6$ POT per GeV. The total integrated flux is \SI{1.52e-04}{/m^{-2}/ POT}, with an average neutrino energy of \SI{0.75}{GeV}.

\FloatBarrier
\newpage
\section{Particle Identification}
\label{sec:pid}

\begin{figure}[h]
  \centering
  \begin{tabular}{@{}cc@{}}
  \includegraphics[width=0.4\textwidth]{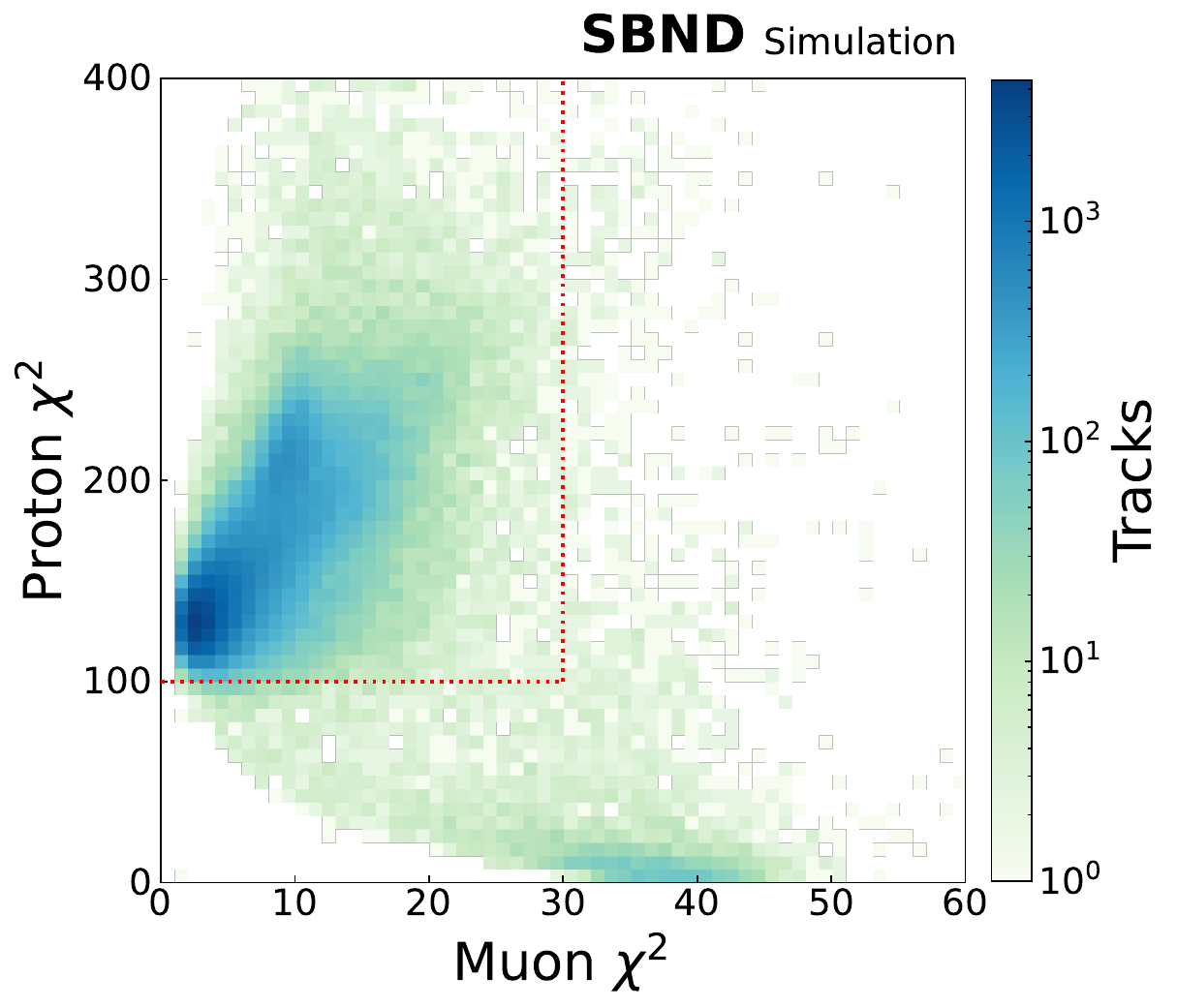} &
  \includegraphics[width=0.4\textwidth]{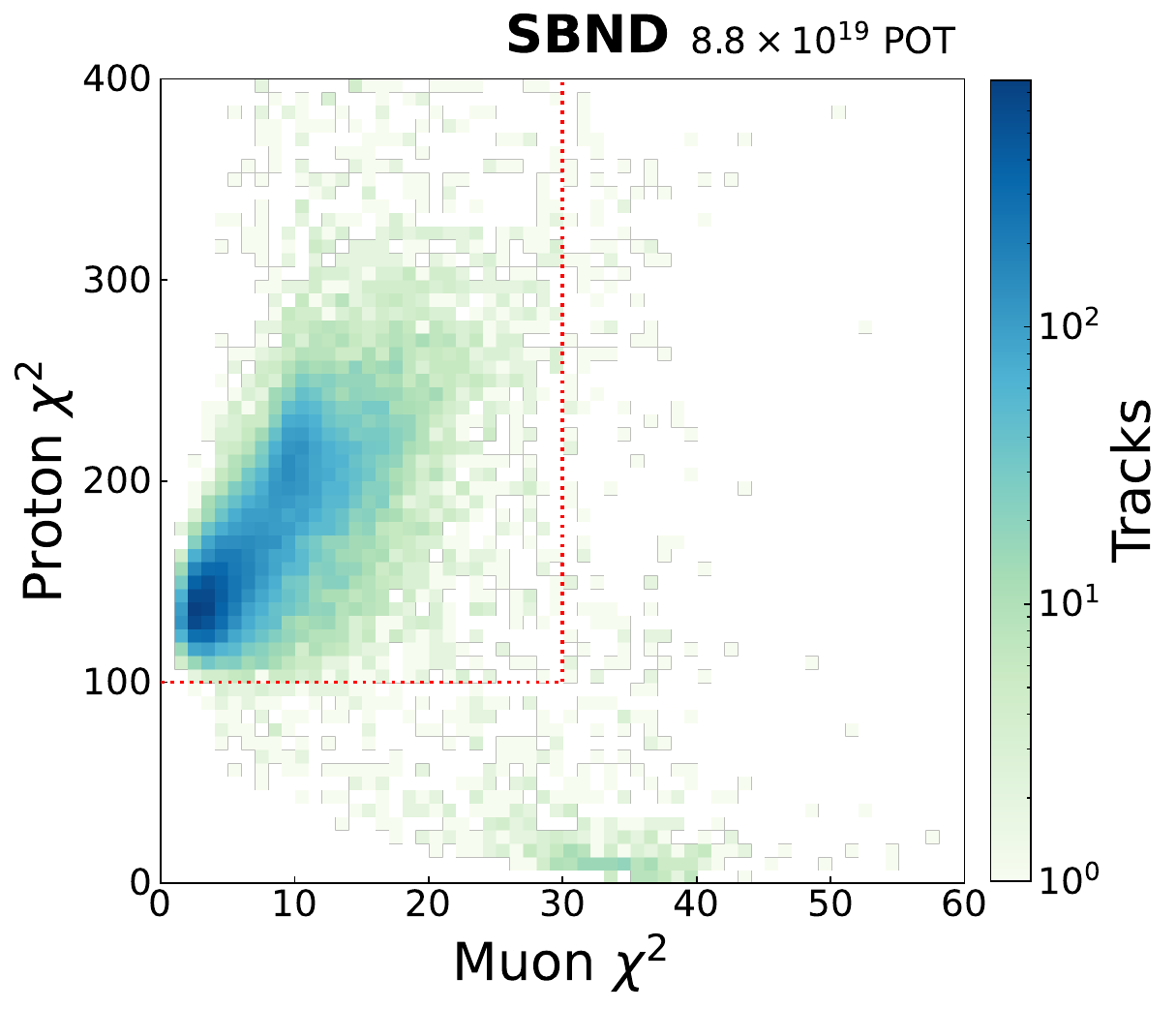} \\
  (a) & (b) \\
  \includegraphics[width=0.4\textwidth]{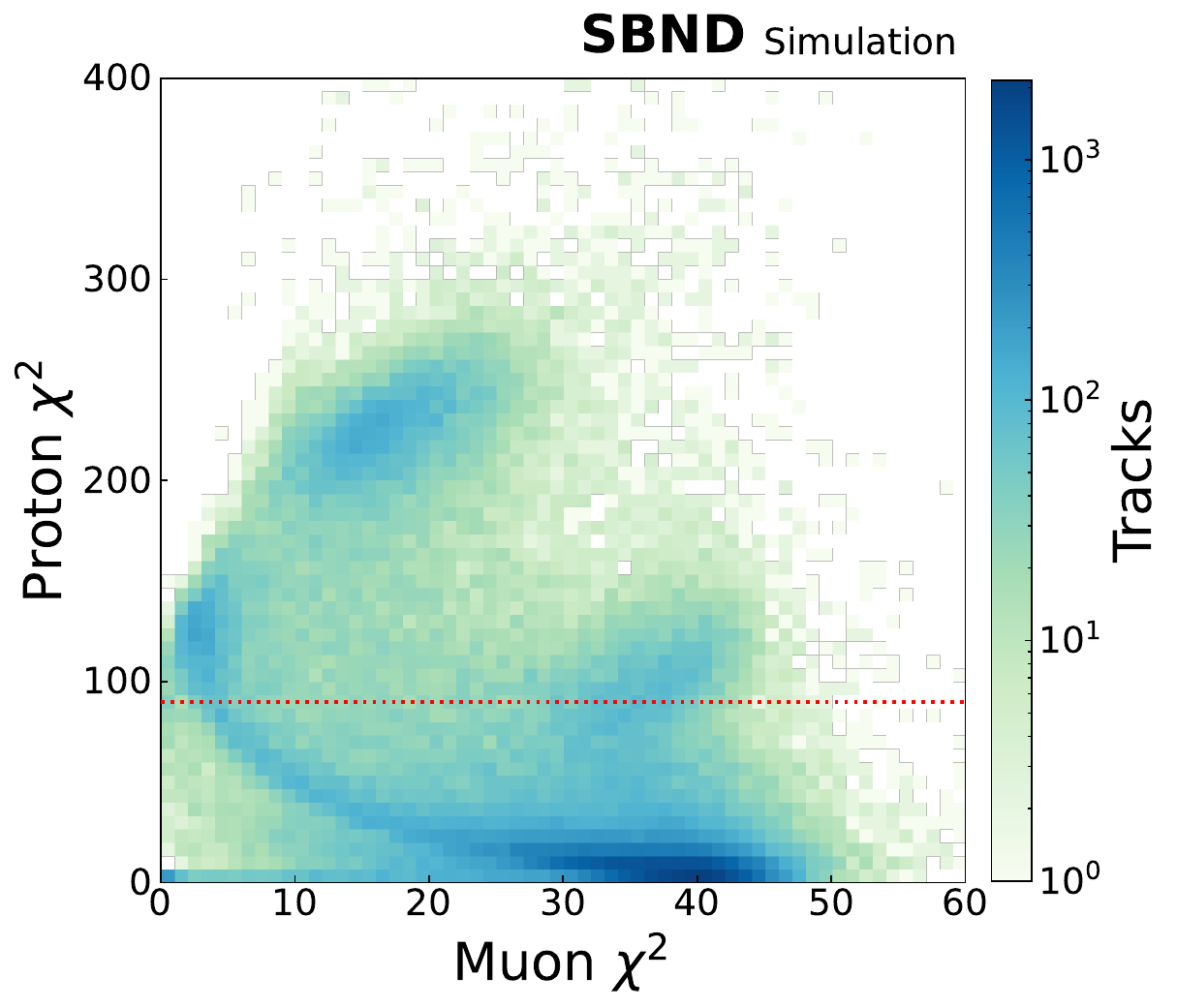} &
  \includegraphics[width=0.4\textwidth]{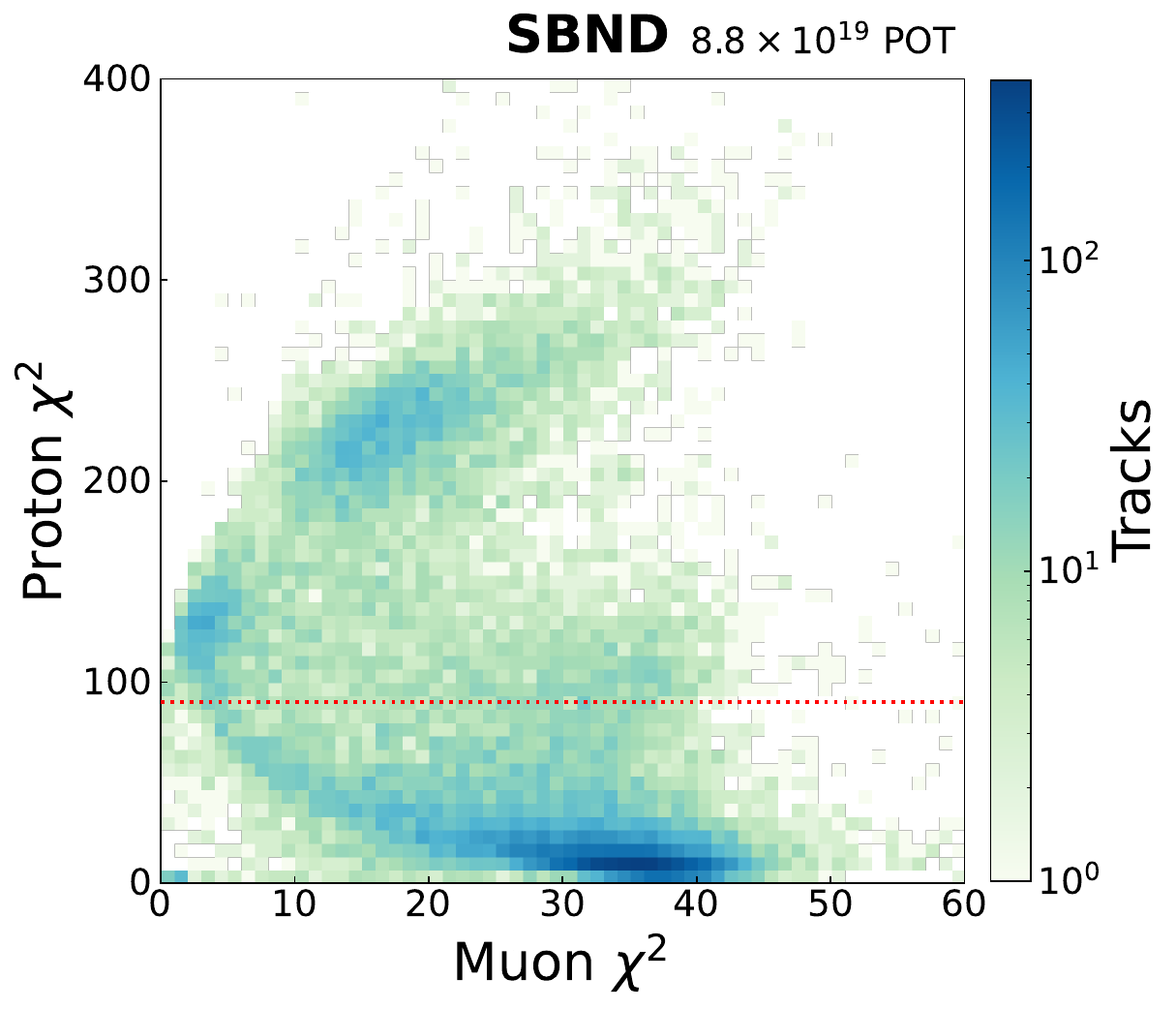} \\
  (c) & (d) \\
  \end{tabular}
  \caption{Plane-averaged muon and proton $\chi^2$ distributions. (a)~Simulation and (b)~data for tracks that pass the muon quality cuts. (c)~Simulation and (d)~data for tracks that are not selected as muon candidates. The dotted line marks the cuts for the muon-candidate and proton-candidate requirements.}
  \label{fig:pid-chi2}
\end{figure}

Figure~\ref{fig:pid-chi2} shows the plane-averaged muon and proton $\chi^2$ distributions for the track selections used in the particle identification for the MC and data. The muon (proton) PID efficiency is $81\%$ ($84\%$) with a purity of $94\%$ ($99\%$). 

\FloatBarrier
\newpage

\section{Selected Event Distribution in Final-State Particle Kinematics}
\label{sec:selected-event-distribution}

\begin{figure}[H]
\centering
\begin{tabular}{@{}ccc@{}}
\includegraphics[width=0.4\textwidth]{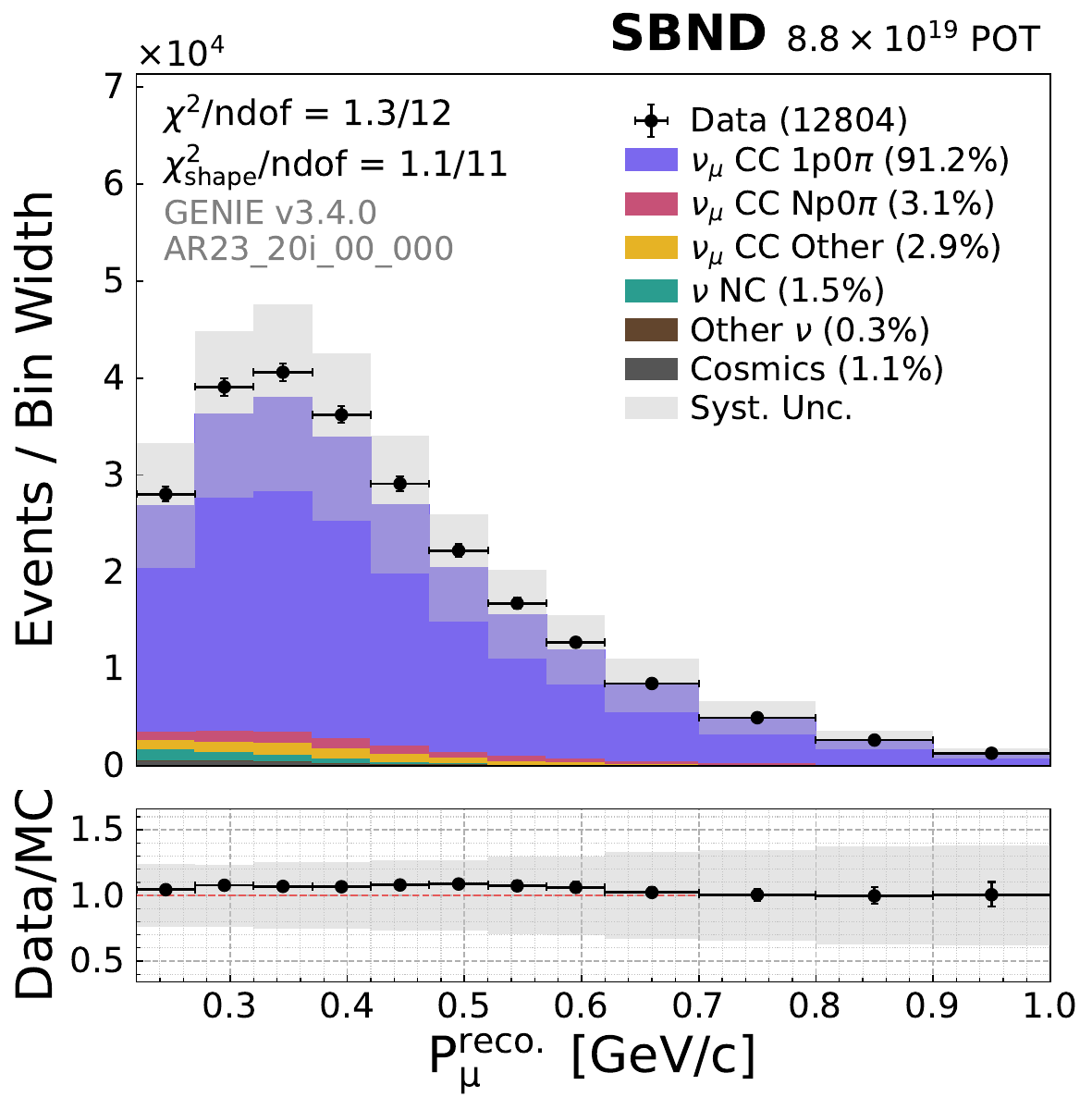} &
\includegraphics[width=0.404\textwidth]{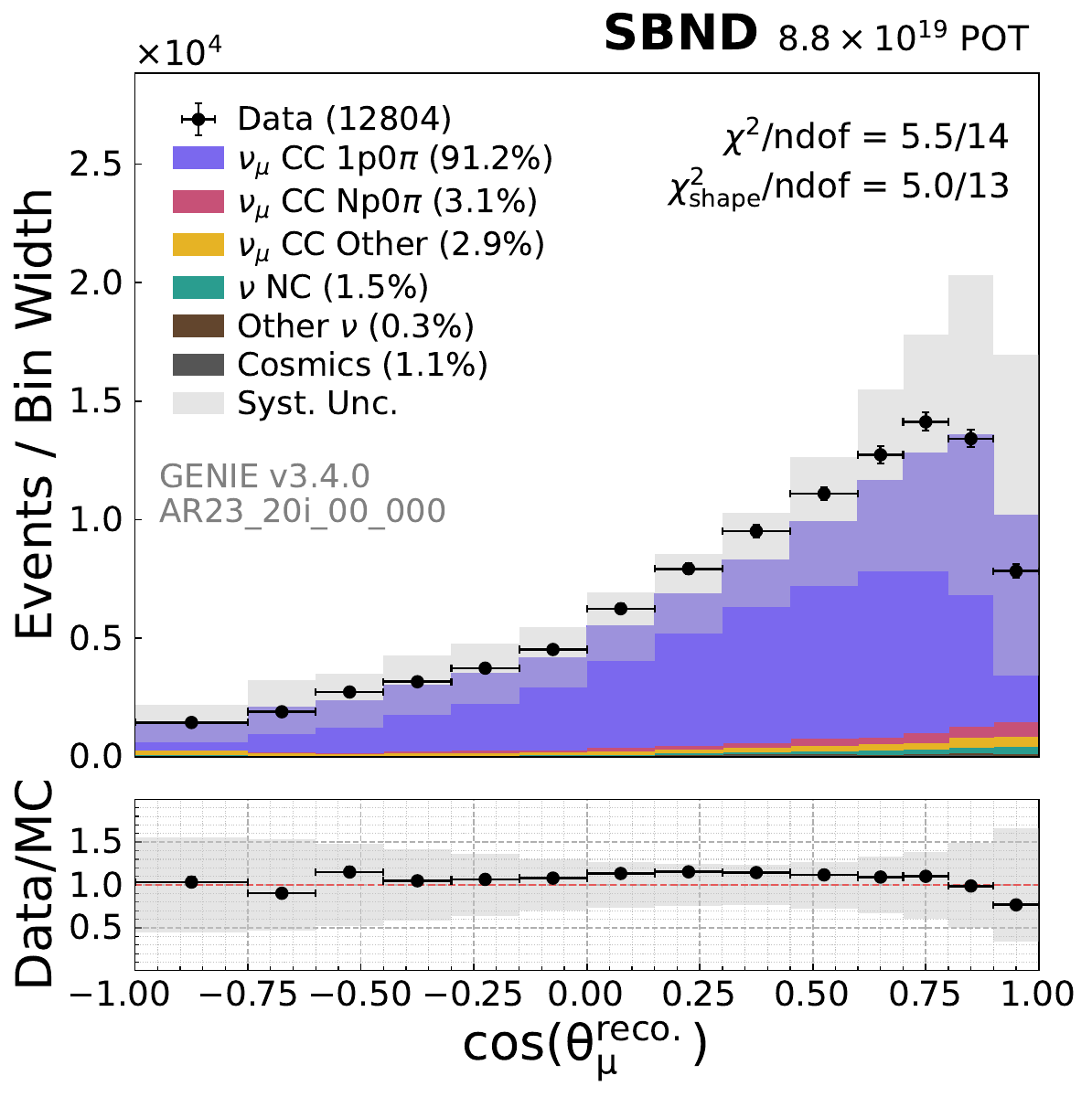} \\
(a) & (b) \\
\includegraphics[width=0.4\textwidth]{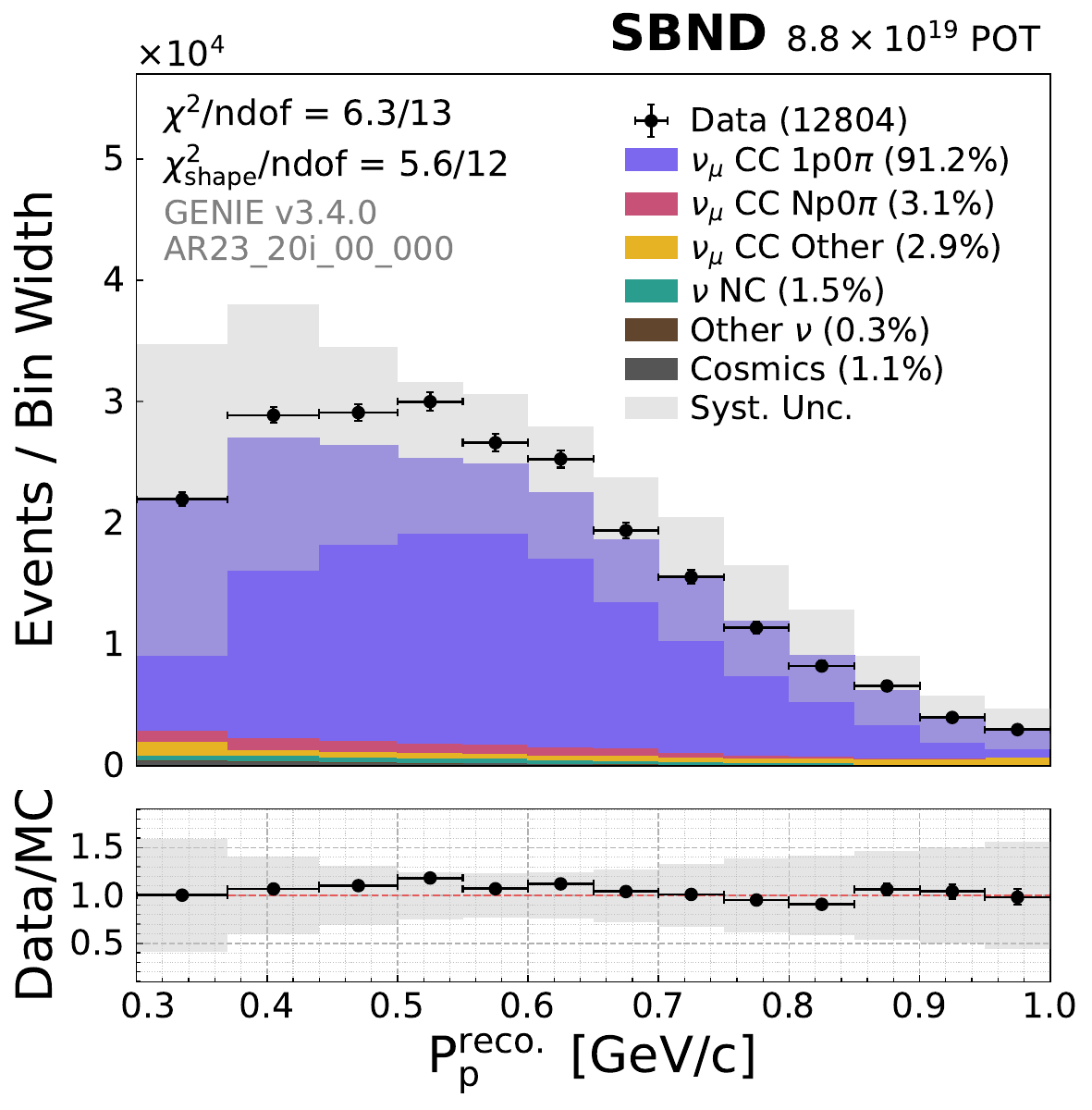} &
\includegraphics[width=0.412\textwidth]{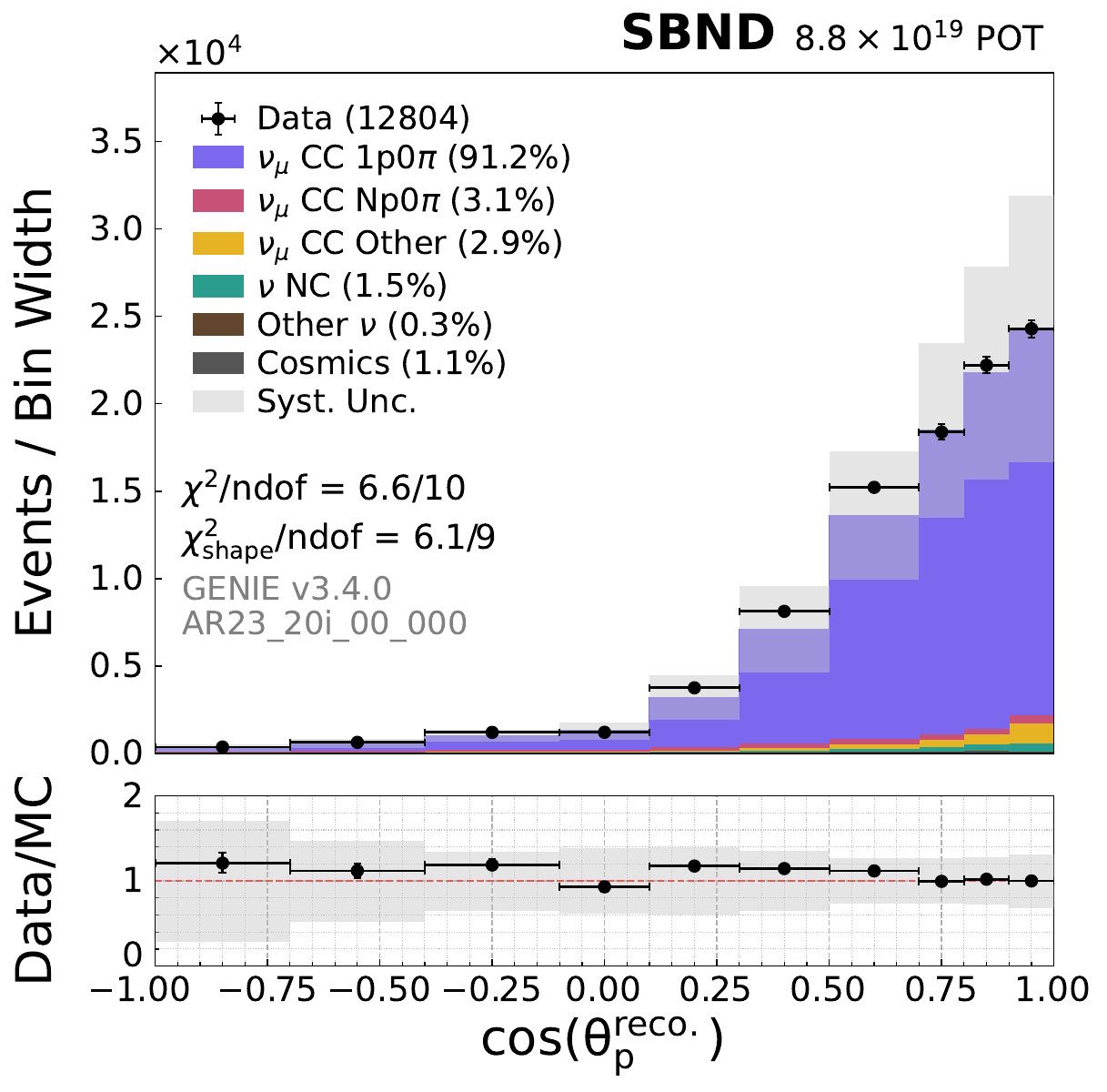} \\
(c) & (d) \\
\end{tabular}
\caption{
Selected data and Monte Carlo (MC) simulation event rates as functions of (a)~muon momentum, (b)~muon direction, (c)~proton momentum, and (d)~proton direction.
Stacked histograms show the MC prediction by event category, and the shaded band shows the total systematic uncertainty.
The $\nu_\mu$ CC events are subdivided by the final-state topology. 
``Other~$\nu$'' includes $\bar{\nu}_\mu$ and $\nu_e$ events.
}
\label{fig:sel_mup}
\end{figure}

Figure~\ref{fig:sel_mup} shows the selected data and Monte Carlo (MC) simulation event rates as functions of muon momentum, muon direction, proton momentum, and proton direction. 

\FloatBarrier
\newpage
\section{Selection efficiency and response matrices}
\label{sec:response}

\begin{figure}[H]
\centering
\begin{tabular}{@{}cc@{}}
\includegraphics[width=0.32\textwidth]{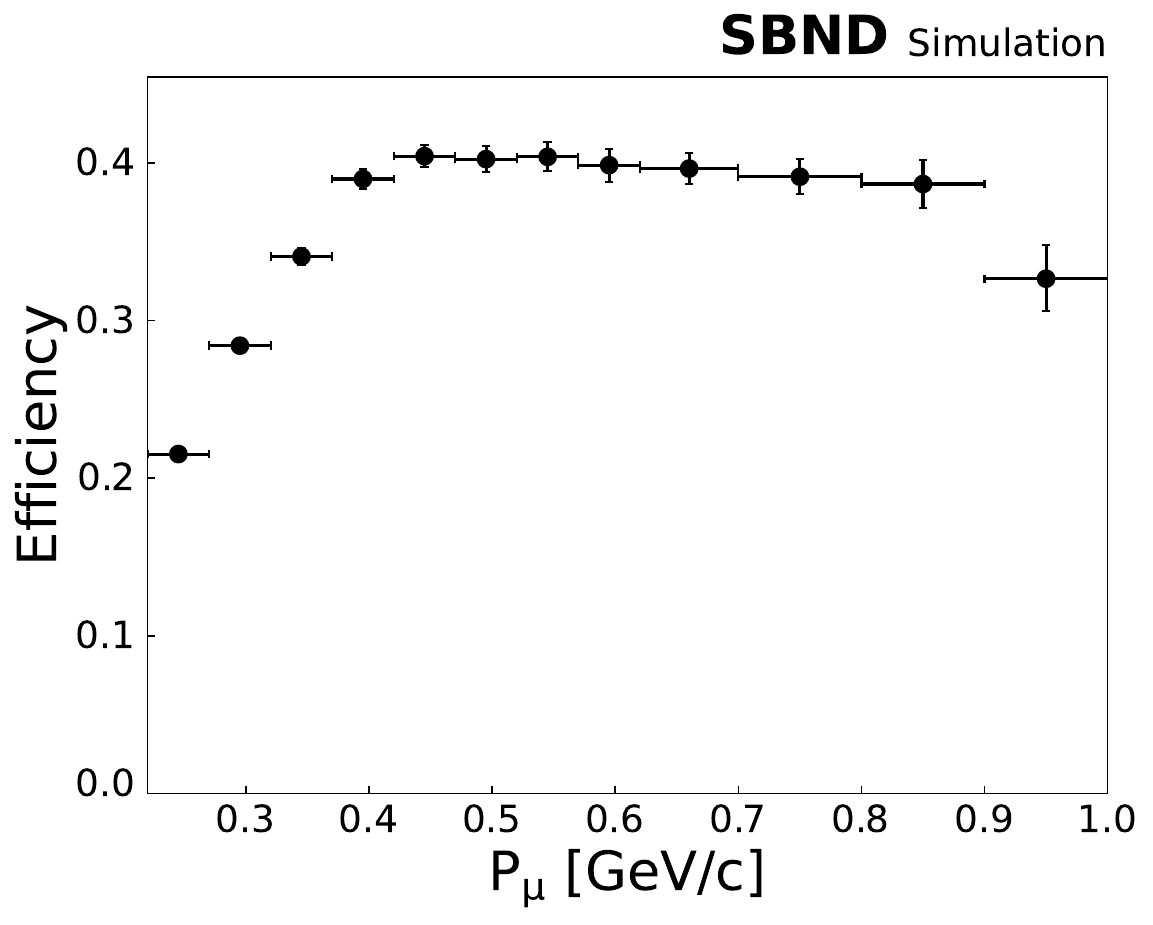} &
\includegraphics[width=0.33\textwidth]{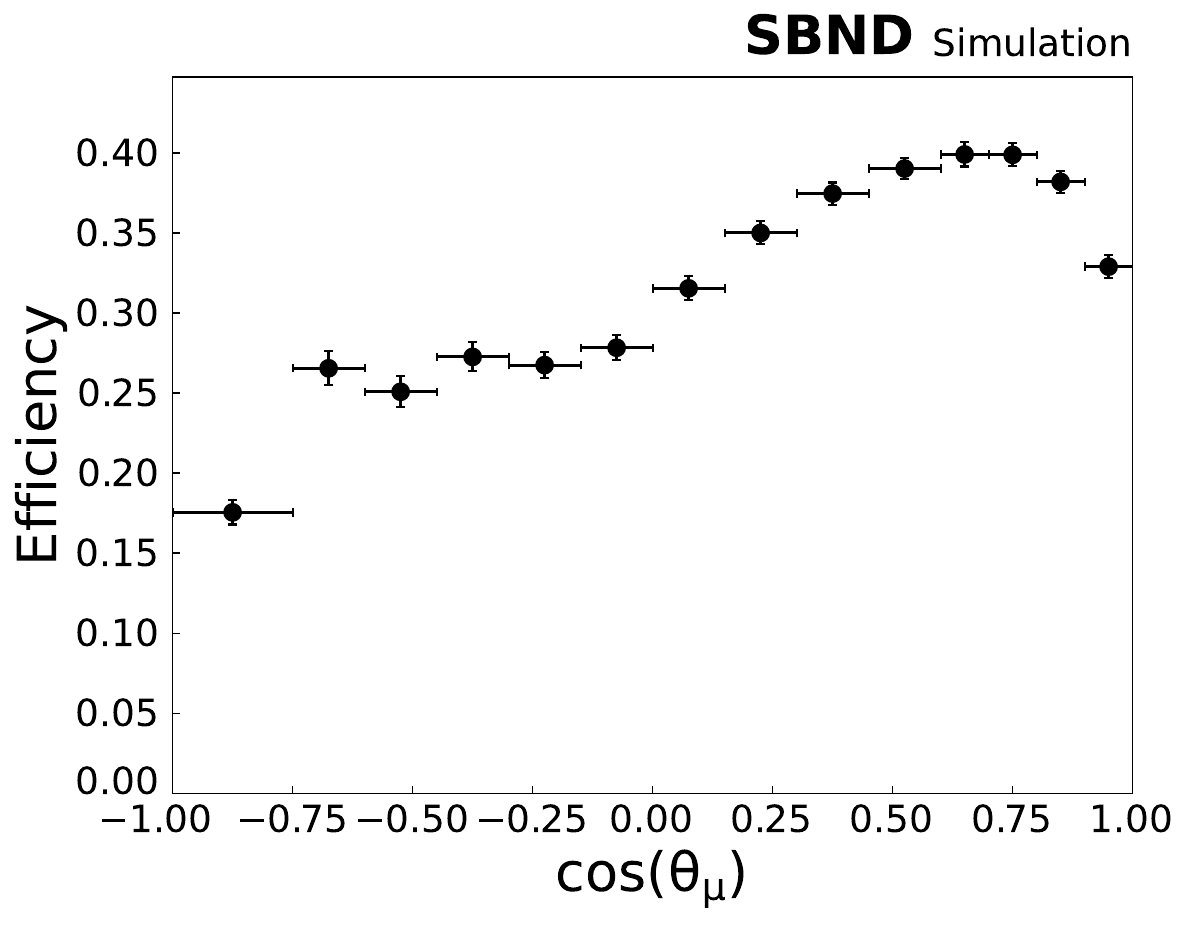}  \\
(a) & (b) \\
\includegraphics[width=0.32\textwidth]{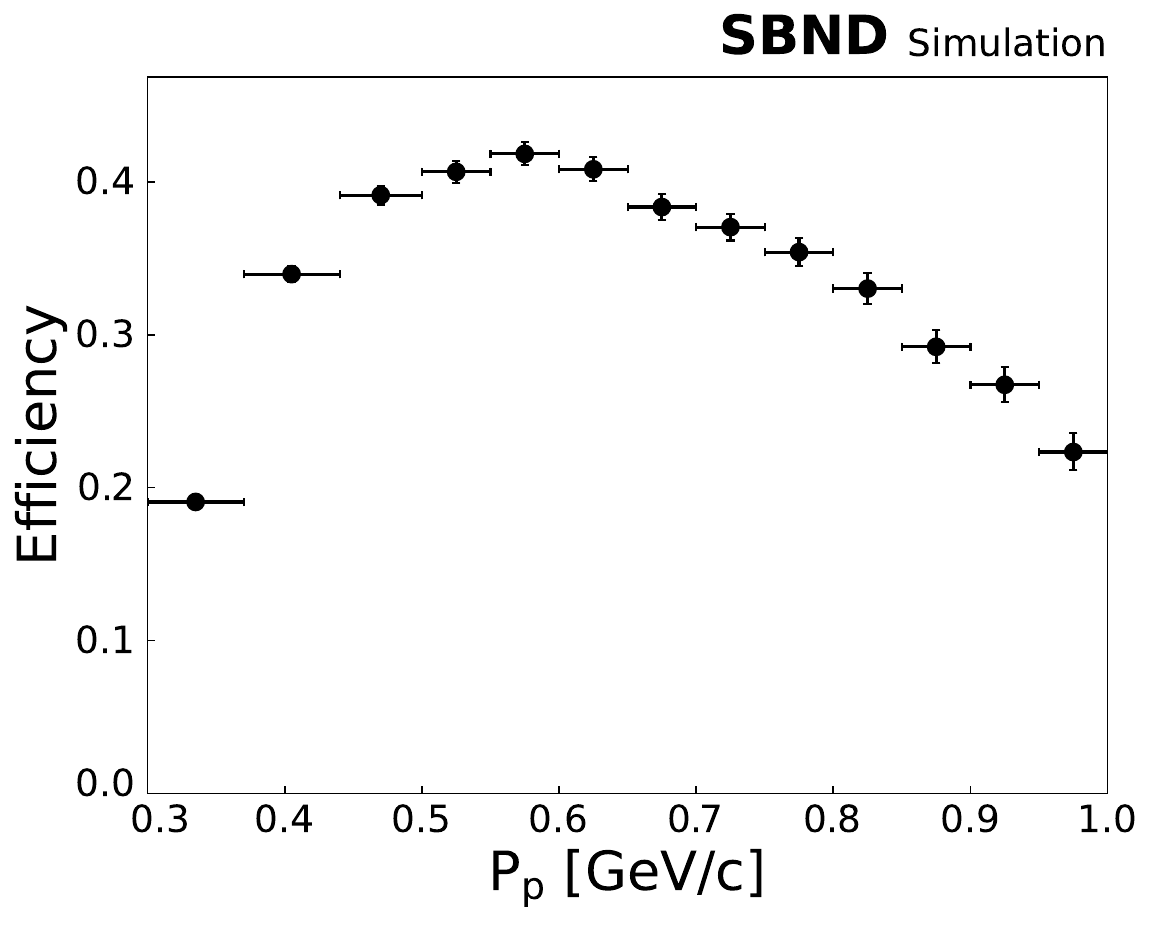} &
\includegraphics[width=0.33\textwidth]{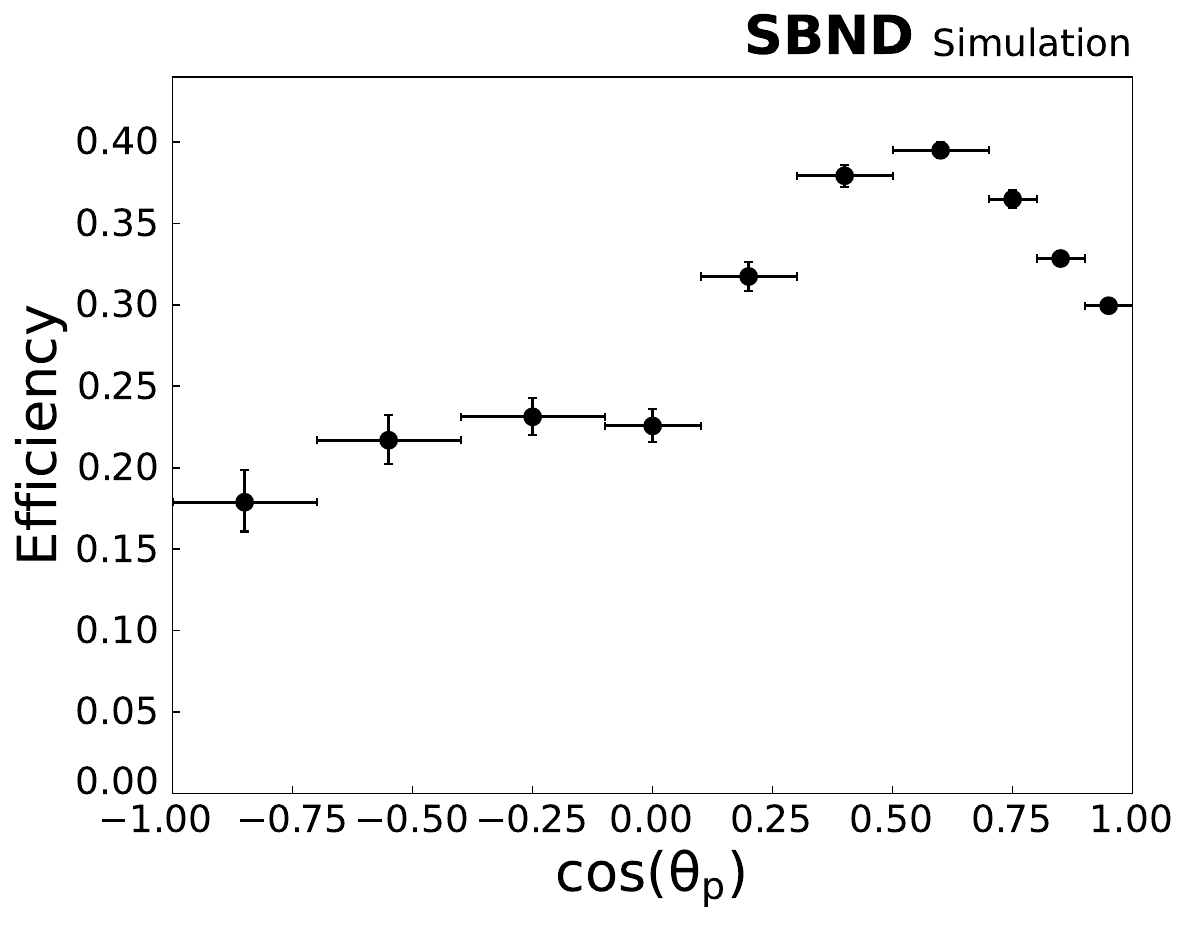} \\
(c) & (d) \\
\end{tabular}
\caption{Selection efficiency as a function of (a)~muon momentum, (b)~muon direction, (c)~proton momentum, and (d)~proton direction.}
\label{fig:efficiency-muon-proton}
\end{figure}

\begin{figure}[H]
\centering
\begin{tabular}{@{}ccc@{}}
\includegraphics[width=0.326\textwidth]{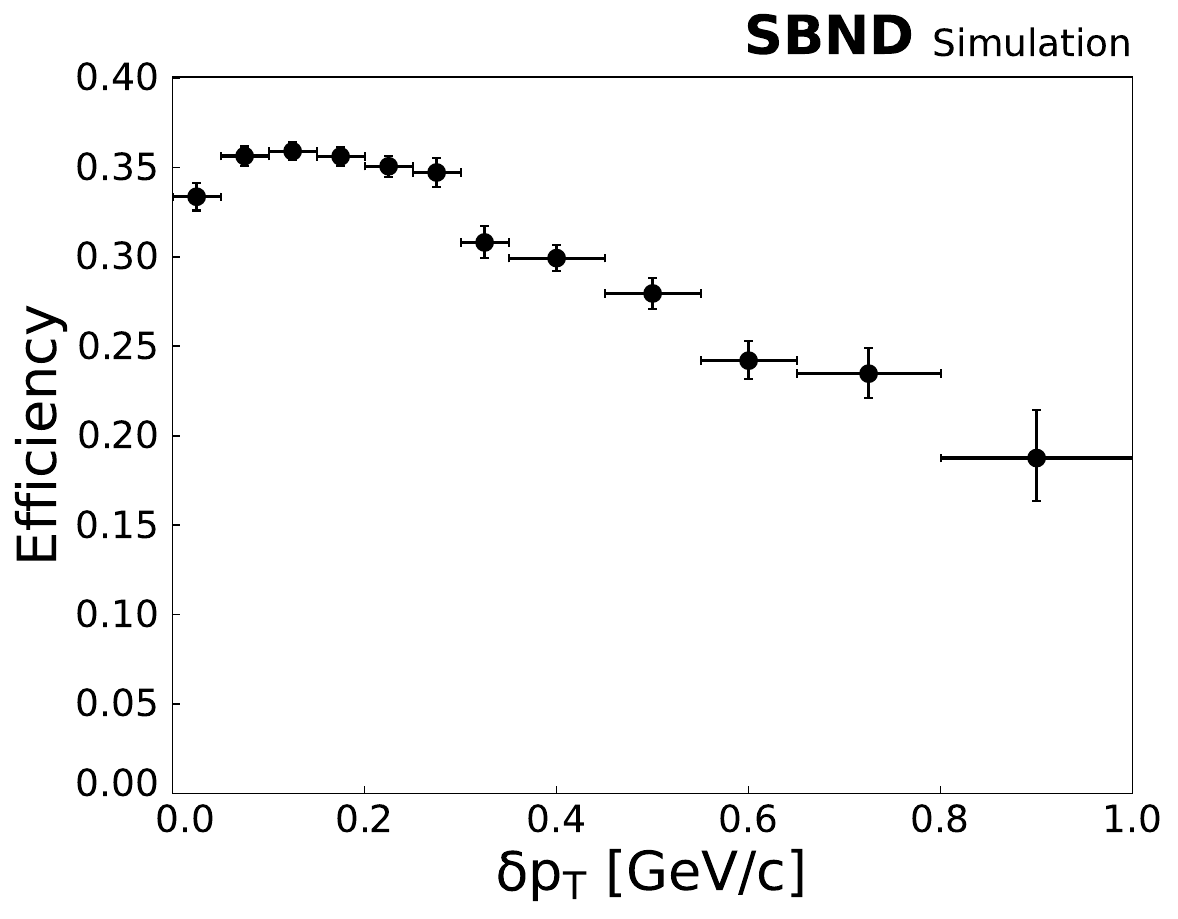} &
\includegraphics[width=0.32\textwidth]{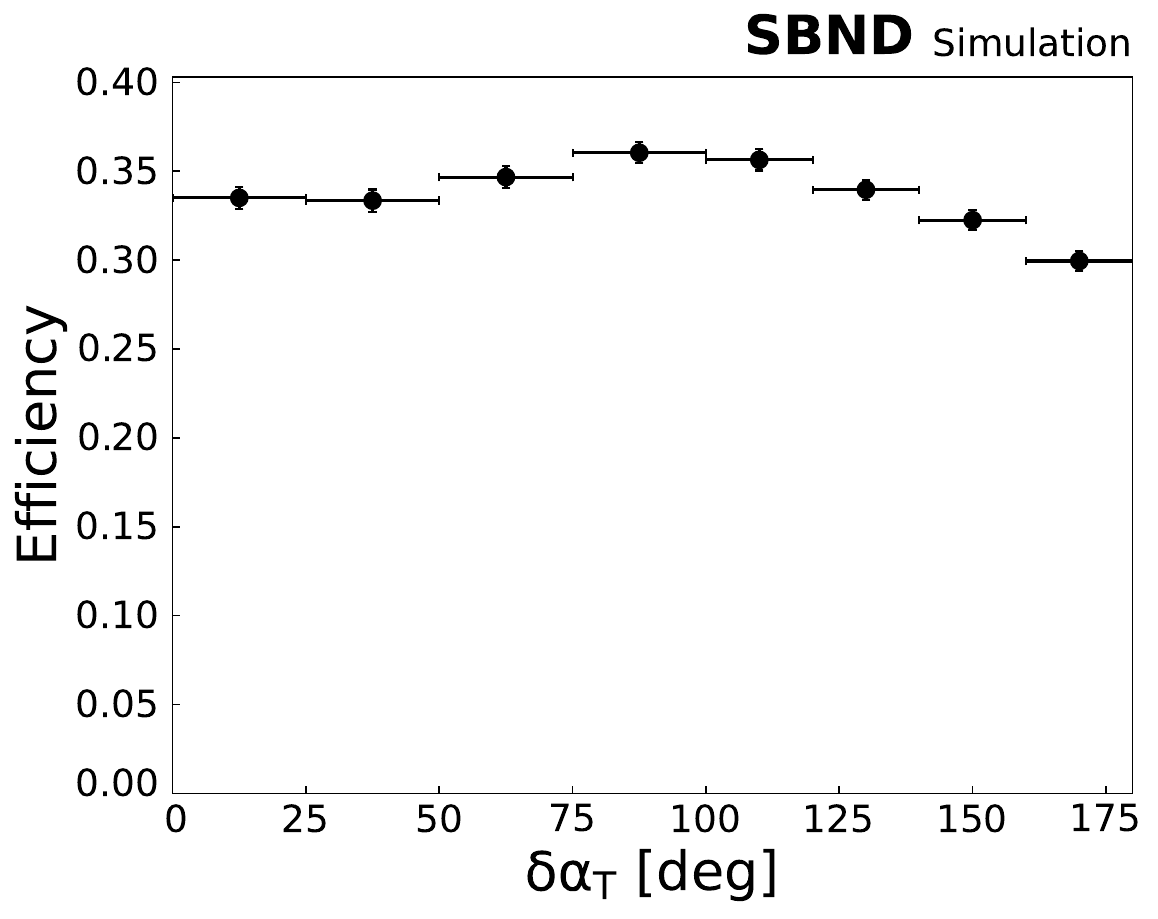} &
\includegraphics[width=0.32\textwidth]{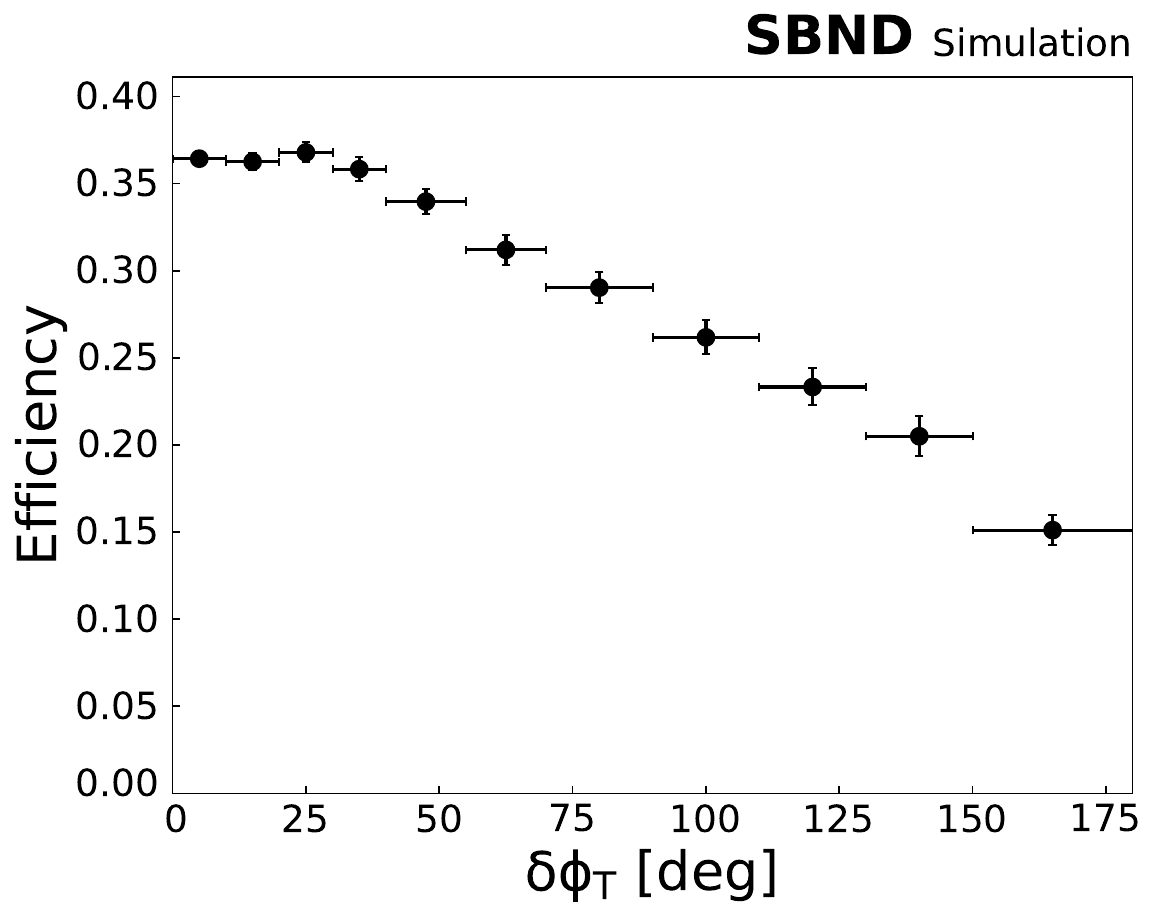} \\
(a) & (b) & (c) \\
\end{tabular}
\caption{Selection efficiency as a function of (a)~$\delta p_T$, (b)~$\delta\alpha_T$, and (c)~$\delta\phi_T$.}
\label{fig:efficiency-tki}
\end{figure}

The selection efficiency in truth bin $i$ is
\begin{equation}
  \varepsilon_{\mathrm{sel},i}
  = \frac{N^{\mathrm{sel}}_{\mathrm{cont},i}}{N^{\mathrm{gen}}_{\mathrm{cont},i}}\,,
\end{equation}
where the numerator and denominator are, respectively, the numbers of selected and generated true-signal events in truth bin~$i$ that satisfy the containment requirement.
Figures~\ref{fig:efficiency-muon-proton} and~\ref{fig:efficiency-tki} show the efficiency of the event selection as functions of muon and proton kinematics and TKI variables, respectively. 

The full detector response, including the acceptance and bin-to-bin migration from reconstruction effects, is encoded as the response matrix: 
\begin{equation} 
  R_{ji} = \frac{N^{\mathrm{sel}}_{\mathrm{cont}, \left(x^{\mathrm{true}}\in i, x^{\mathrm{reco}}\in j\right)}}{N^{\mathrm{gen}}_{i}},
\end{equation}
where $N^{\mathrm{sel}}_{\mathrm{cont}, \left(x^{\mathrm{true}}\in i, x^{\mathrm{reco}}\in j\right)}$ is the number of selected contained signal events with true kinematics in bin~$i$ and reconstructed kinematics in bin~$j$, and $N^{\mathrm{gen}}_{i}$ is the number of generated signal events in bin~$i$ with the neutrino vertex in the fiducial volume but without the containment requirement. The selected event distribution is unfolded using this response. 

The variations of response across the differential bins (e.g. low selection efficiency for backward-going tracks and low acceptance efficiency at high muon momentum) are modeled from simulation. As explained in the main text, we perform fake data tests with the alternate generator \texttt{GiBUU} to assess the robustness to the model dependence of our MC, including acceptance and detector response effects.

\begin{figure}[H]
\centering
\begin{tabular}{@{}cc@{}}
\includegraphics[width=0.32\textwidth]{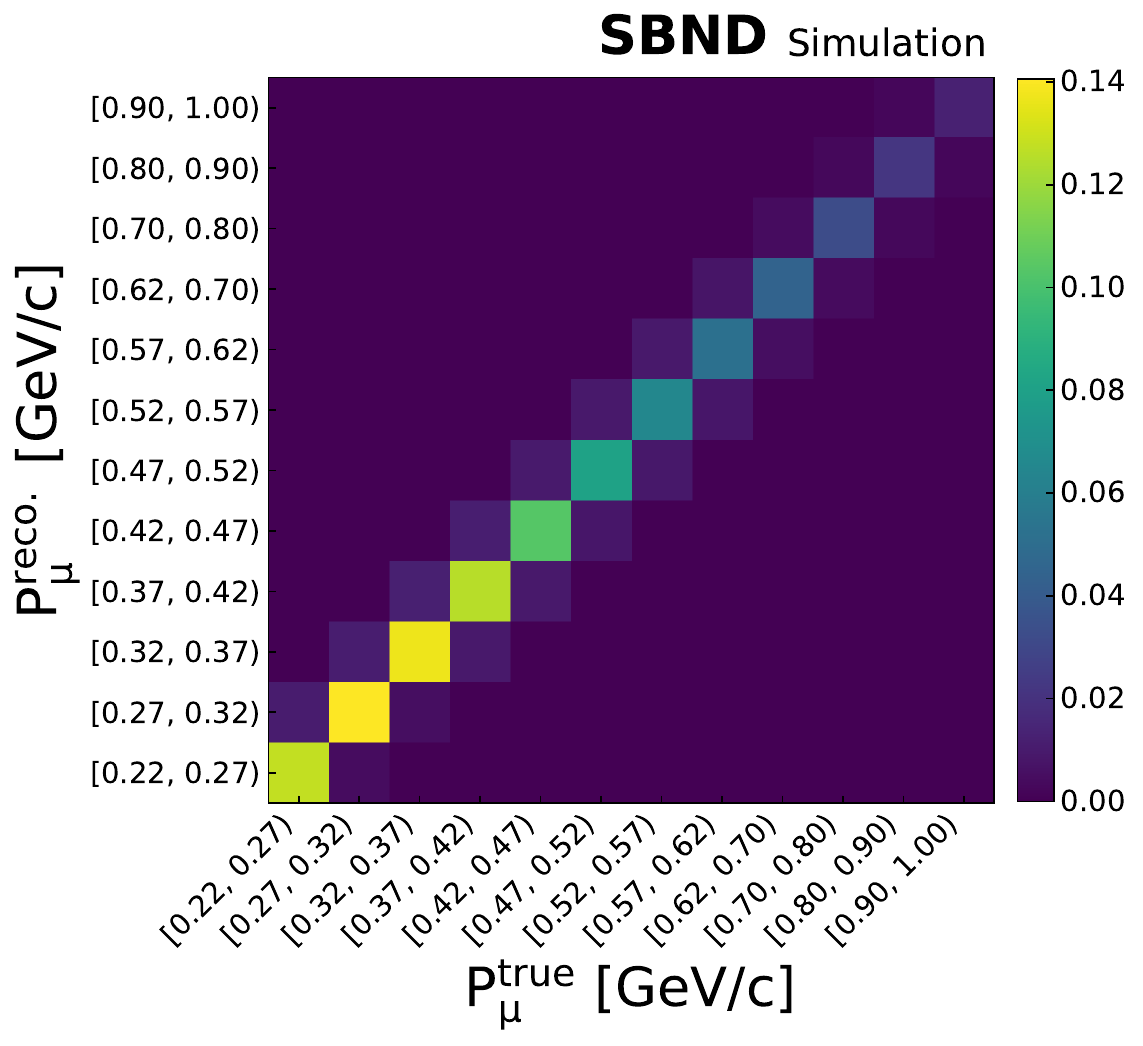} &
\includegraphics[width=0.32\textwidth]{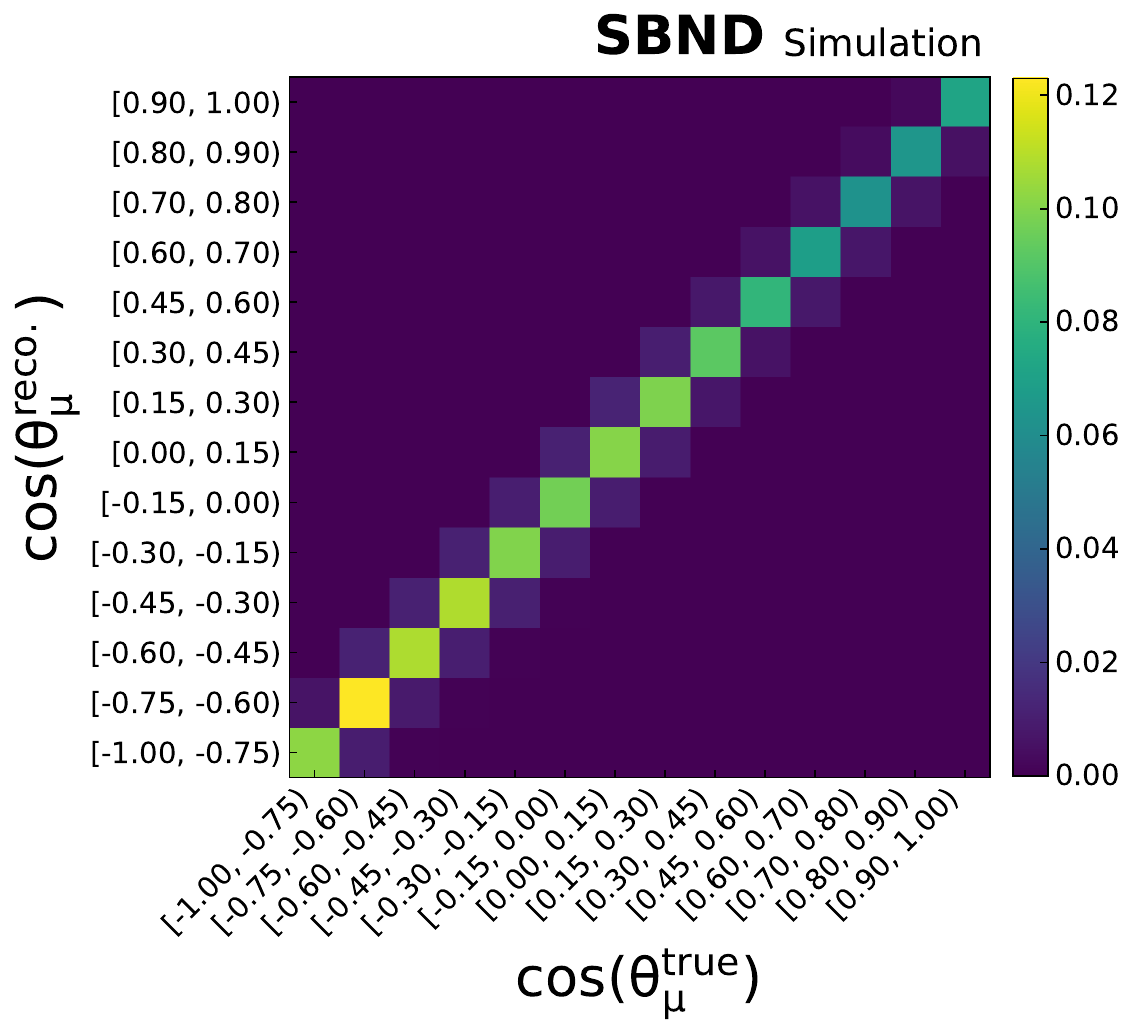}  \\
(a) & (b) \\
\includegraphics[width=0.32\textwidth]{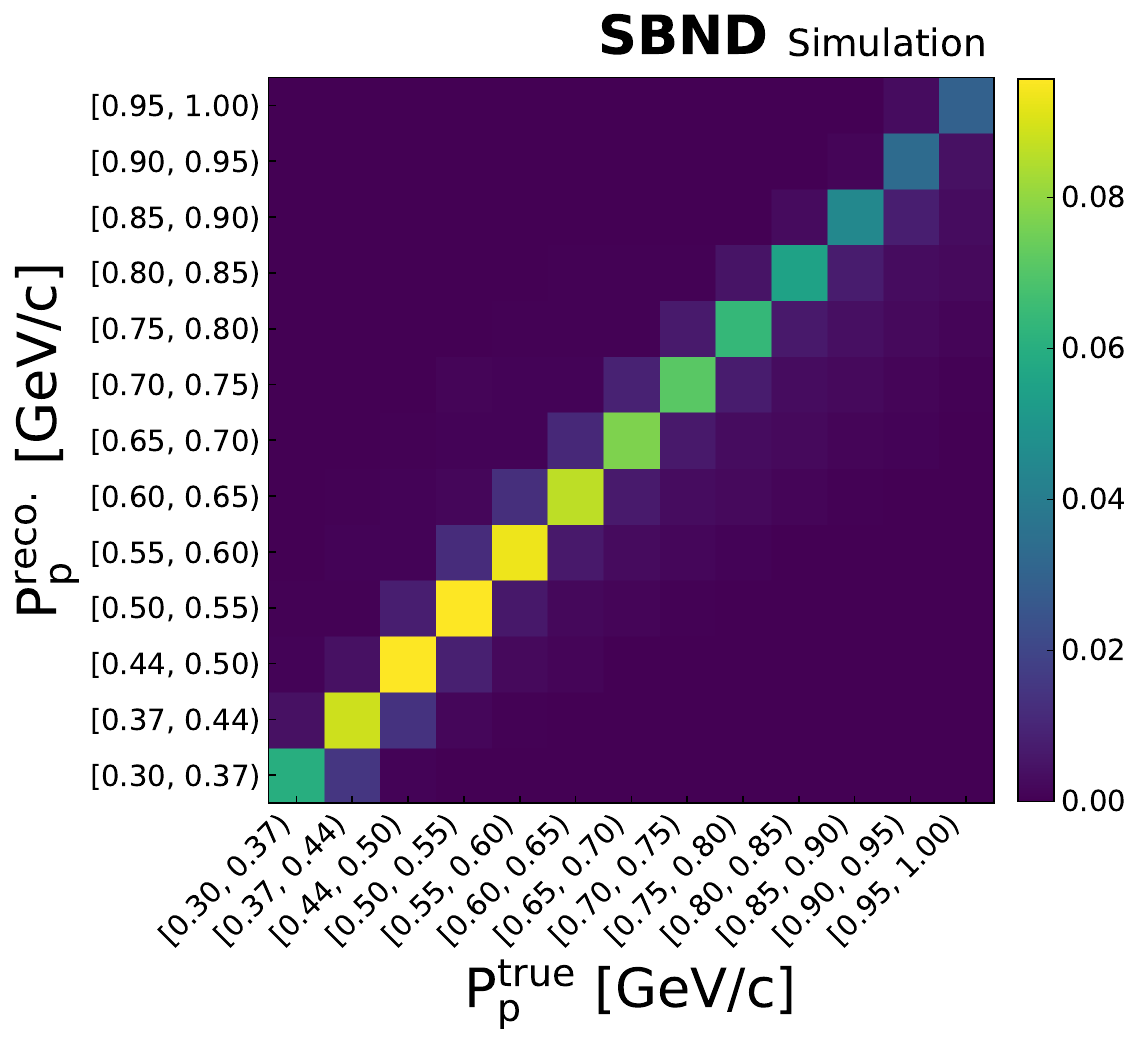} &
\includegraphics[width=0.32\textwidth]{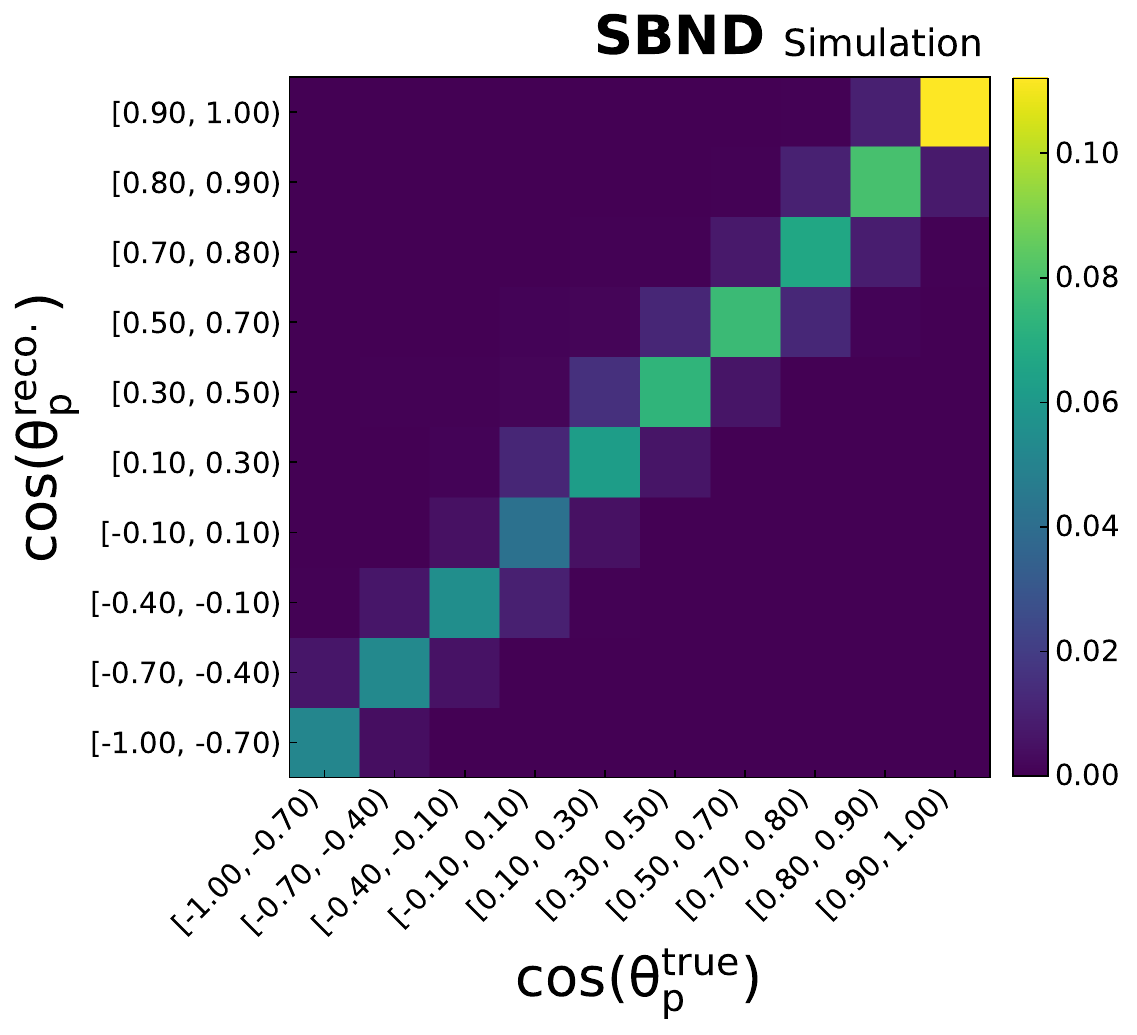} \\
(c) & (d) \\
\end{tabular}
\caption{Response matrices for all true signal events interacting in the FV, mapping true to reconstructed values for (a)~muon momentum, (b)~muon direction, (c)~proton momentum, and (d)~proton direction.}
\label{fig:response-muon-proton}
\end{figure}

\begin{figure}[H]
\centering
\begin{tabular}{@{}ccc@{}}
\includegraphics[width=0.32\textwidth]{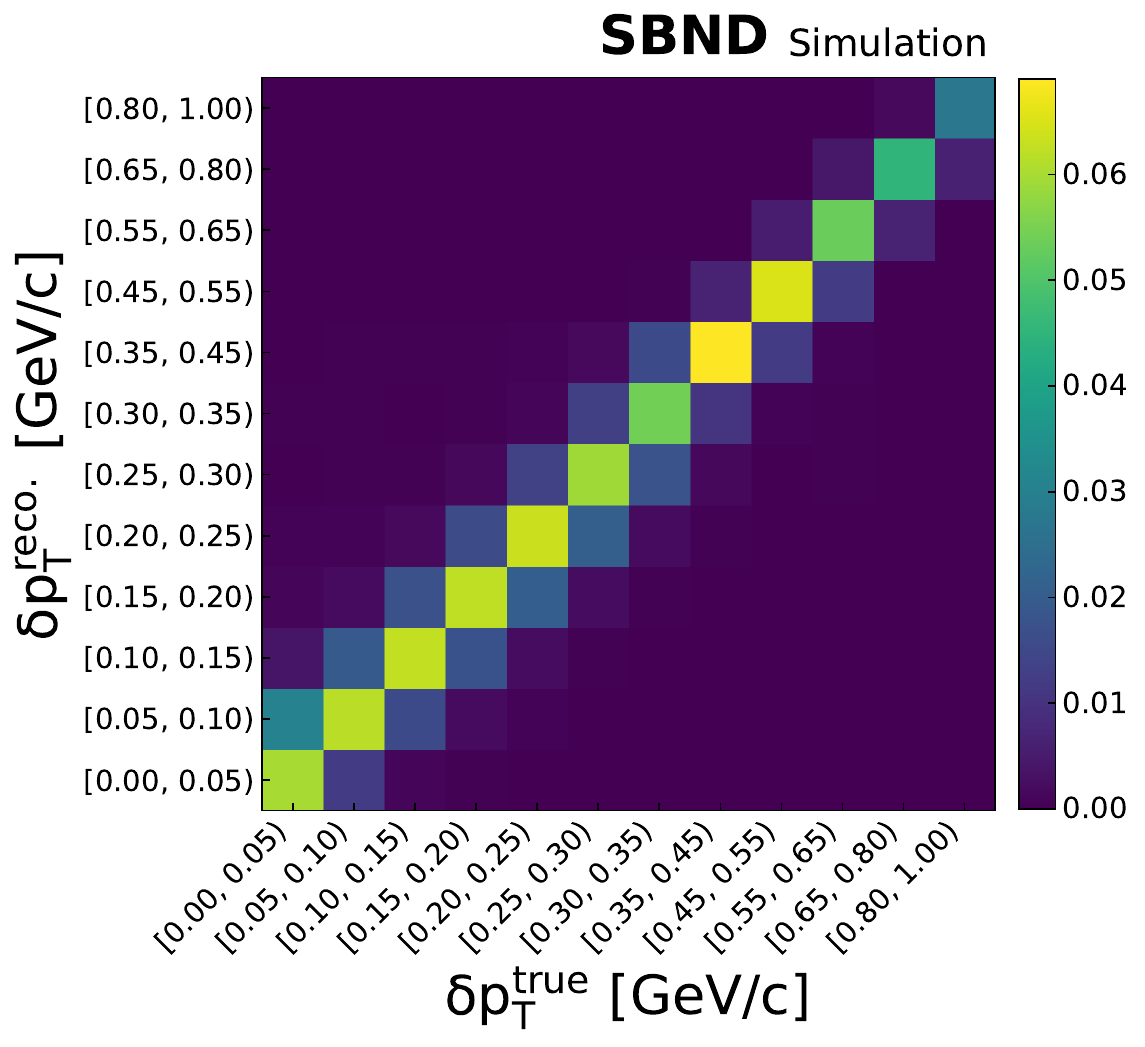} &
\includegraphics[width=0.32\textwidth]{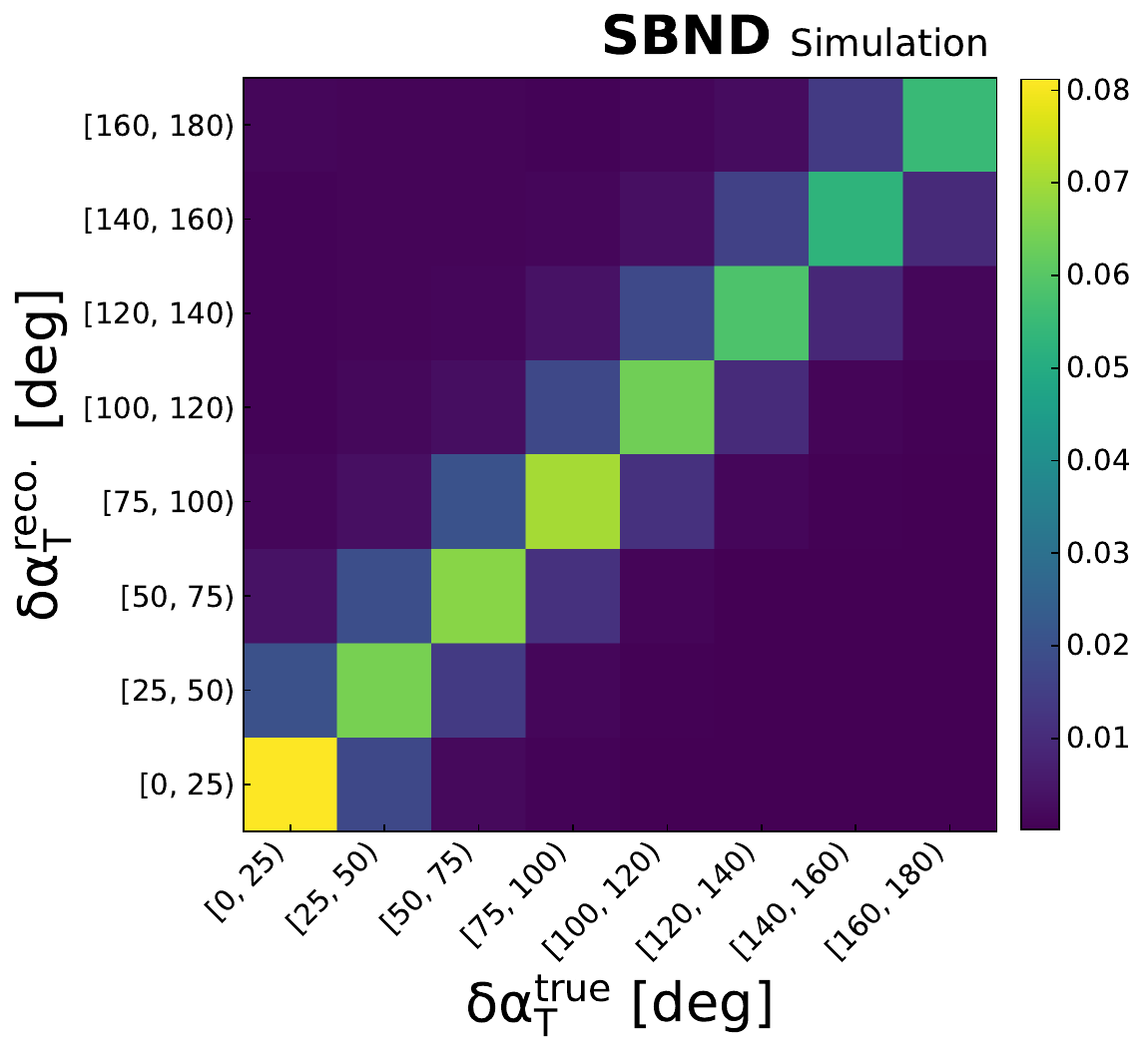} &
\includegraphics[width=0.32\textwidth]{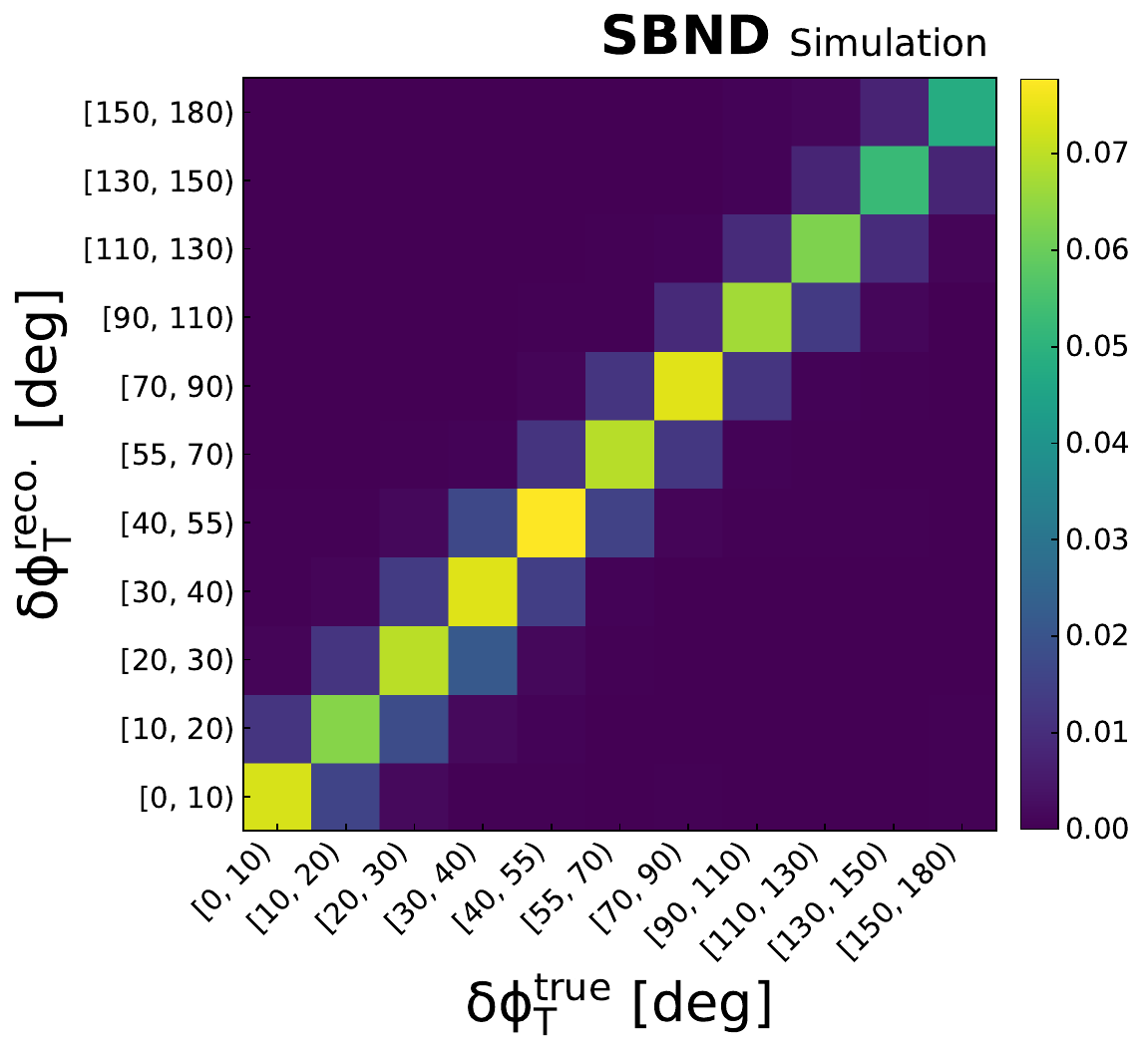} \\
(a) & (b) & (c) \\
\end{tabular}
\caption{Response matrices for all true signal events interacting in the FV, mapping true to reconstructed values for (a)~$\delta p_T$, (b)~$\delta\alpha_T$, and (c)~$\delta\phi_T$.}
\label{fig:response-tki}
\end{figure}

Figures~\ref{fig:response-muon-proton} and~\ref{fig:response-tki} show the corresponding response matrices for all true signal events interacting in the FV.

\FloatBarrier
\newpage
\section{Systematic uncertainties}
\label{sec:systematic}

\begin{figure}[H]
\centering
\begin{tabular}{@{}cc@{}}
\includegraphics[width=0.32\textwidth]{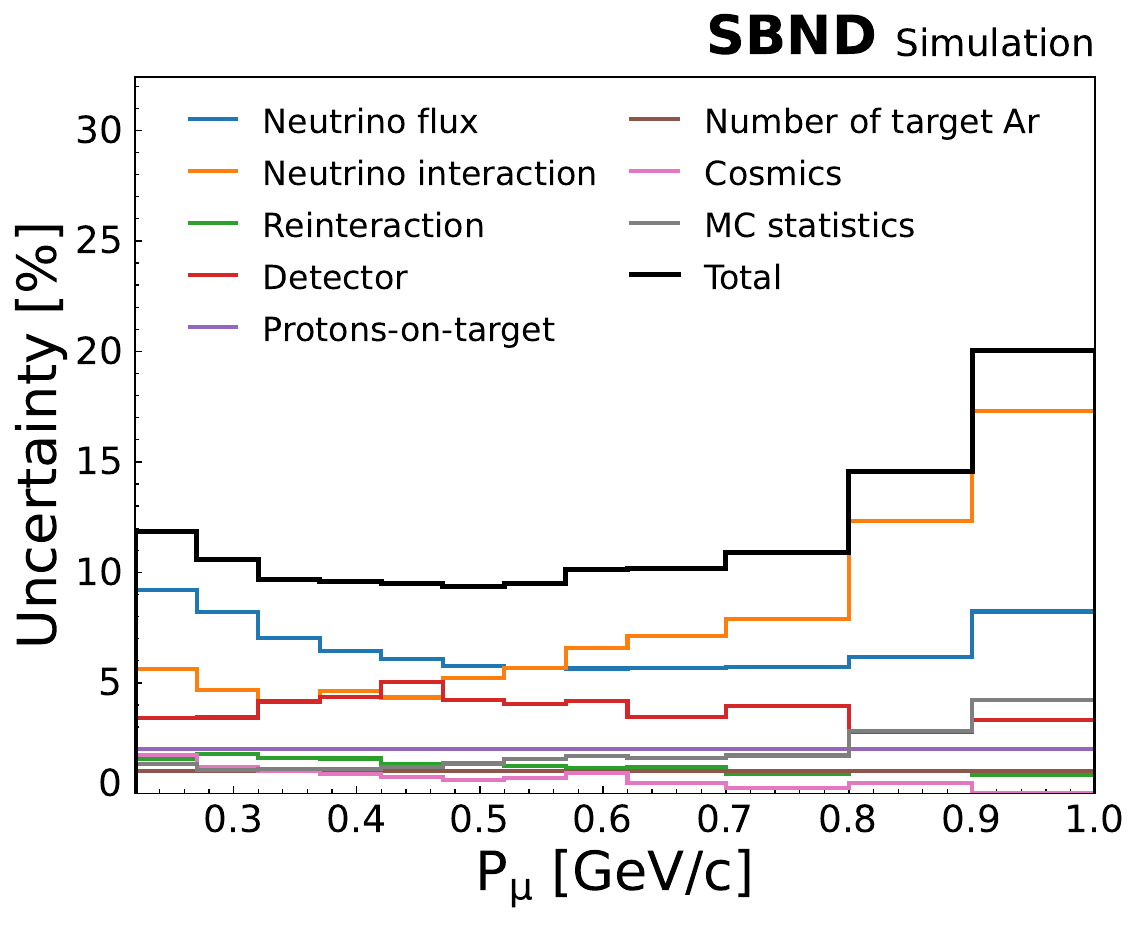} &
\includegraphics[width=0.323\textwidth]{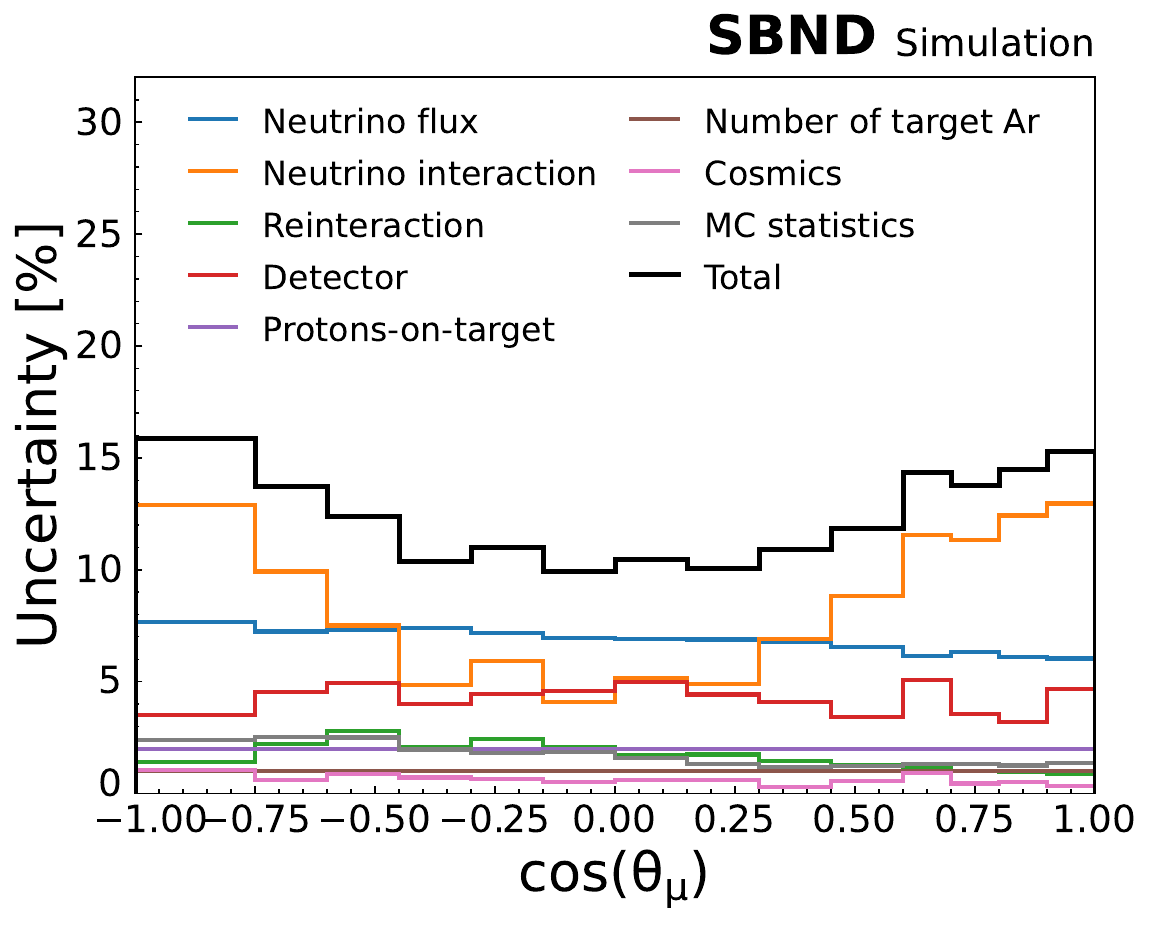}  \\
(a) & (b) \\
\includegraphics[width=0.32\textwidth]{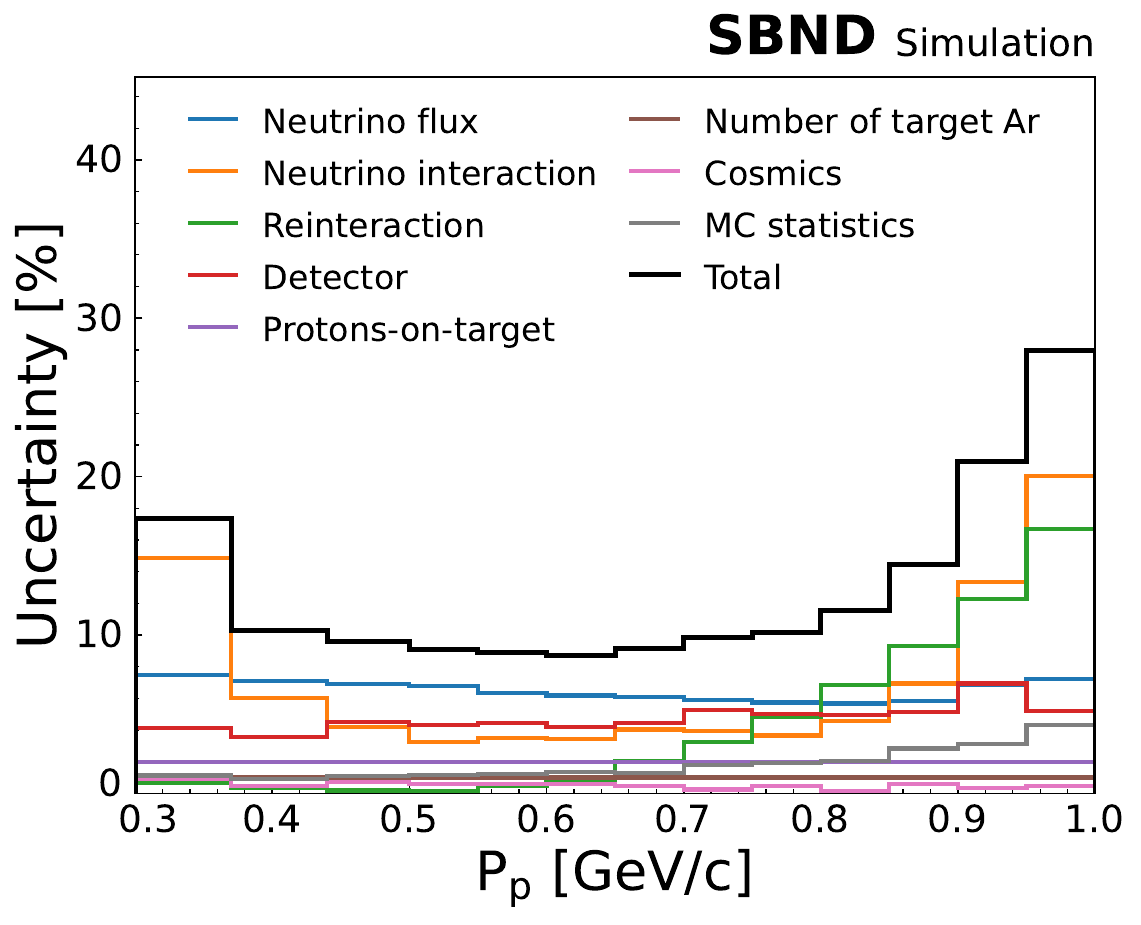} &
\includegraphics[width=0.323\textwidth]{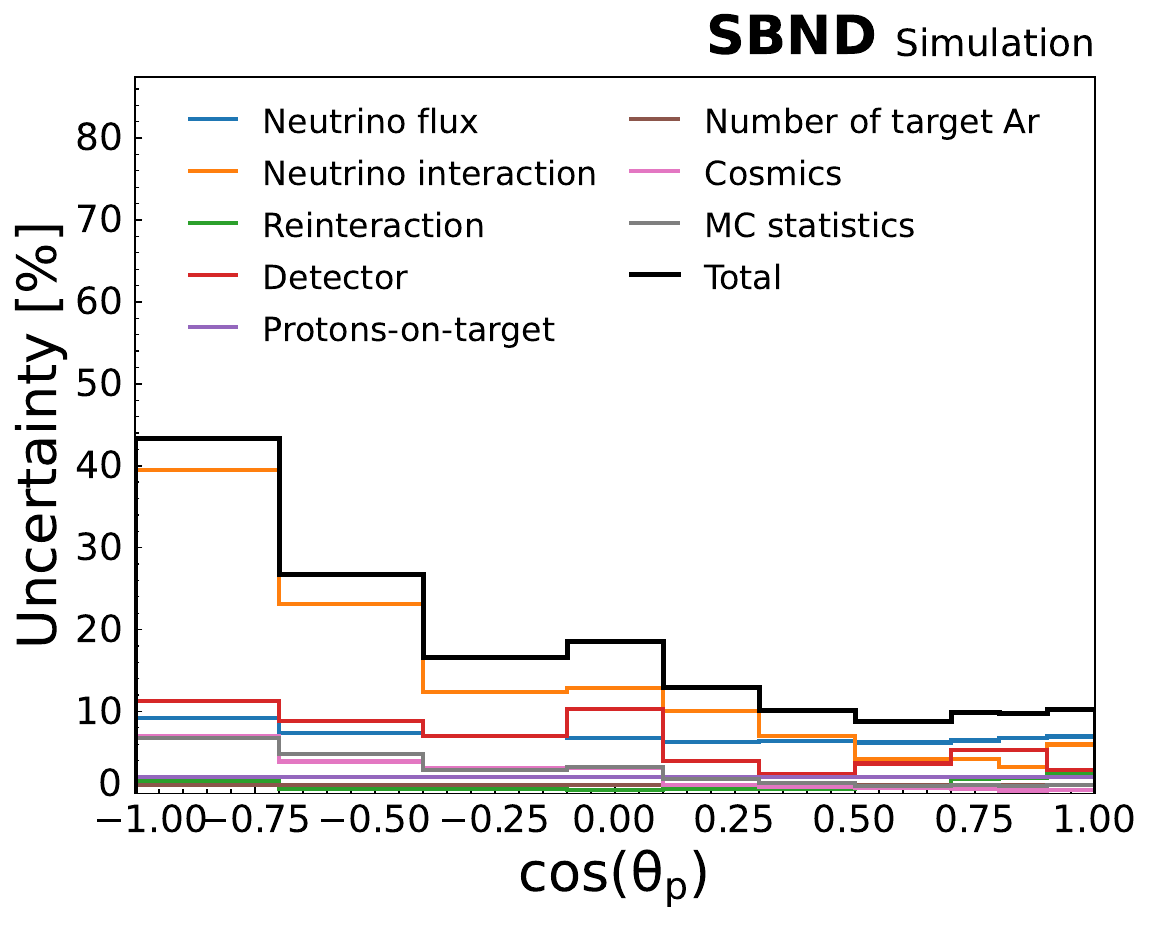} \\
(c) & (d) \\
\end{tabular}
\caption{Fractional uncertainties on the cross section-measurement as functions of (a)~muon momentum, (b)~muon direction, (c)~proton momentum, and (d)~proton direction, broken down by systematic source.}
\label{fig:systematic-muon-proton}
\end{figure}

\begin{figure}[H]
\centering
\begin{tabular}{@{}ccc@{}}
\includegraphics[width=0.32\textwidth]{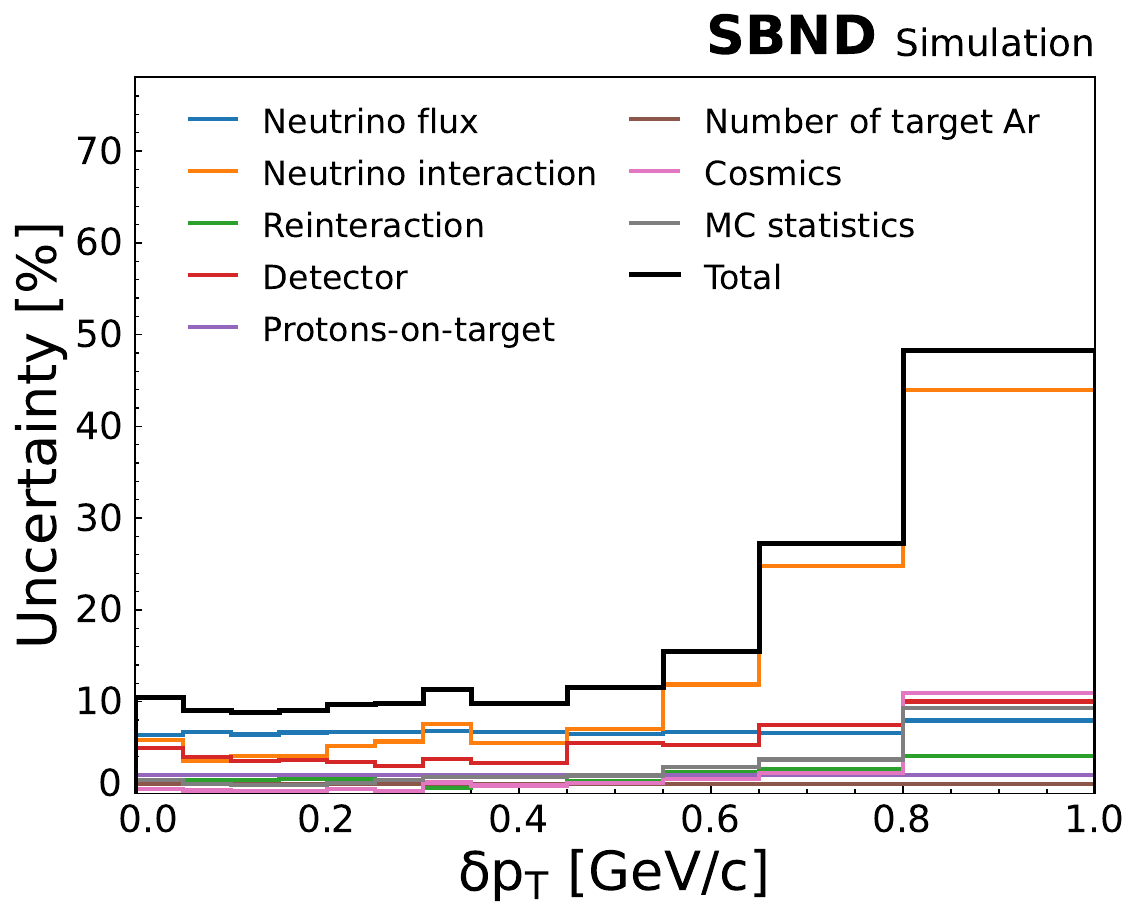} &
\includegraphics[width=0.325\textwidth]{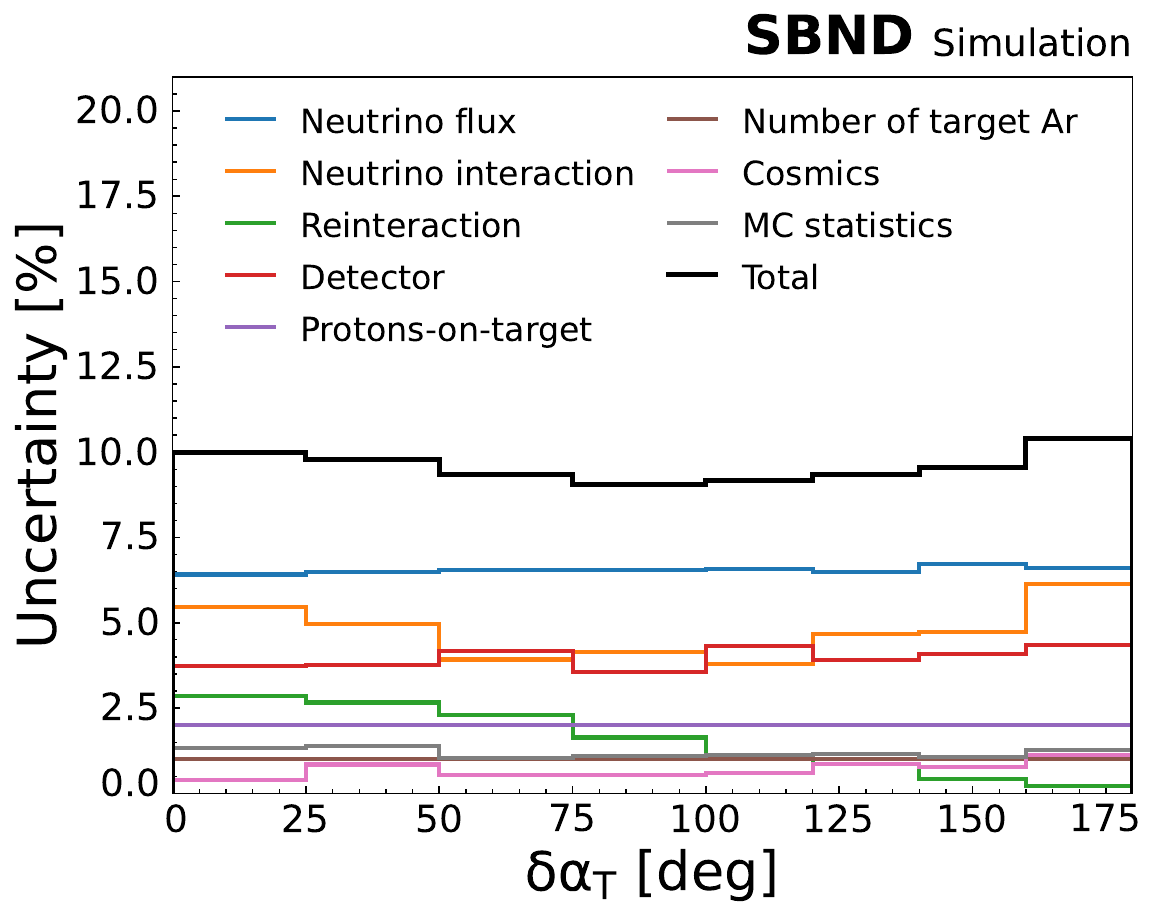} &
\includegraphics[width=0.318\textwidth]{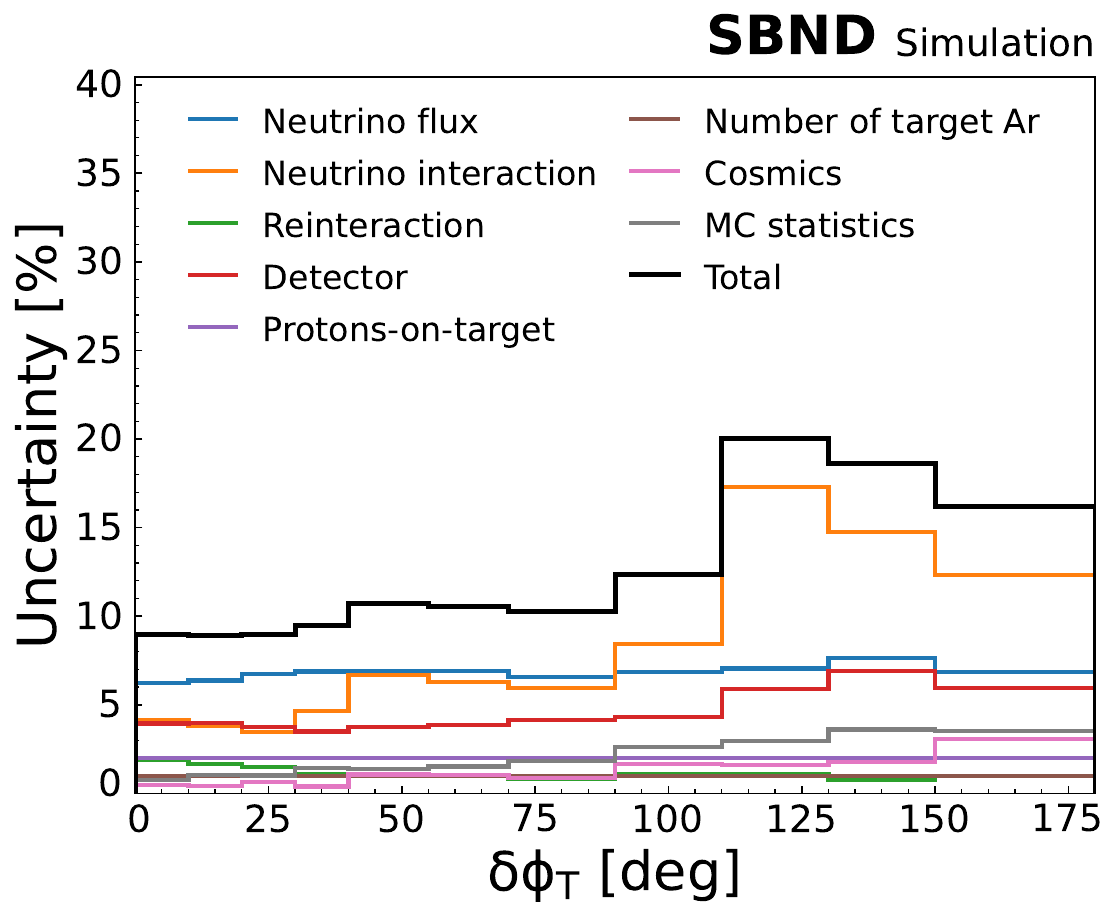} \\
(a) & (b) & (c) \\
\end{tabular}
\caption{Fractional uncertainties on the cross-section measurement as functions of (a)~$\delta p_T$, (b)~$\delta\alpha_T$, and (c)~$\delta\phi_T$, broken down by systematic source.}
\label{fig:systematic-tki}
\end{figure}

Figures~\ref{fig:systematic-muon-proton} and~\ref{fig:systematic-tki} show the breakdown of fractional uncertainties on the extracted differential cross sections as functions of muon and proton kinematics and of the TKI variables, respectively.
The figures show bin-by-bin uncertainty contributions from the neutrino flux, neutrino-nucleus interaction modeling, secondary interactions, detector response, protons-on-target and number of argon targets, cosmic-ray contamination, and Monte Carlo statistics, together with the total systematic uncertainty. 
In particular, the neutrino-nucleus interaction modeling uncertainties vary significantly as functions of the kinematic variables, with notable increases in bins with large background content.

\begin{figure}[H]
\centering
\begin{tabular}{@{}cc@{}}
\includegraphics[width=0.32\textwidth]{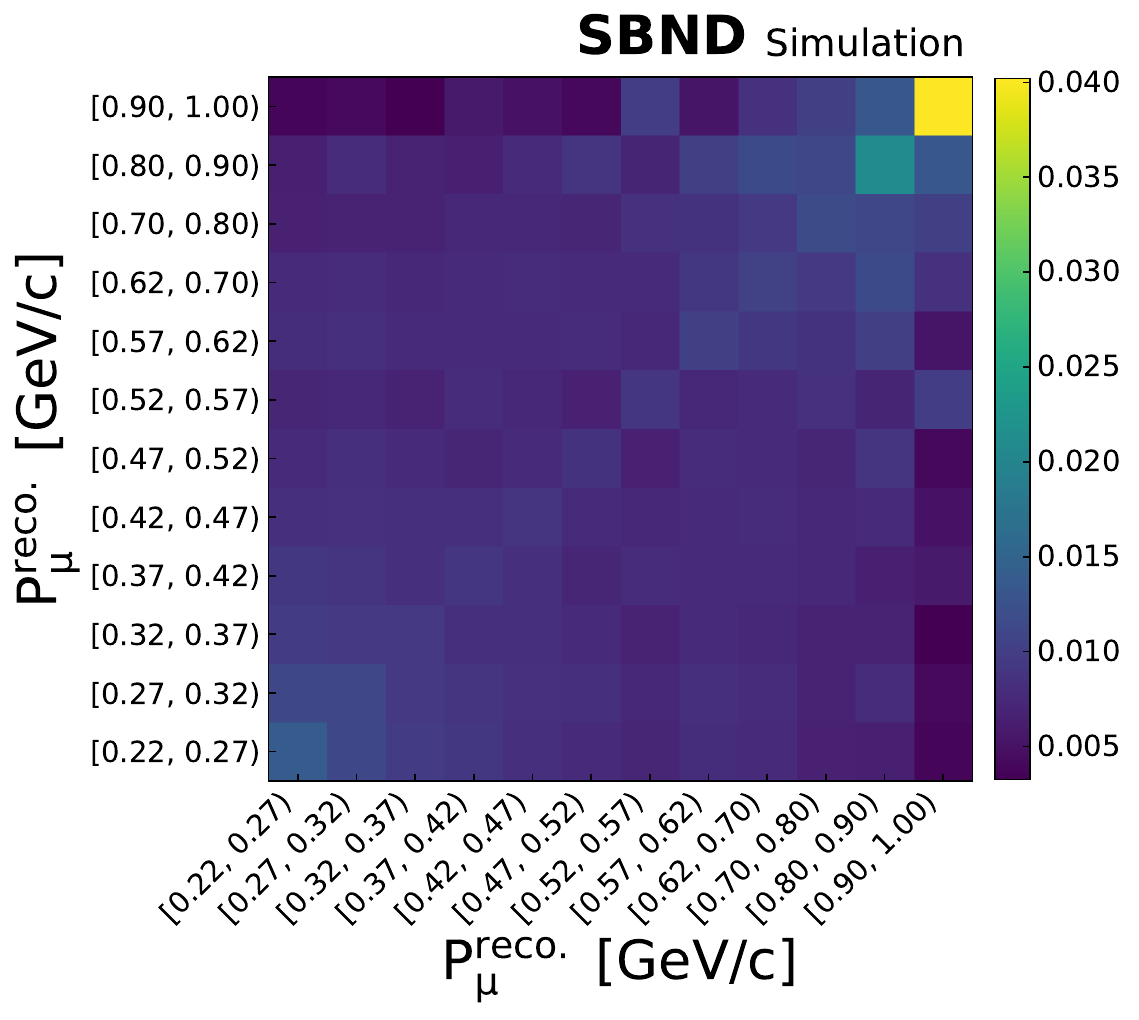} &
\includegraphics[width=0.32\textwidth]{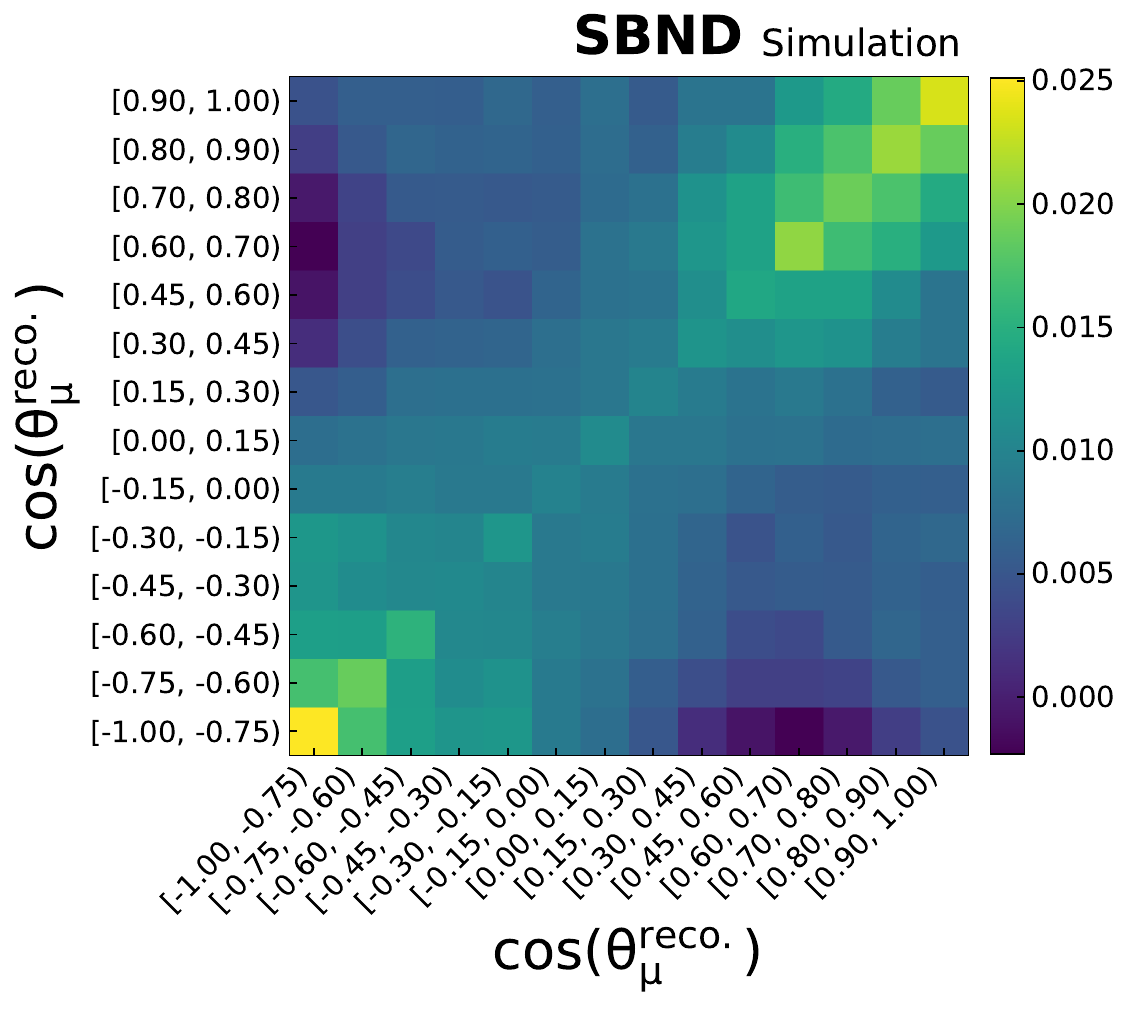}  \\
(a) & (b) \\
\includegraphics[width=0.32\textwidth]{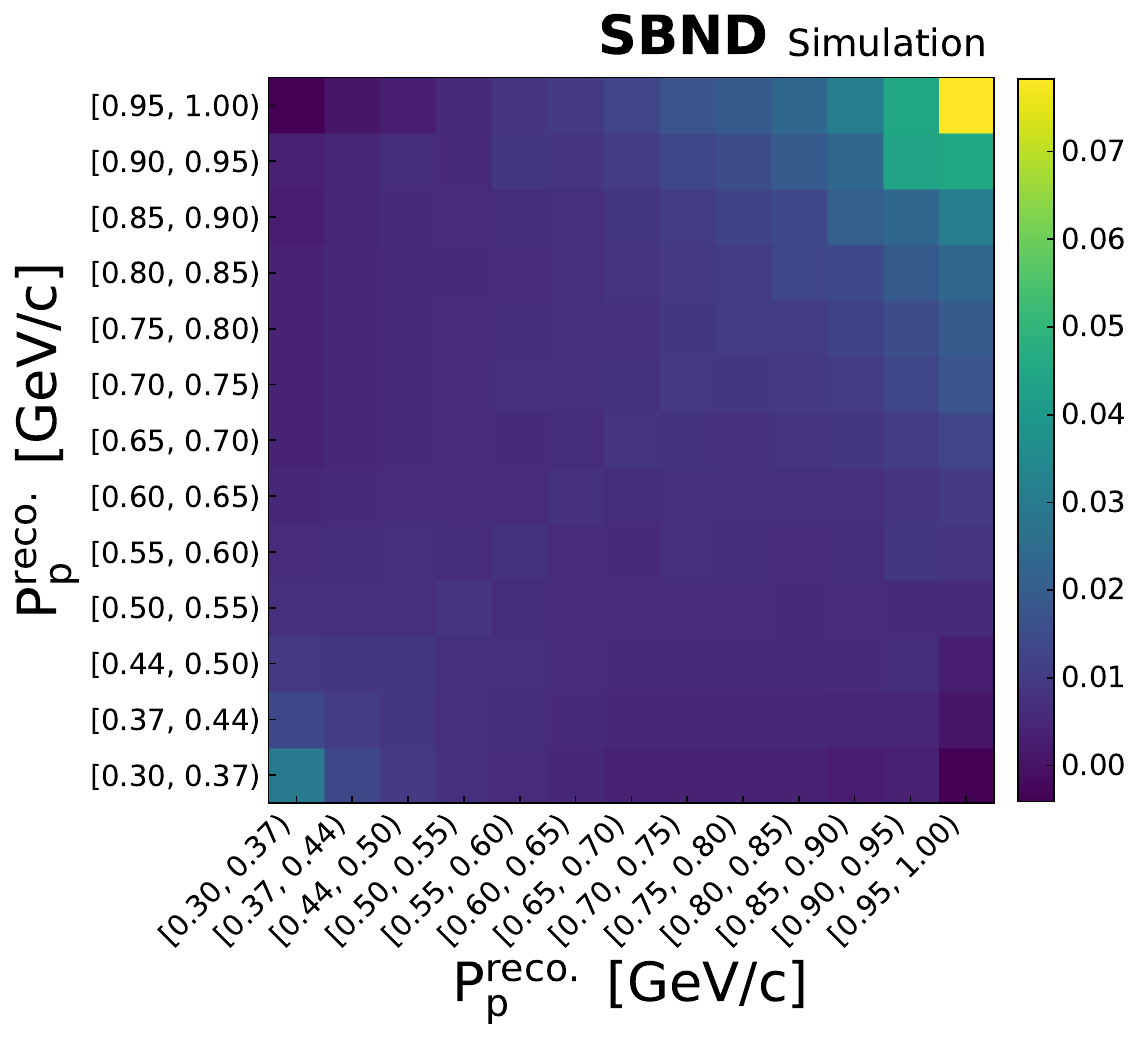} &
\includegraphics[width=0.32\textwidth]{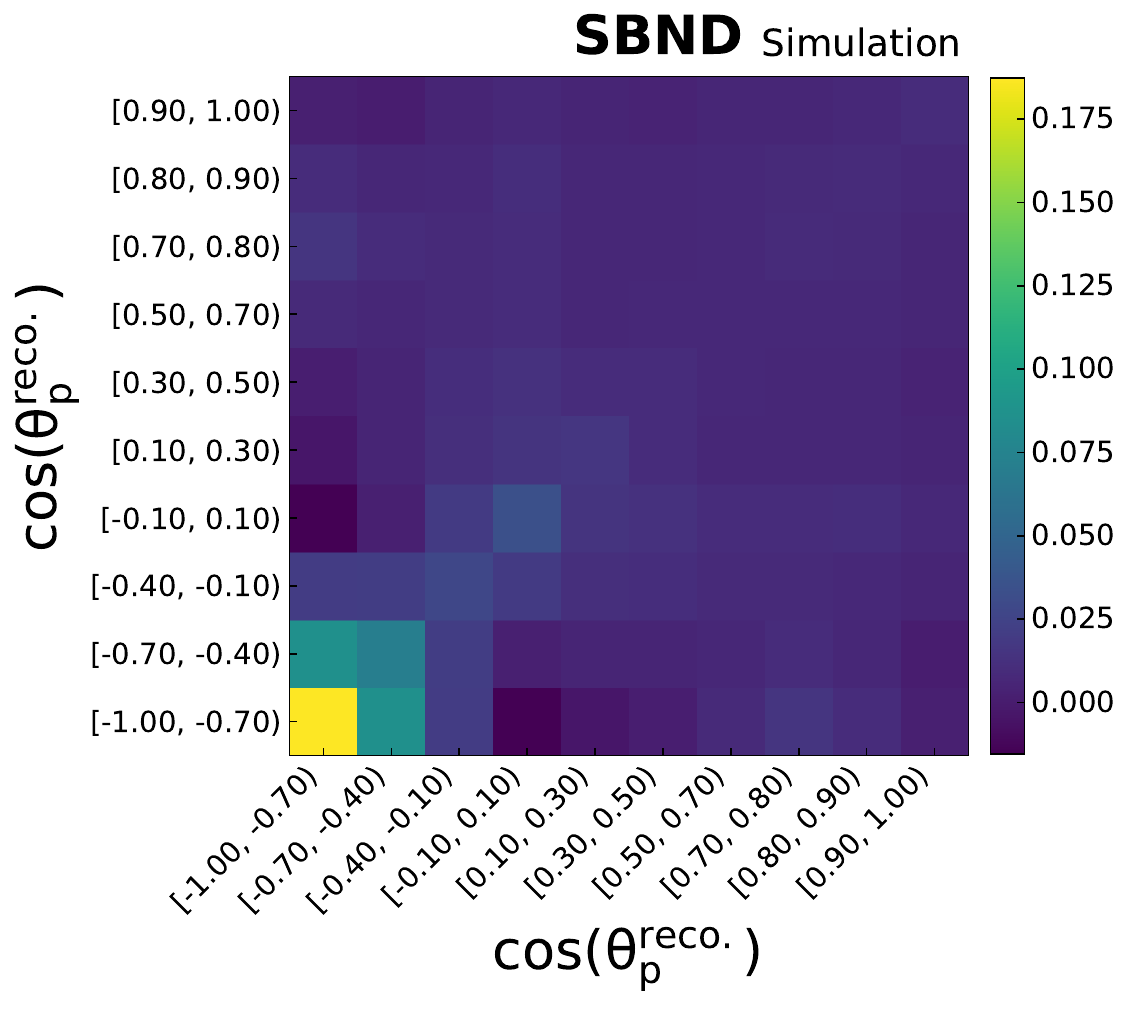} \\
(c) & (d) \\
\end{tabular}
\caption{Fractional covariance matrices of the cross-section measurement for (a)~muon momentum, (b)~muon direction, (c)~proton momentum, and (d)~proton direction.}
\label{fig:covariance-muon-proton}
\end{figure}

\begin{figure}[H]
\centering
\begin{tabular}{@{}ccc@{}}
\includegraphics[width=0.32\textwidth]{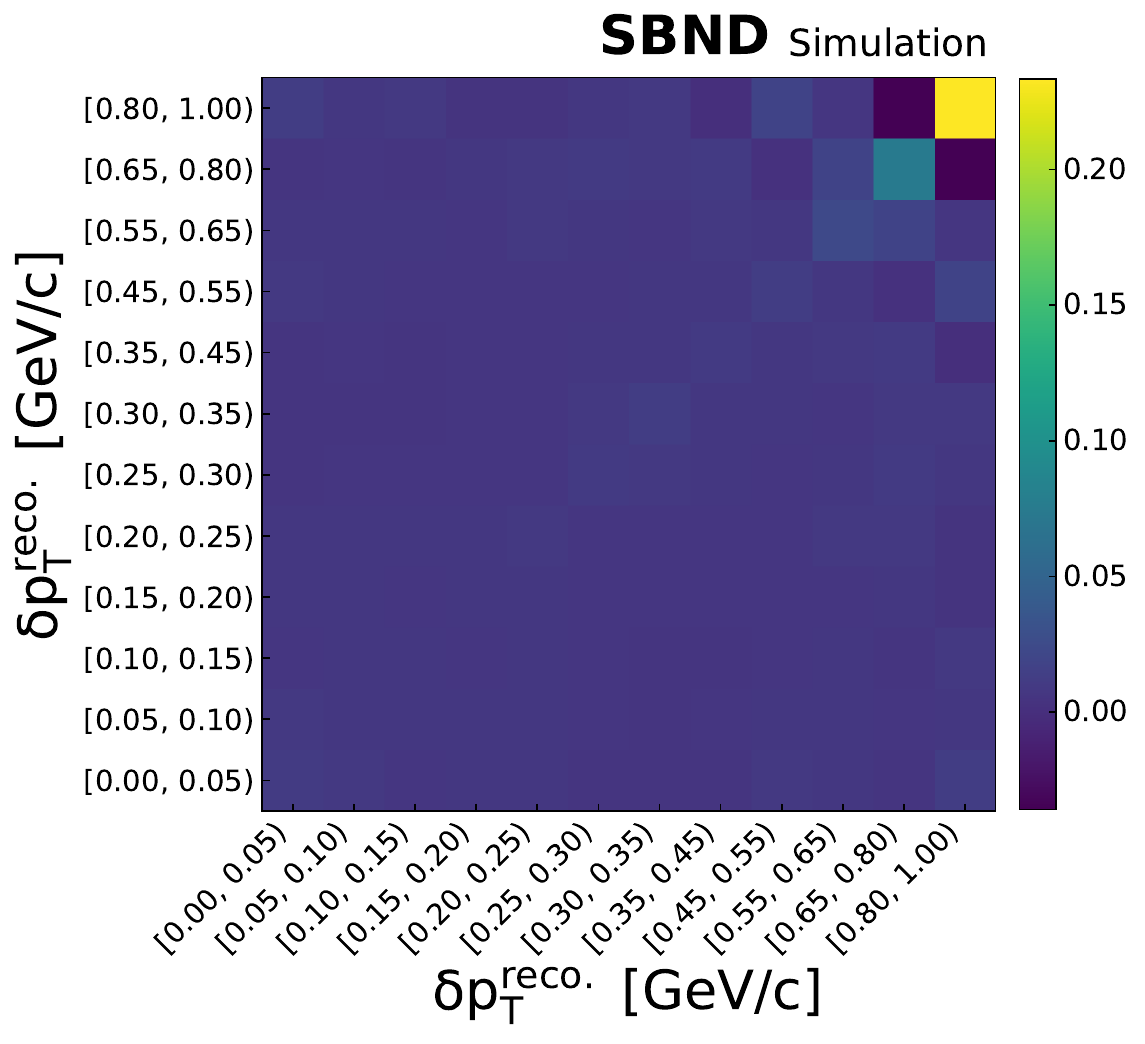} &
\includegraphics[width=0.32\textwidth]{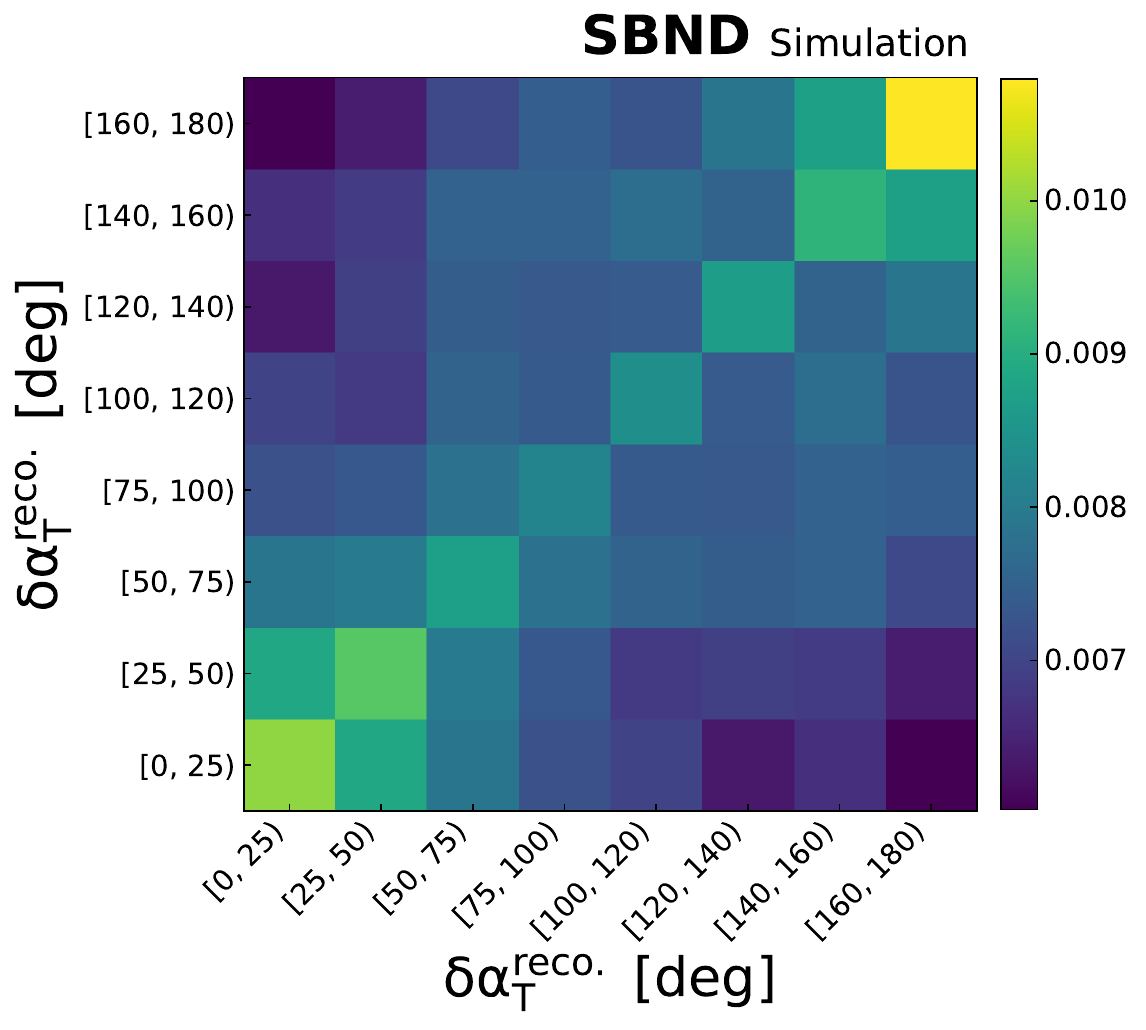} &
\includegraphics[width=0.32\textwidth]{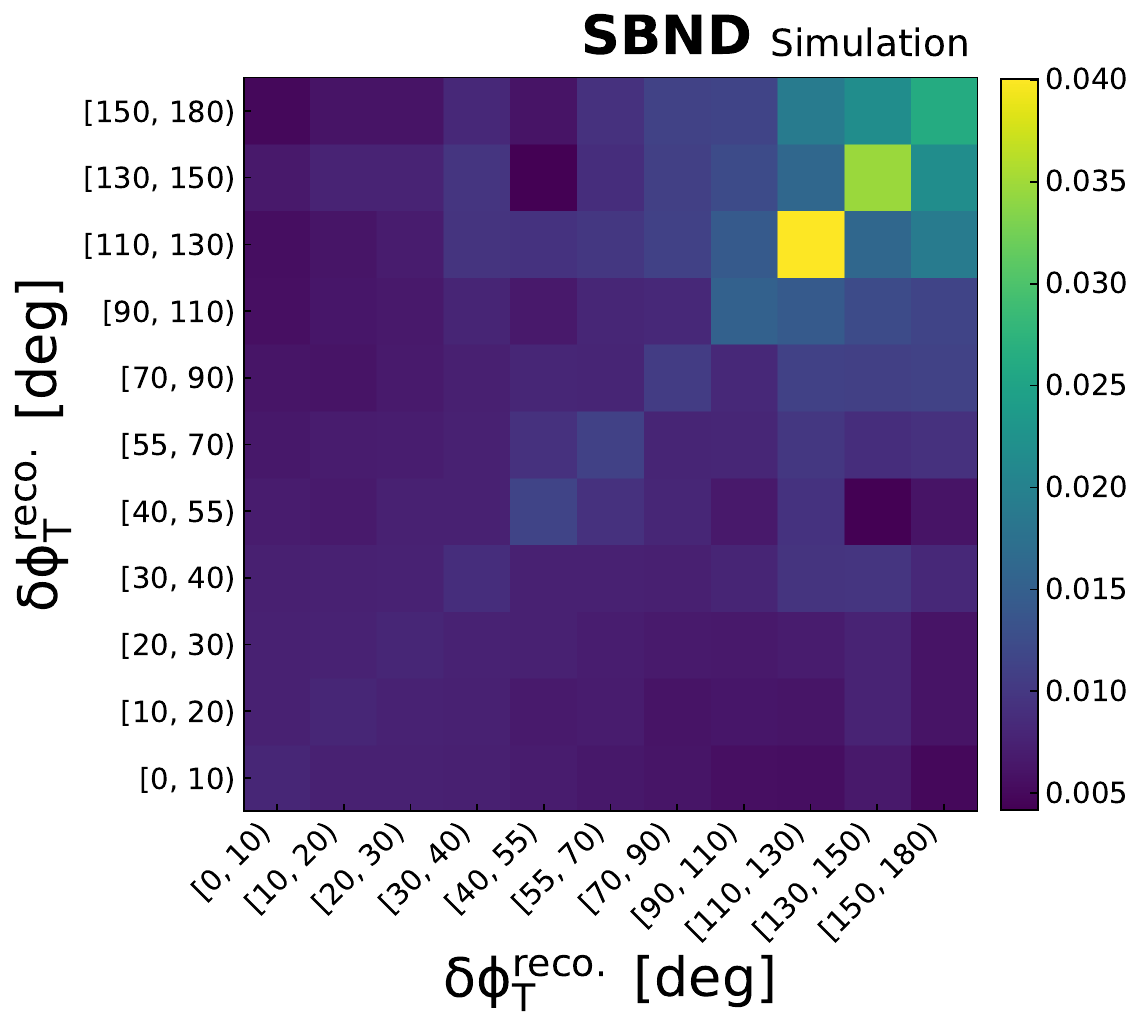} \\
(a) & (b) & (c) \\
\end{tabular}
\caption{Fractional covariance matrices of the cross-section measurement for (a)~$\delta p_T$, (b)~$\delta\alpha_T$, and (c)~$\delta\phi_T$.}
\label{fig:covariance-tki}
\end{figure}

Figures~\ref{fig:covariance-muon-proton} and~\ref{fig:covariance-tki} show the corresponding total fractional covariance matrices, which encode the full bin-to-bin correlations.

\FloatBarrier
\newpage
\section{Data-driven conditional-constraint validation tests}
\label{data_driven_tests}

We perform data-driven conditional-constraint validation tests to test whether our model and systematic uncertainties adequately describe the data, following the procedure introduced in Ref.~\cite{Validation}. 
In particular, we validate that the model constrained by muon kinematics predicts the proton kinematics well.
Both muon momentum and muon direction are used as constraining variables. The constraint procedure is applied separately to predict the proton momentum and proton direction distributions and their systematic covariance matrices.
After applying the constraints, we find that both the proton momentum and direction distributions remain consistent with the data, indicating that our model sufficiently covers the kinematic phase space of the final-state particles.

\begin{figure}[H]
\centering
\begin{tabular}{@{}cc@{}}
\includegraphics[width=0.475\textwidth]{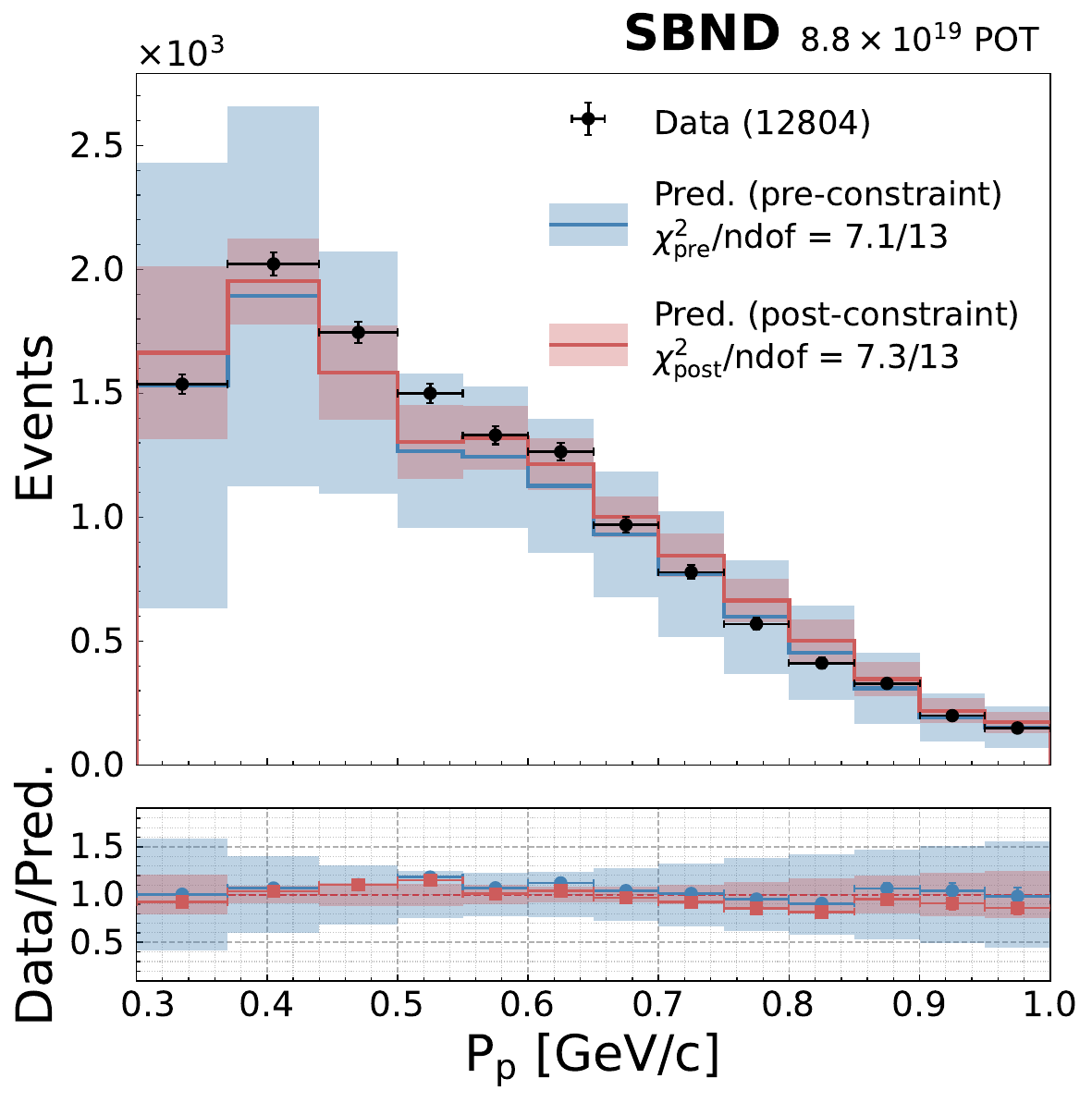} &
\includegraphics[width=0.48\textwidth]{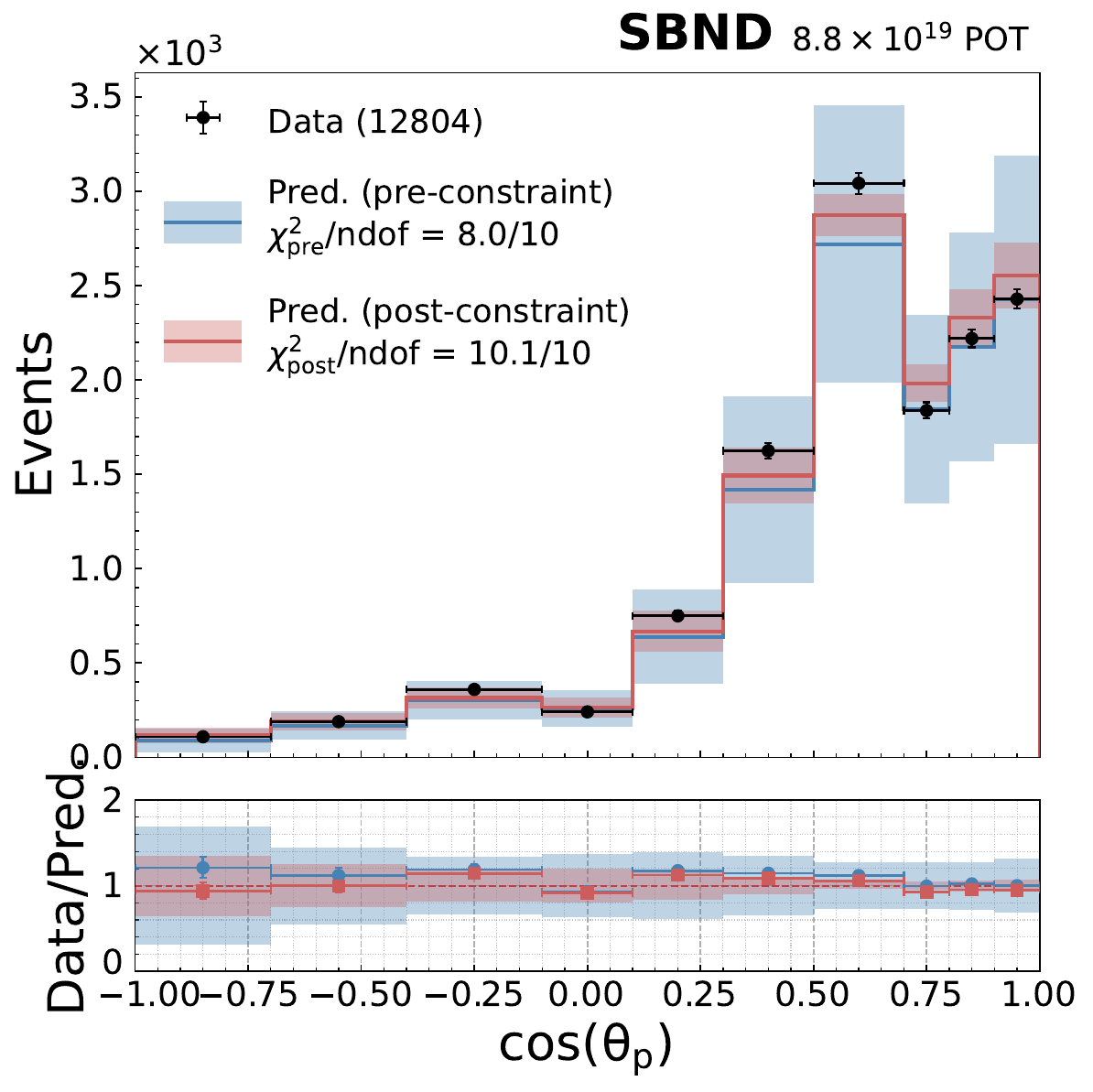} \\
(a) & (b) \\
\end{tabular}
\caption{The conditional-constraint validation tests for the proton momentum and direction. The pre-constraint and post-constraint spectra and systematic uncertainties on the proton momentum (left) and direction (right) are shown in blue and red, respectively.}
\label{fig:data-driven-validation}
\end{figure}

\FloatBarrier
\newpage
\section{Fake Data Tests}
\label{sec:fake_data_tests}

\begin{figure}[H]
\centering
\begin{tabular}{@{}ccc@{}}
\includegraphics[width=0.337\textwidth]{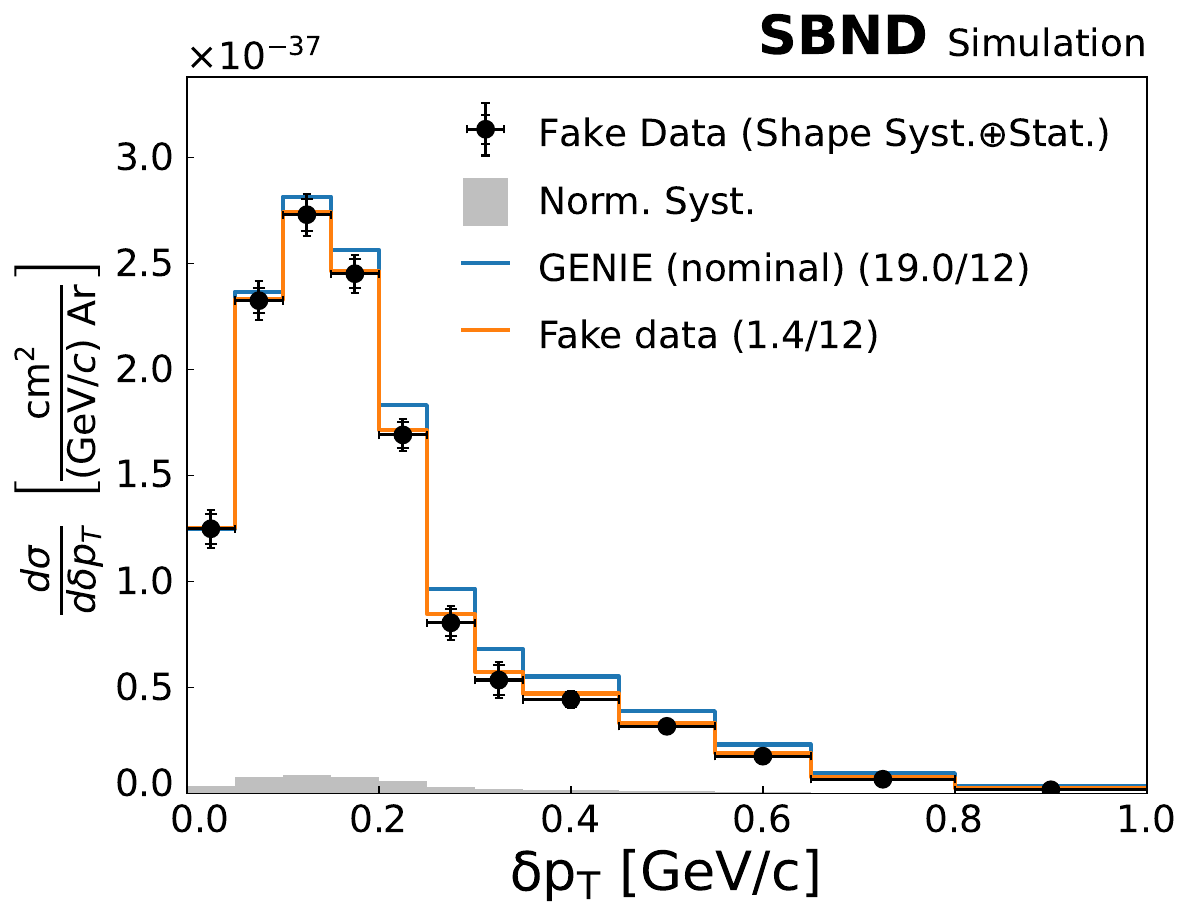} &
\includegraphics[width=0.32\textwidth]{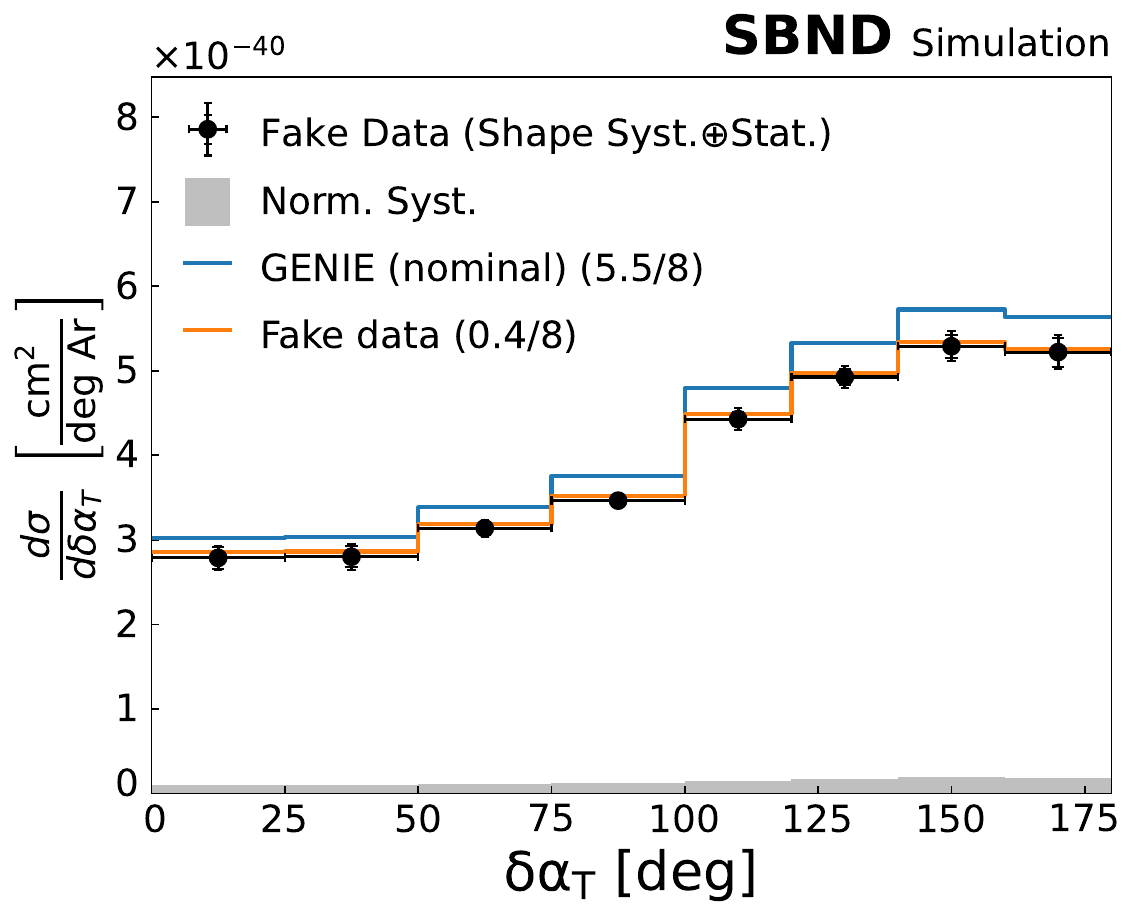} &
\includegraphics[width=0.33\textwidth]{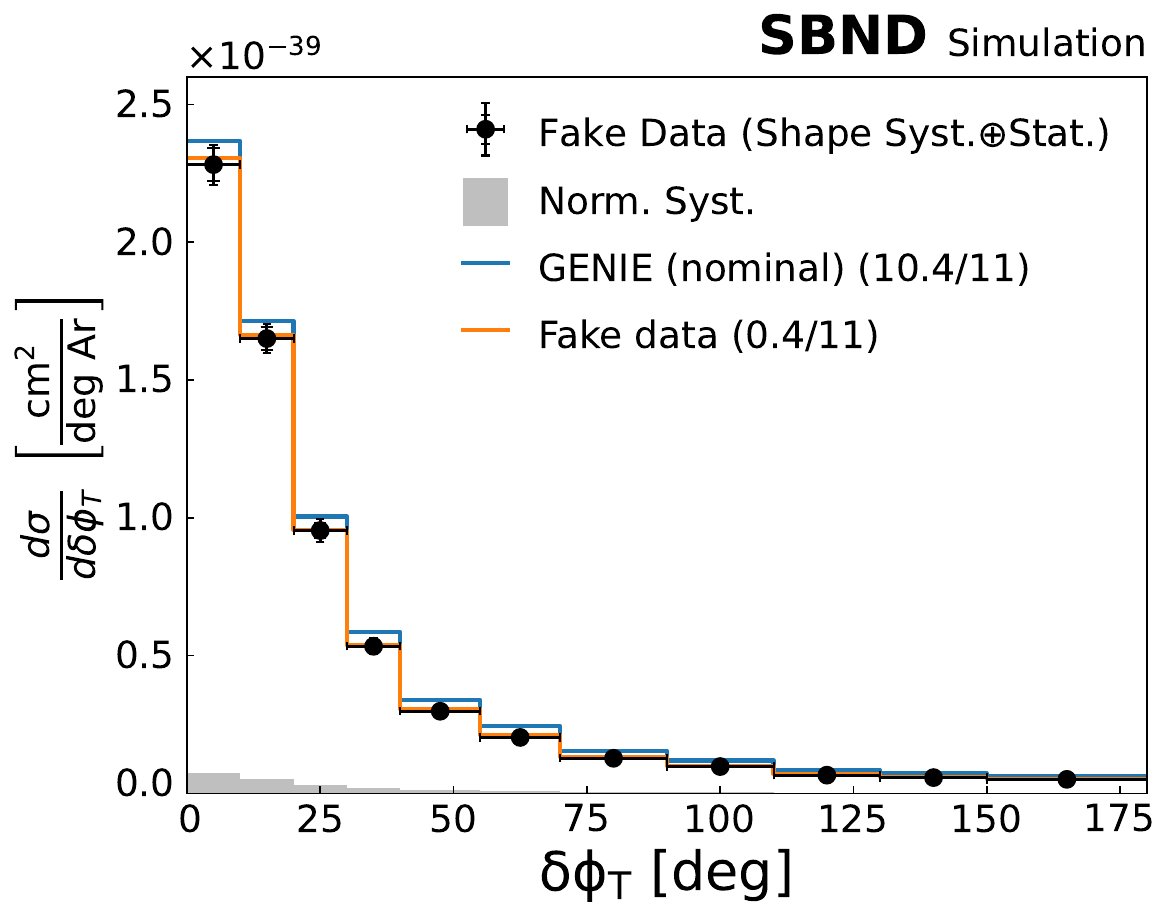} \\
(a) & (b) & (c) \\
\end{tabular}
\caption{Fake data test for the $\delta p_T$, $\delta\alpha_T$, and $\delta\phi_T$ where MEC events are reweighted by a factor of 0.5.}
\label{fig:fake-data-mec}
\end{figure}

\begin{figure}[H]
\centering
\begin{tabular}{@{}ccc@{}}
\includegraphics[width=0.335\textwidth]{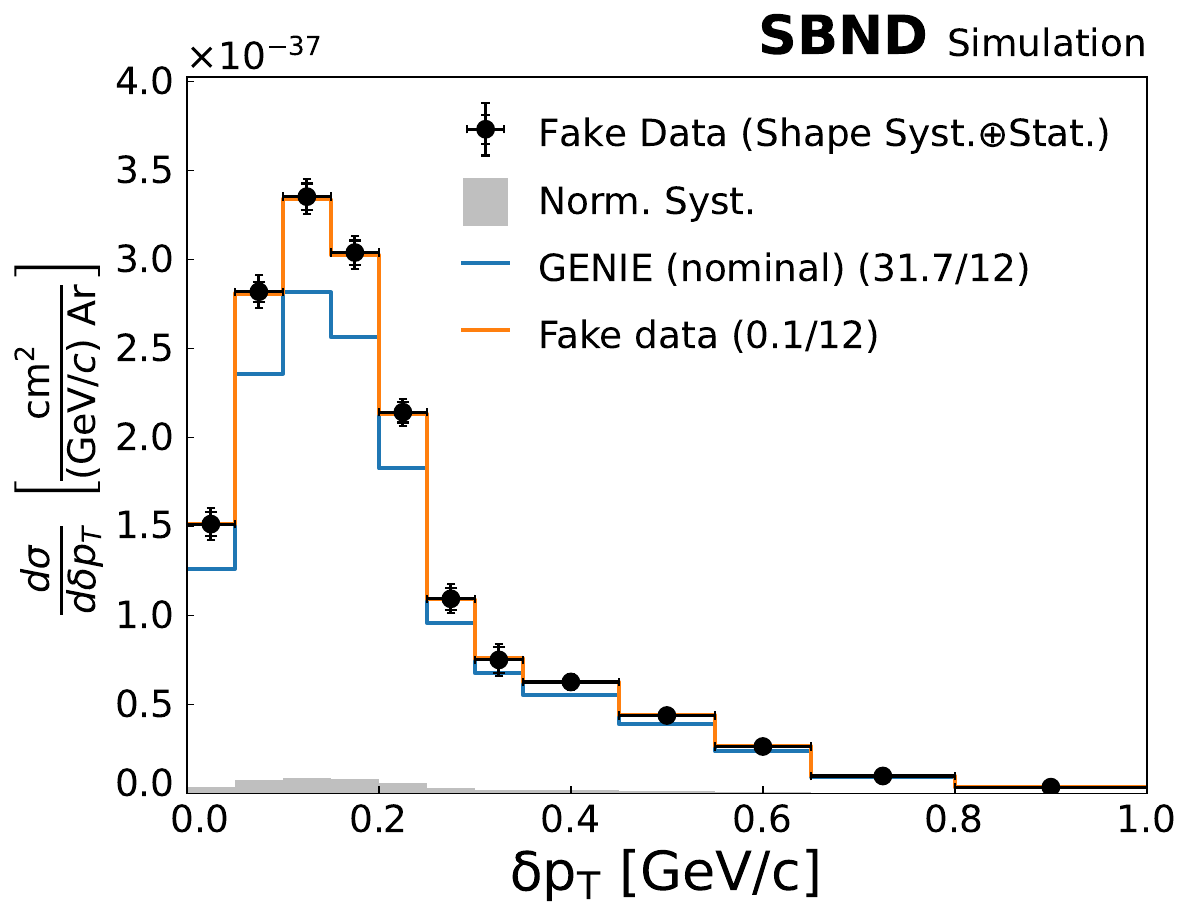} &
\includegraphics[width=0.319\textwidth]{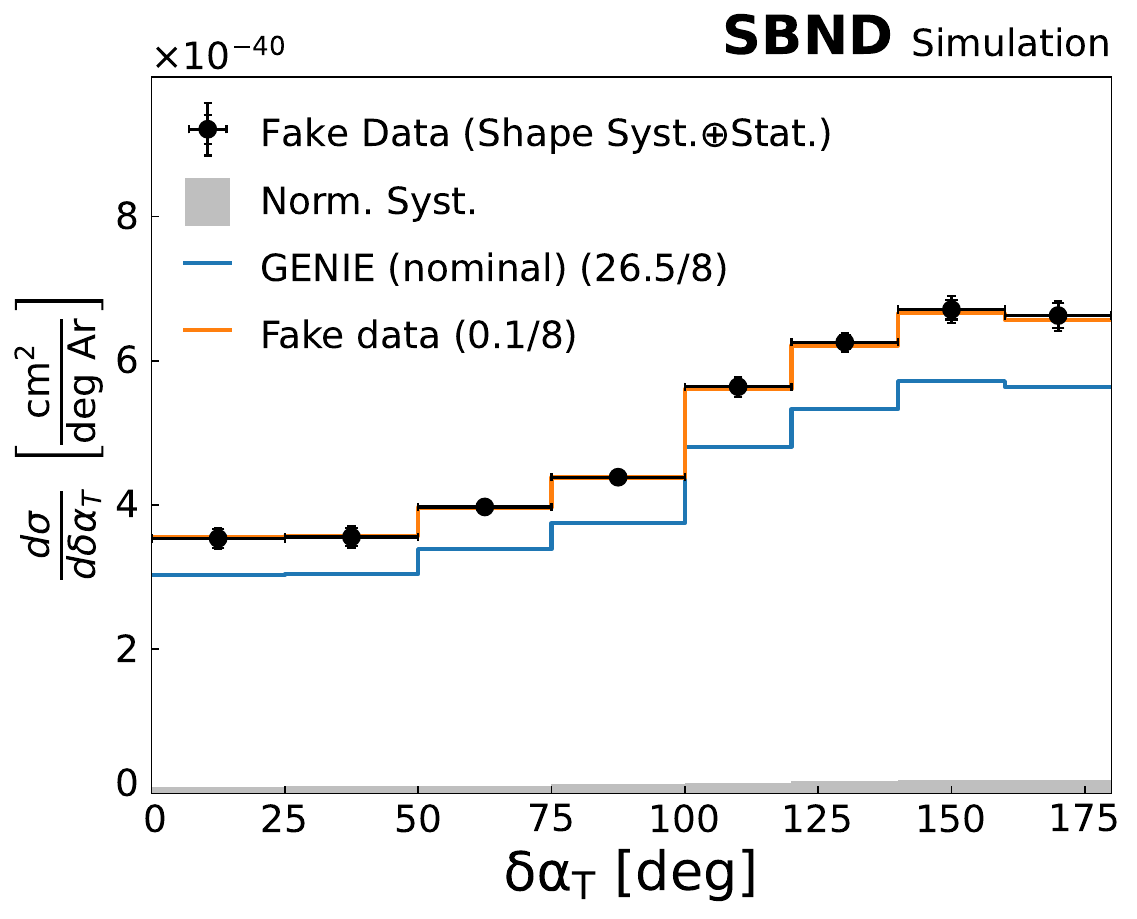} &
\includegraphics[width=0.328\textwidth]{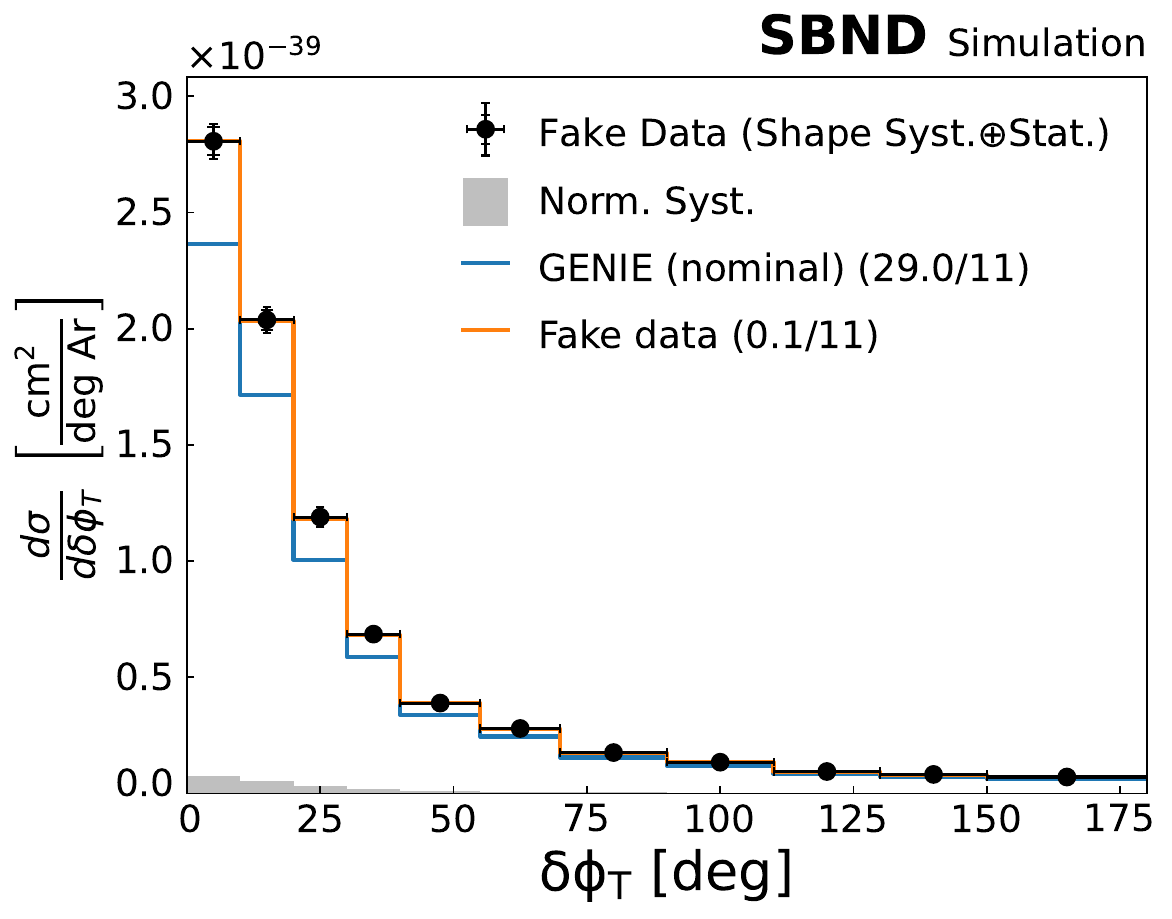} \\
(a) & (b) & (c) \\
\end{tabular}
\caption{Fake data test for the $\delta p_T$, $\delta\alpha_T$, and $\delta\phi_T$ where CCQE events are reweighted by a factor of 1.2.}
\label{fig:fake-data-qe}
\end{figure}

\begin{figure}[H]
\centering
\begin{tabular}{@{}ccc@{}}
\includegraphics[width=0.336\textwidth]{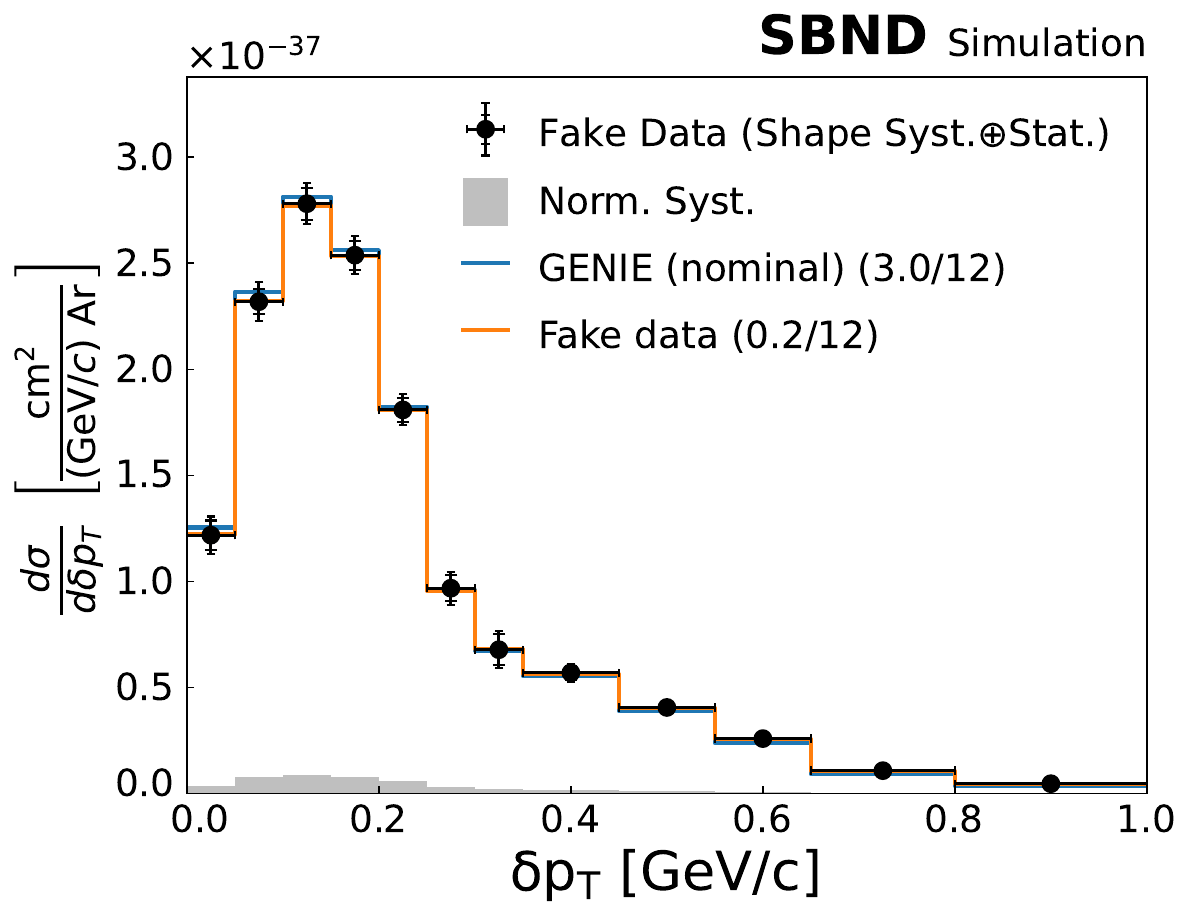} &
\includegraphics[width=0.32\textwidth]{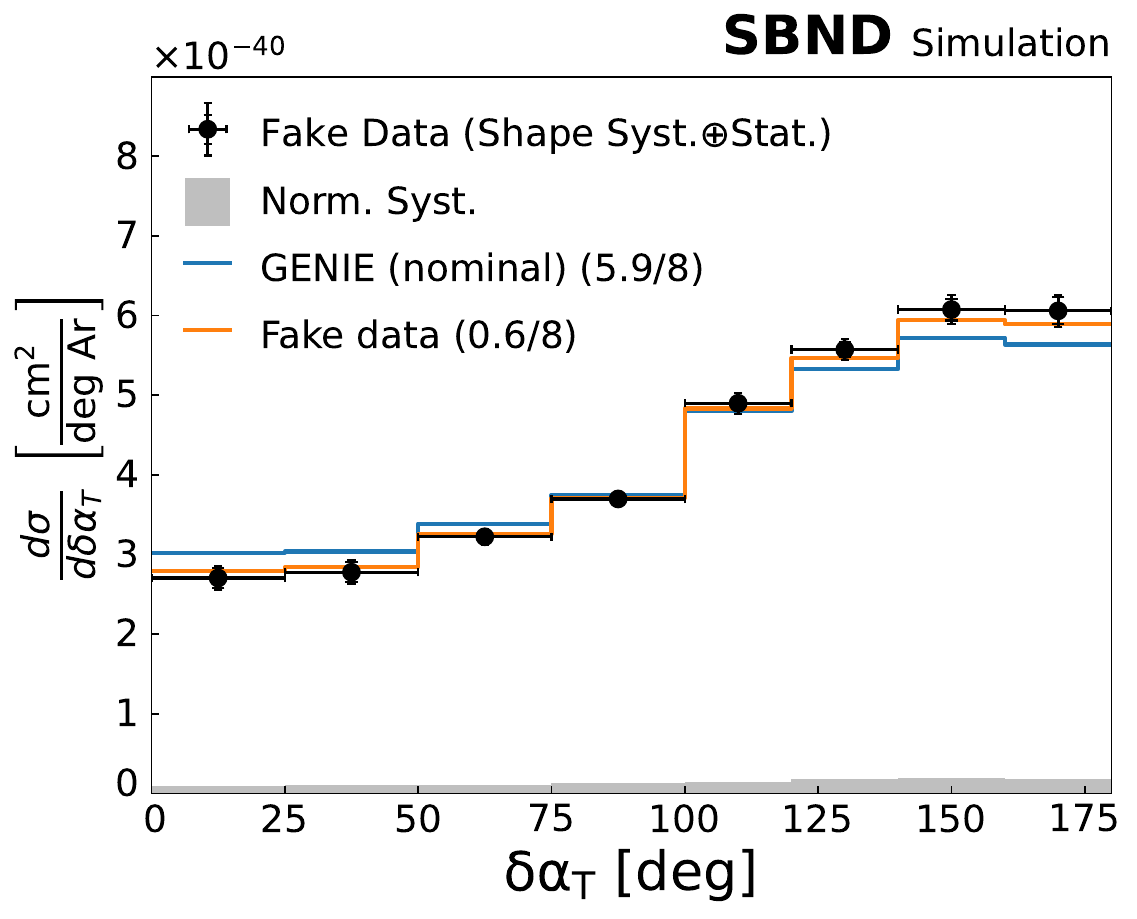} &
\includegraphics[width=0.33\textwidth]{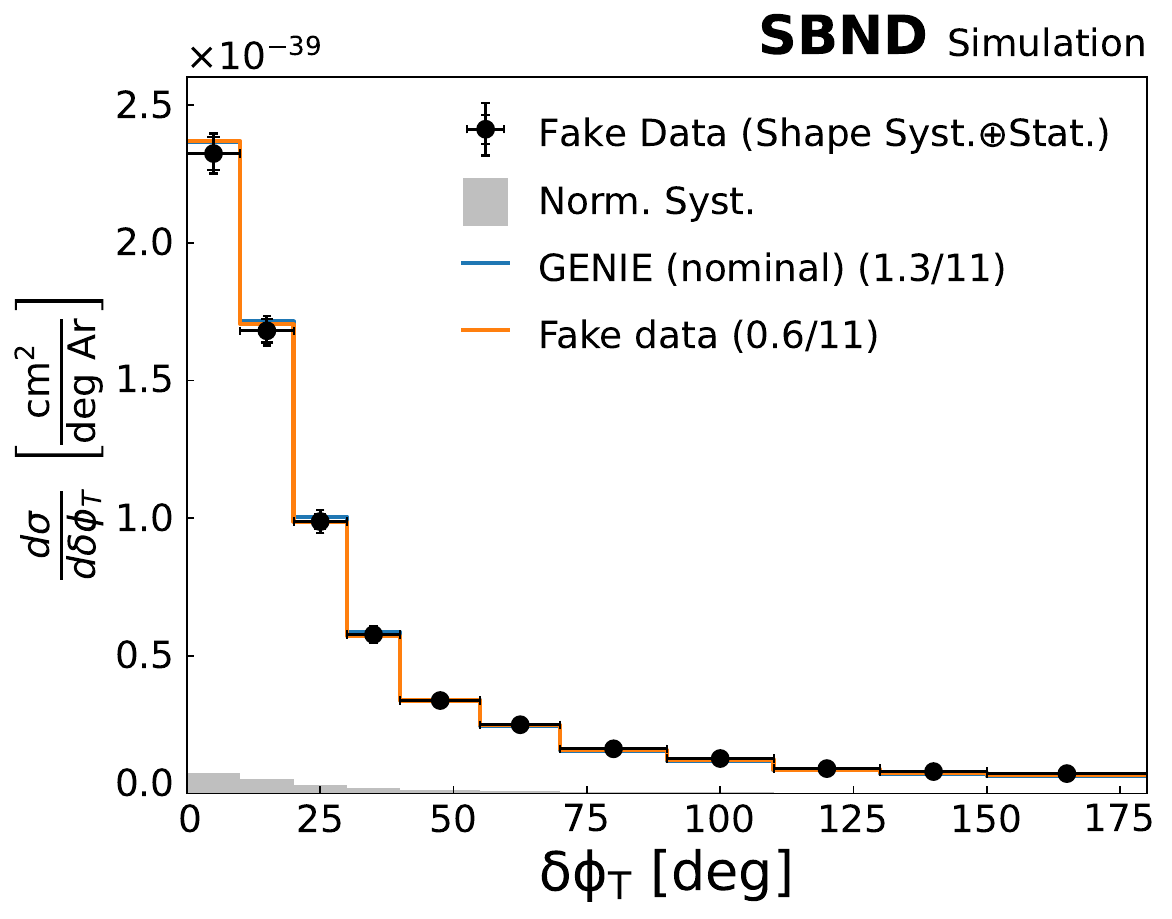} \\
(a) & (b) & (c) \\
\end{tabular}
\caption{Fake data test for the $\delta p_T$, $\delta\alpha_T$, and $\delta\phi_T$ where events are reweighted by a linear tilt in $Q^2$, $w = 1 + 0.3 \frac{Q^2 - \langle Q^2 \rangle}{\langle Q^2 \rangle}$.}
\label{fig:fake-data-q2}
\end{figure}

\begin{figure}[H]
\centering
\begin{tabular}{@{}ccc@{}}
\includegraphics[width=0.336\textwidth]{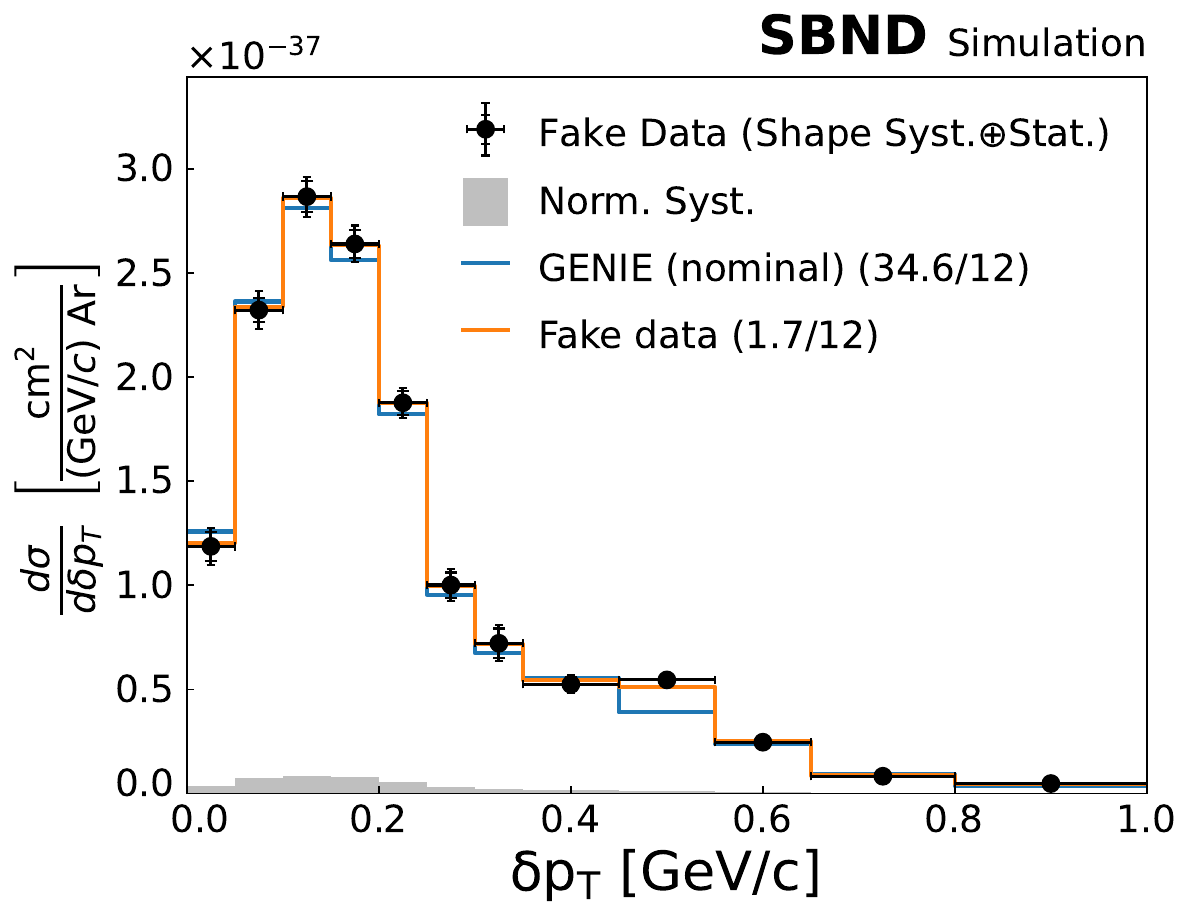} &
\includegraphics[width=0.32\textwidth]{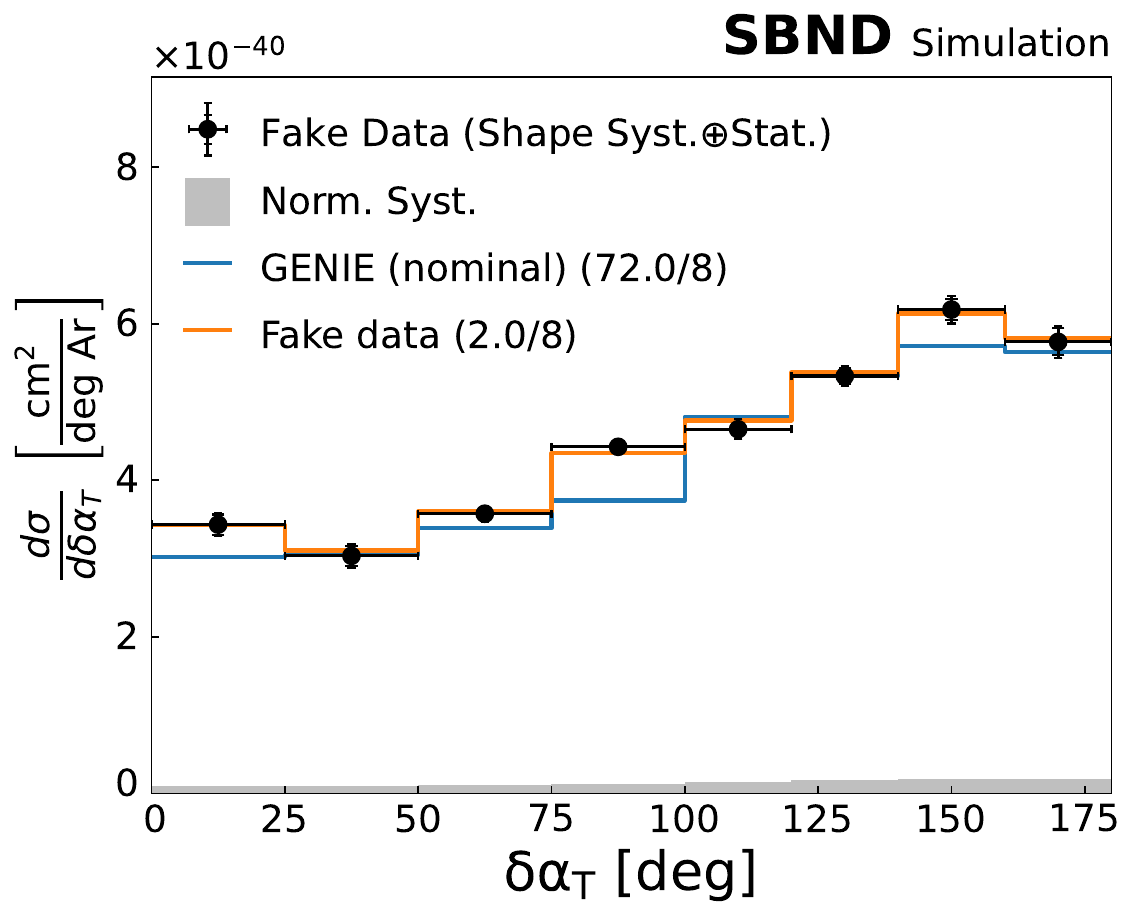} &
\includegraphics[width=0.33\textwidth]{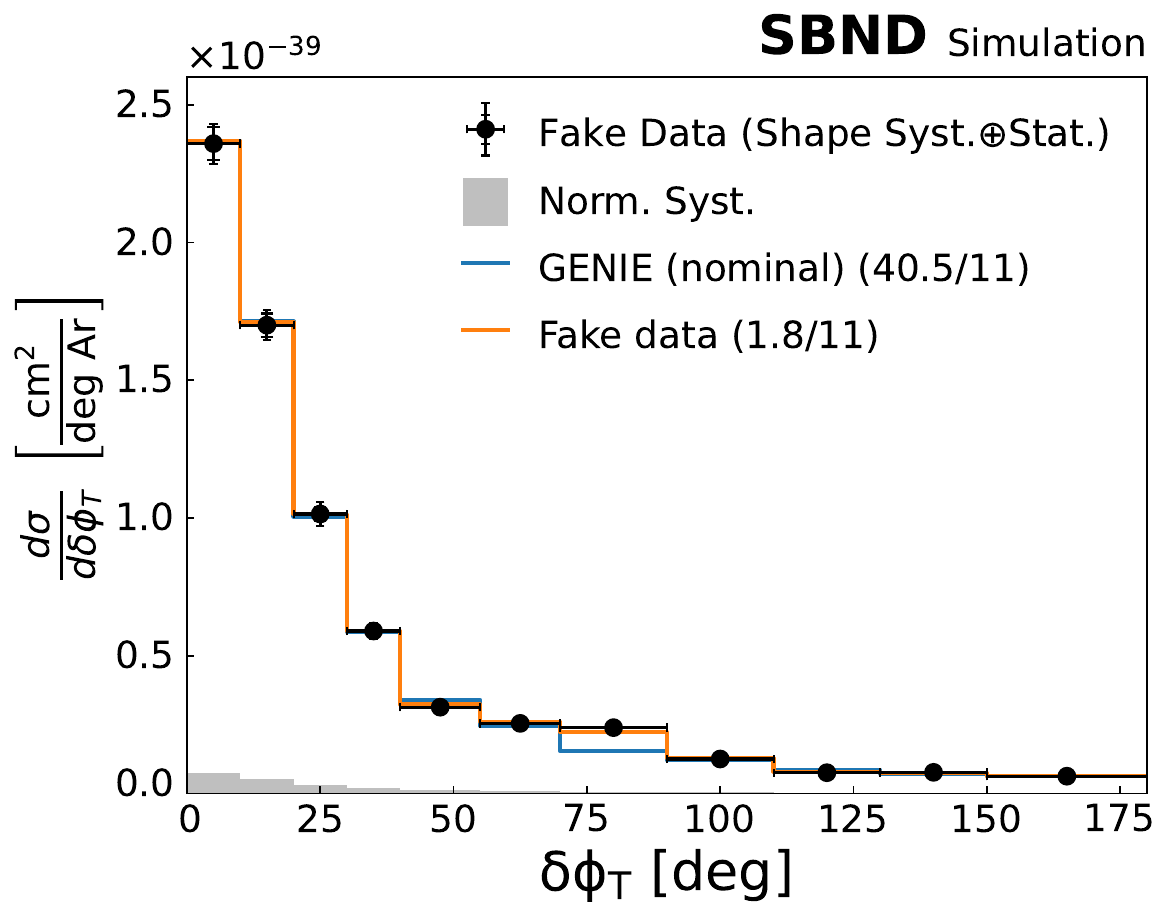} \\
(a) & (b) & (c) \\
\end{tabular}
\caption{Fake data test for the $\delta p_T$, $\delta\alpha_T$, and $\delta\phi_T$ where a Gaussian bump is injected in the center bin, with integrated area equal to half of the bin area.}
\label{fig:fake-data-bump}
\end{figure}

\begin{figure}[H]
\centering
\begin{tabular}{@{}ccc@{}}
\includegraphics[width=0.338\textwidth]{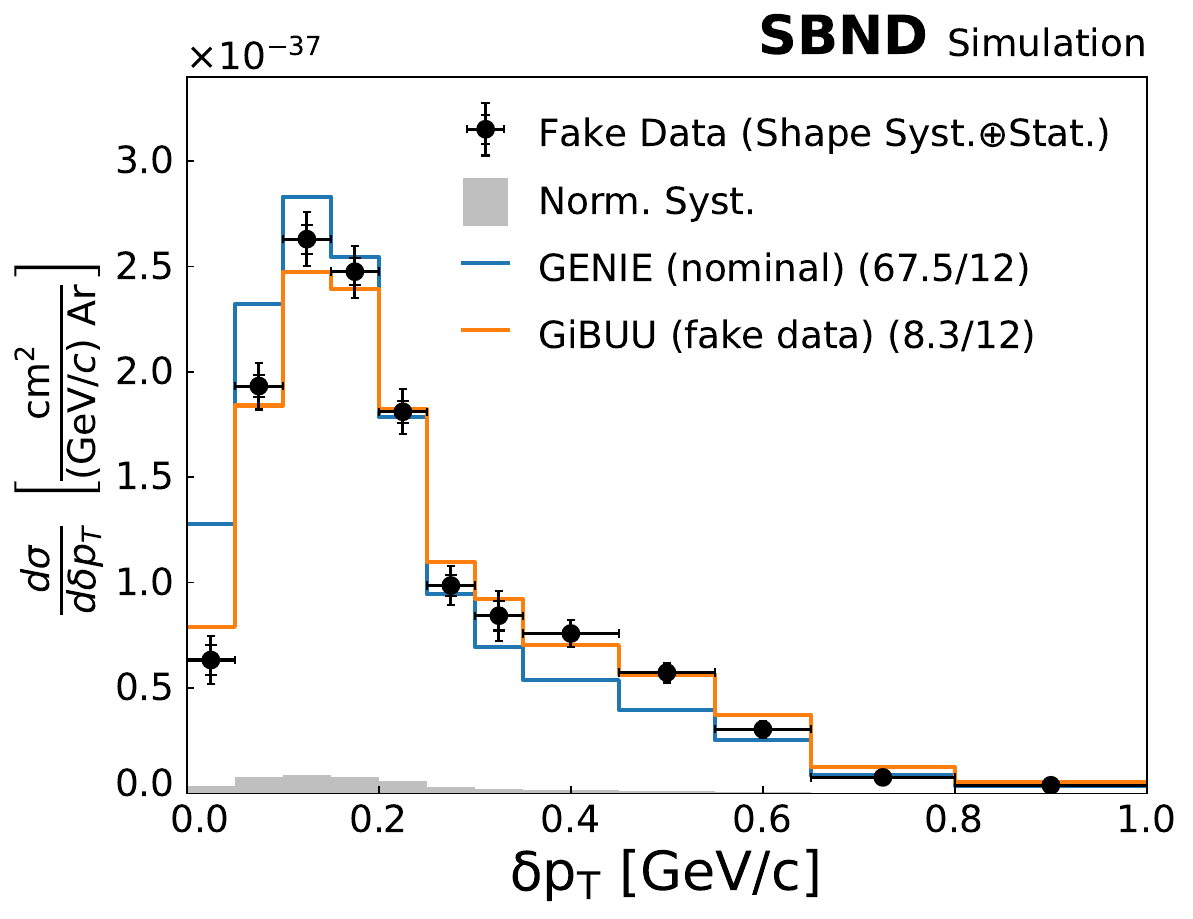} &
\includegraphics[width=0.322\textwidth]{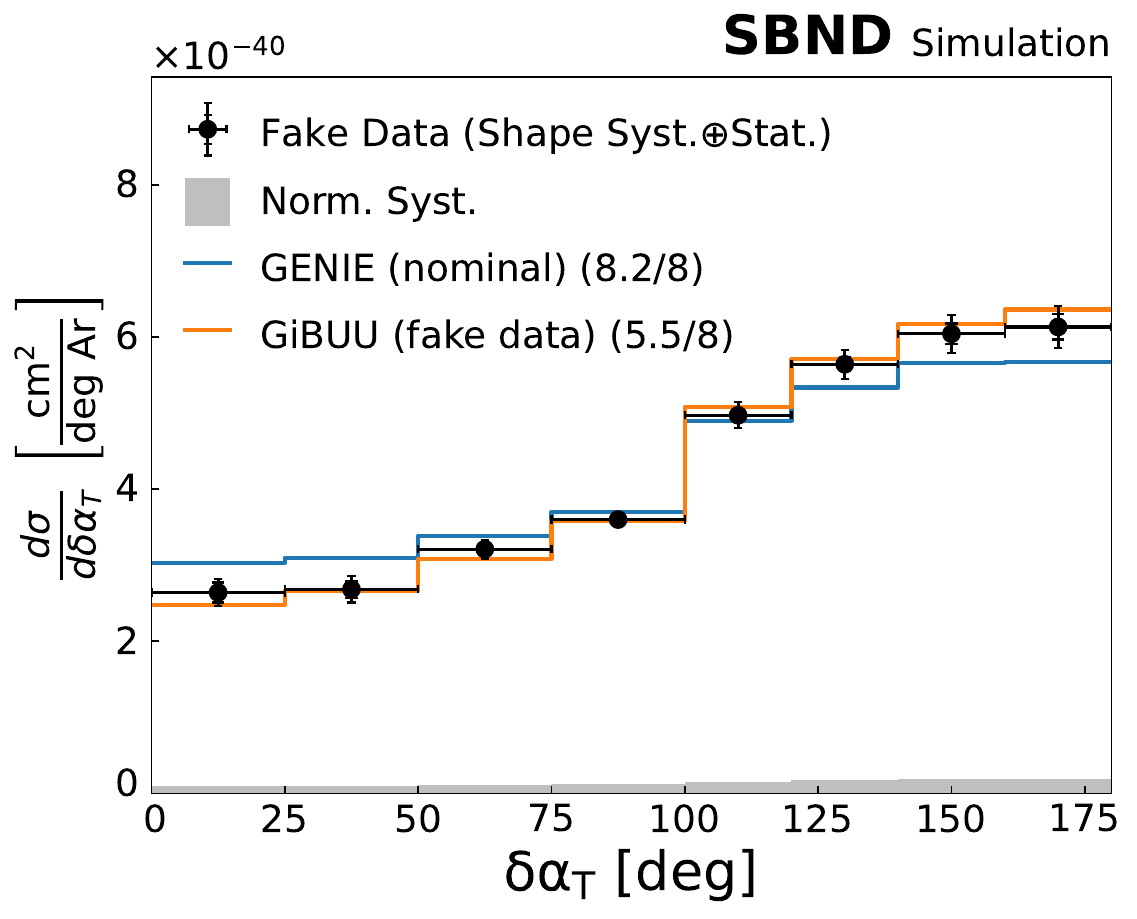} &
\includegraphics[width=0.332\textwidth]{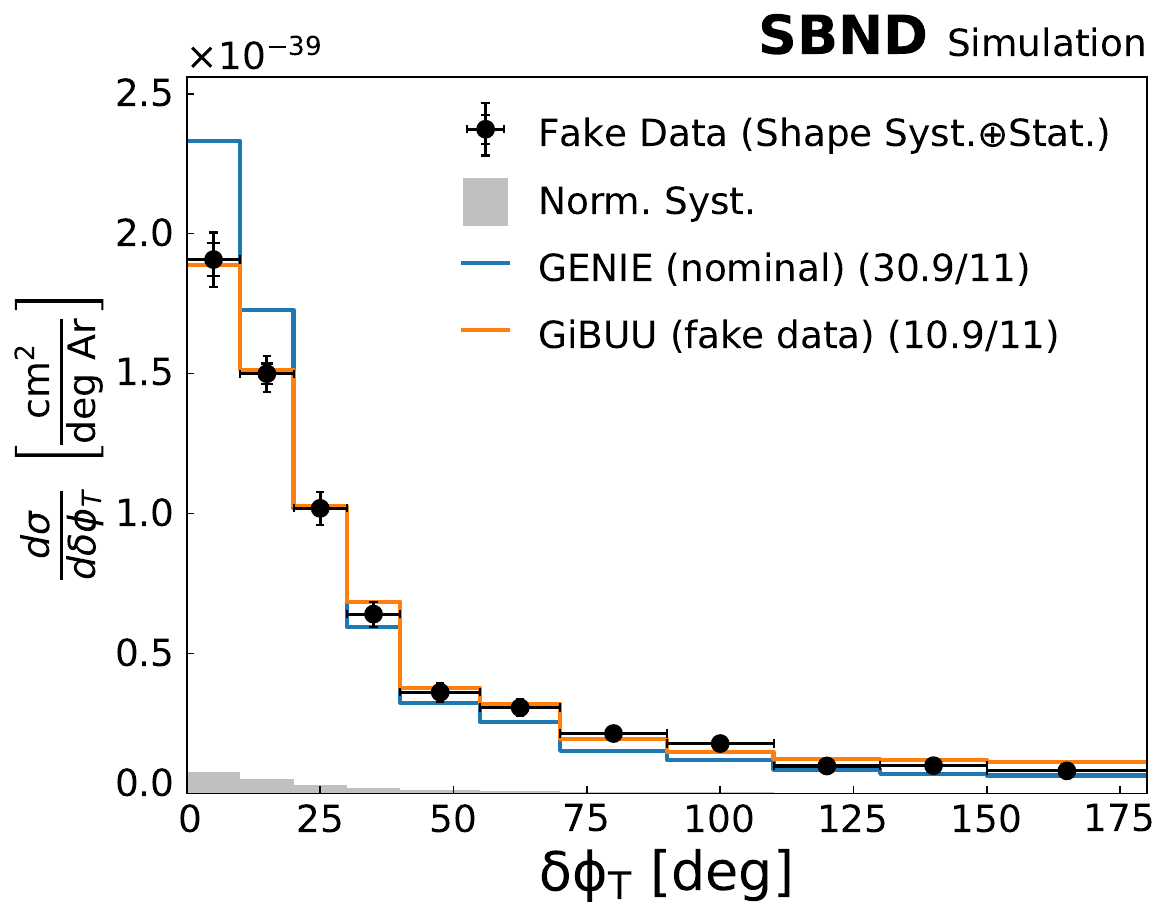} \\
(a) & (b) & (c) \\
\end{tabular}
\caption{Fake data test for the $\delta p_T$, $\delta\alpha_T$, and $\delta\phi_T$ where the GiBUU generator predictions are used as fake data.}
\label{fig:fake-data-gibuu}
\end{figure}

We perform fake data tests to test the robustness of the unfolding procedure. Fake datasets are created with the following variations: rescalings of rates for certain interaction modes (Figures~\ref{fig:fake-data-mec} and~\ref{fig:fake-data-qe}), shape tilts as a function of $Q^2$ (Fig.~\ref{fig:fake-data-q2}), localized Gaussian bumps (Fig.~\ref{fig:fake-data-bump}), and GiBUU generator predictions (Fig.~\ref{fig:fake-data-gibuu}).
The systematic uncertainties included in these tests are the MC statistics and neutrino interaction modeling uncertainties.
Across the broad range of tests the unfolding procedure recovers the injected truth without unphysical bias. These tests confirm that the cross-section extraction procedure remains robust for mismodeling scenarios. The figures show the unfolded results for the $\delta p_T$, $\delta\alpha_T$, and $\delta\phi_T$ variables, and Table~\ref{tab:fake-data-chi2} summarizes the $\chi^2/\mathrm{ndof}$ for all variables.

\begin{table}[H]
\centering
\caption{$\chi^2/\mathrm{ndof}$ of the unfolded fake-data spectra relative to the injected truth.}
\label{tab:fake-data-chi2}
\scriptsize
\begin{ruledtabular}
\begin{tabular}{lccccccc}
Test & $p_\mu$ & $\cos\theta_\mu$ & $p_p$ & $\cos\theta_p$ & $\delta p_T$ & $\delta\alpha_T$ & $\delta\phi_T$ \\
\hline
MEC scale & 0.3/12 & 0.3/14 & 0.4/13 & 1.1/10 & 1.4/12 & 0.4/8 & 0.4/11 \\
QE scale & 0.2/12 & 0.1/14 & 0.1/13 & 0.2/10 & 0.1/12 & 0.1/8 & 0.1/11 \\
$Q^2$ tilt & 5.1/12 & 1.5/14 & 1.8/13 & 1.5/10 & 0.2/12 & 0.6/8 & 0.6/11 \\
Local bump & 2.0/12 & 2.3/14 & 2.9/13 & 1.3/10 & 1.7/12 & 2.0/8 & 1.8/11 \\
GiBUU & 12.2/12 & 10.2/14 & 7.3/13 & 9.8/10 & 8.3/12 & 5.5/8 & 10.9/11 \\
\end{tabular}
\end{ruledtabular}
\end{table}

\FloatBarrier
\newpage
\section{Unfolding and Regularization}
\label{sec:regularization}

\begin{figure}[H]
\centering
\begin{tabular}{@{}cc@{}}
\includegraphics[width=0.32\textwidth]{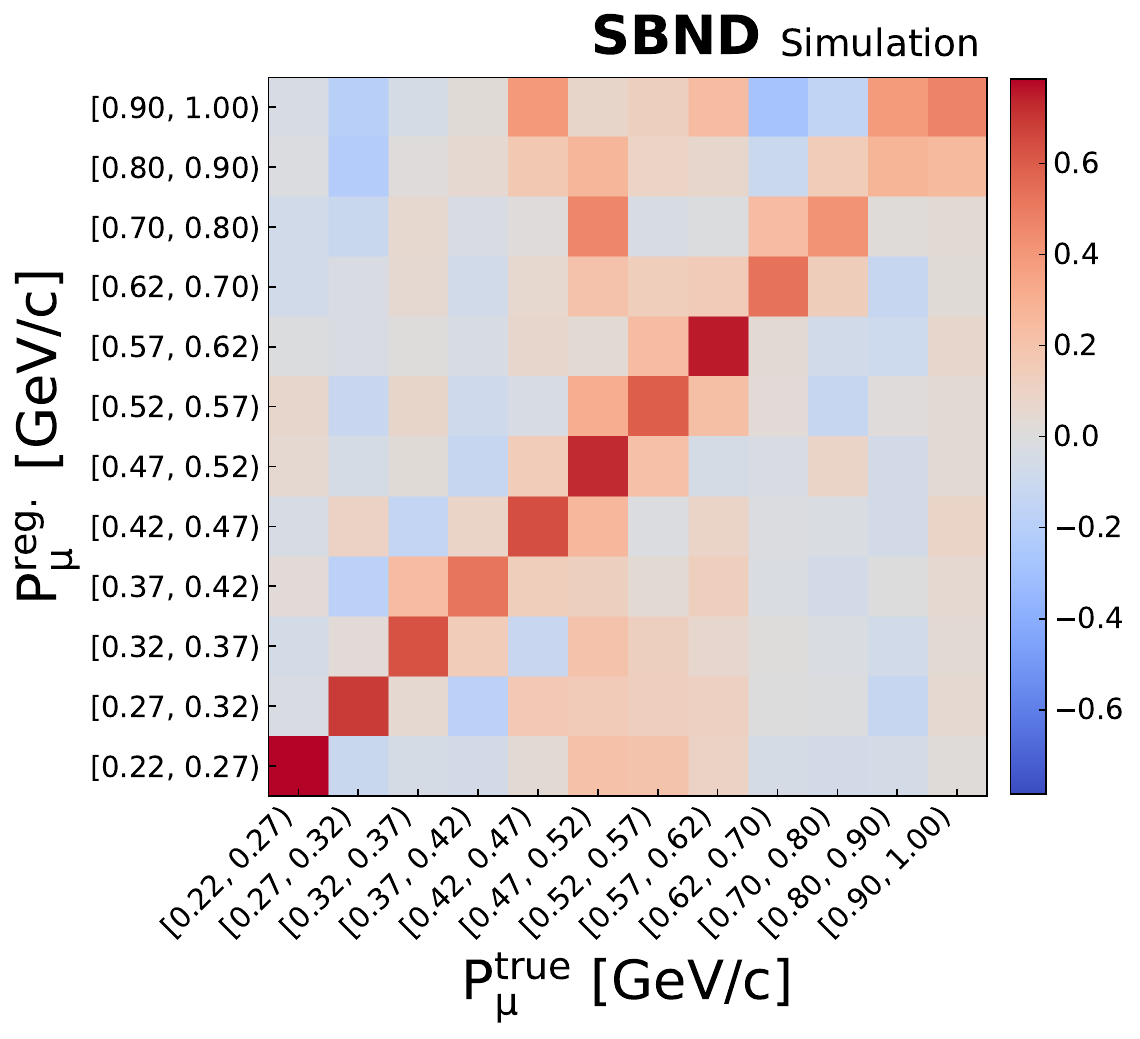} &
\includegraphics[width=0.32\textwidth]{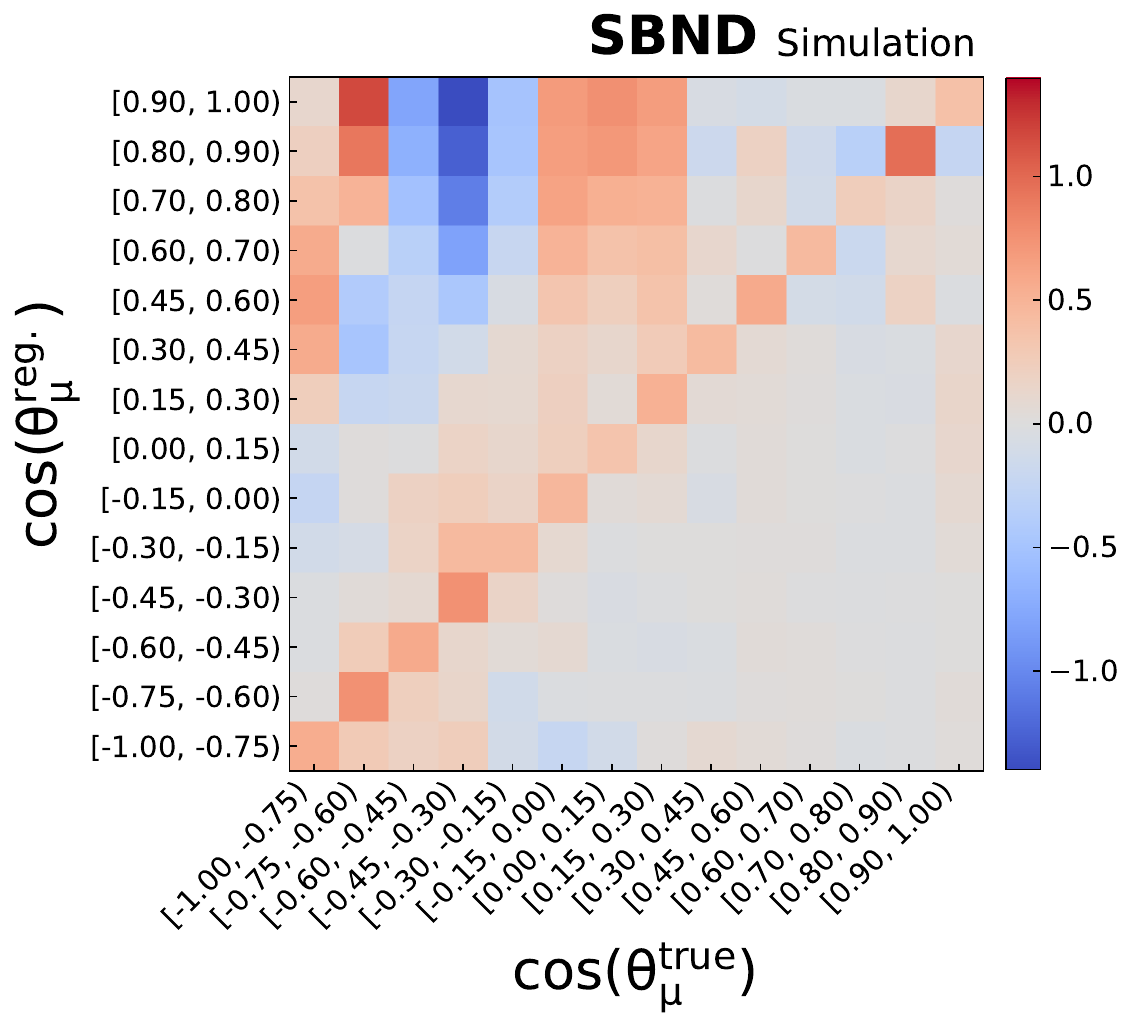}  \\
(a) & (b) \\
\includegraphics[width=0.32\textwidth]{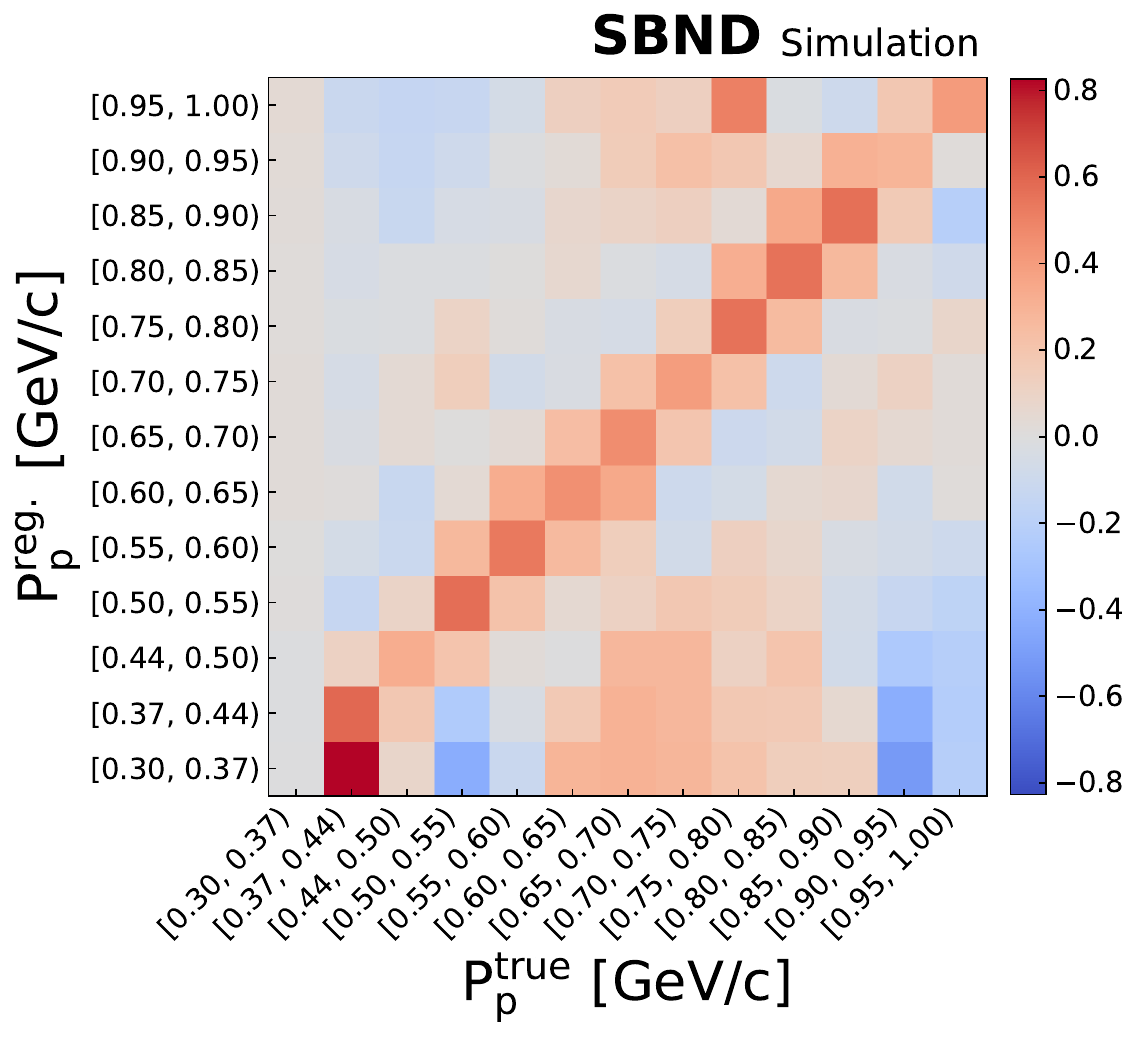} &
\includegraphics[width=0.32\textwidth]{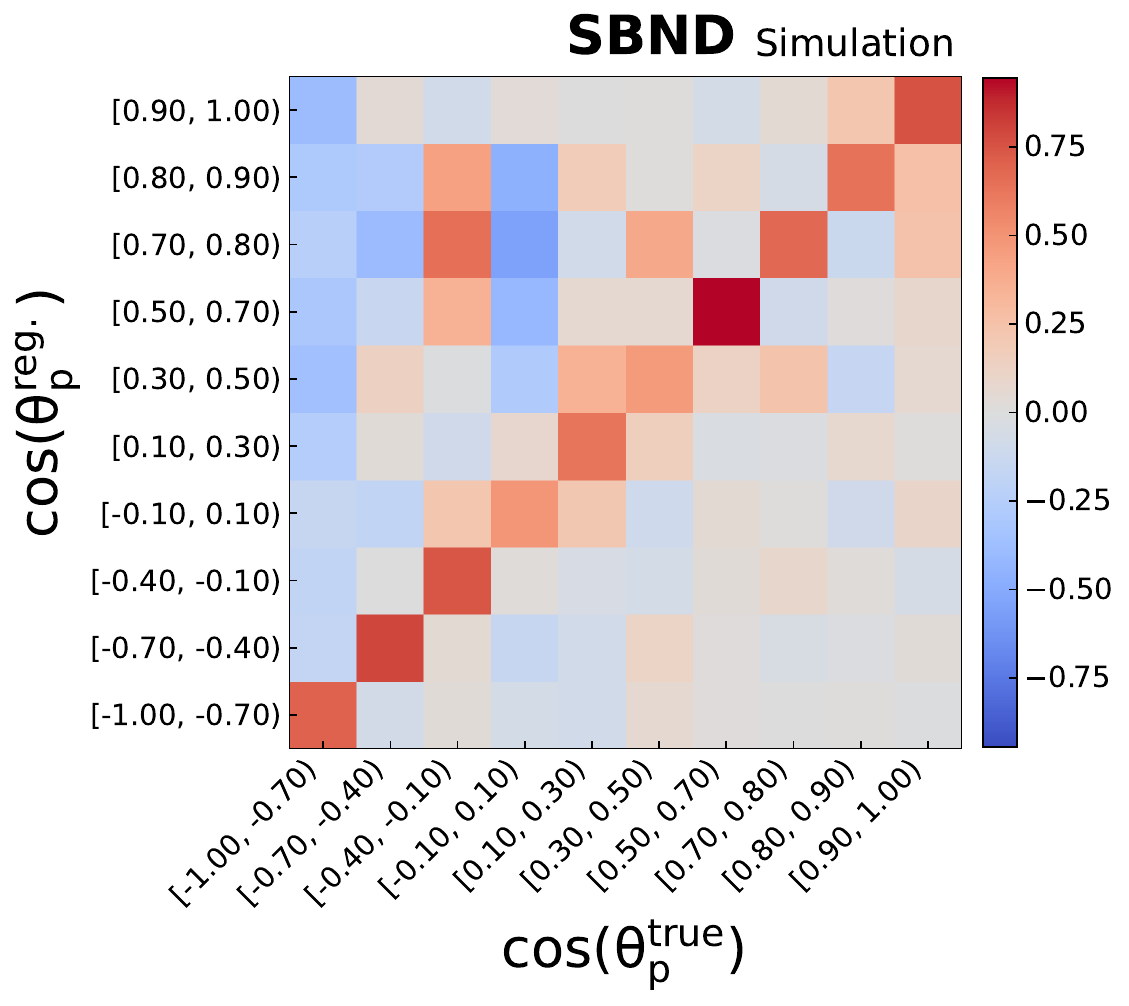} \\
(c) & (d) \\
\end{tabular}
\caption{Regularization smearing matrices $A_C$ for (a)~muon momentum, (b)~muon direction, (c)~proton momentum, and (d)~proton direction.}
\label{fig:ac-muon-proton}
\end{figure}

\begin{figure}[H]
\centering
\begin{tabular}{@{}ccc@{}}
\includegraphics[width=0.32\textwidth]{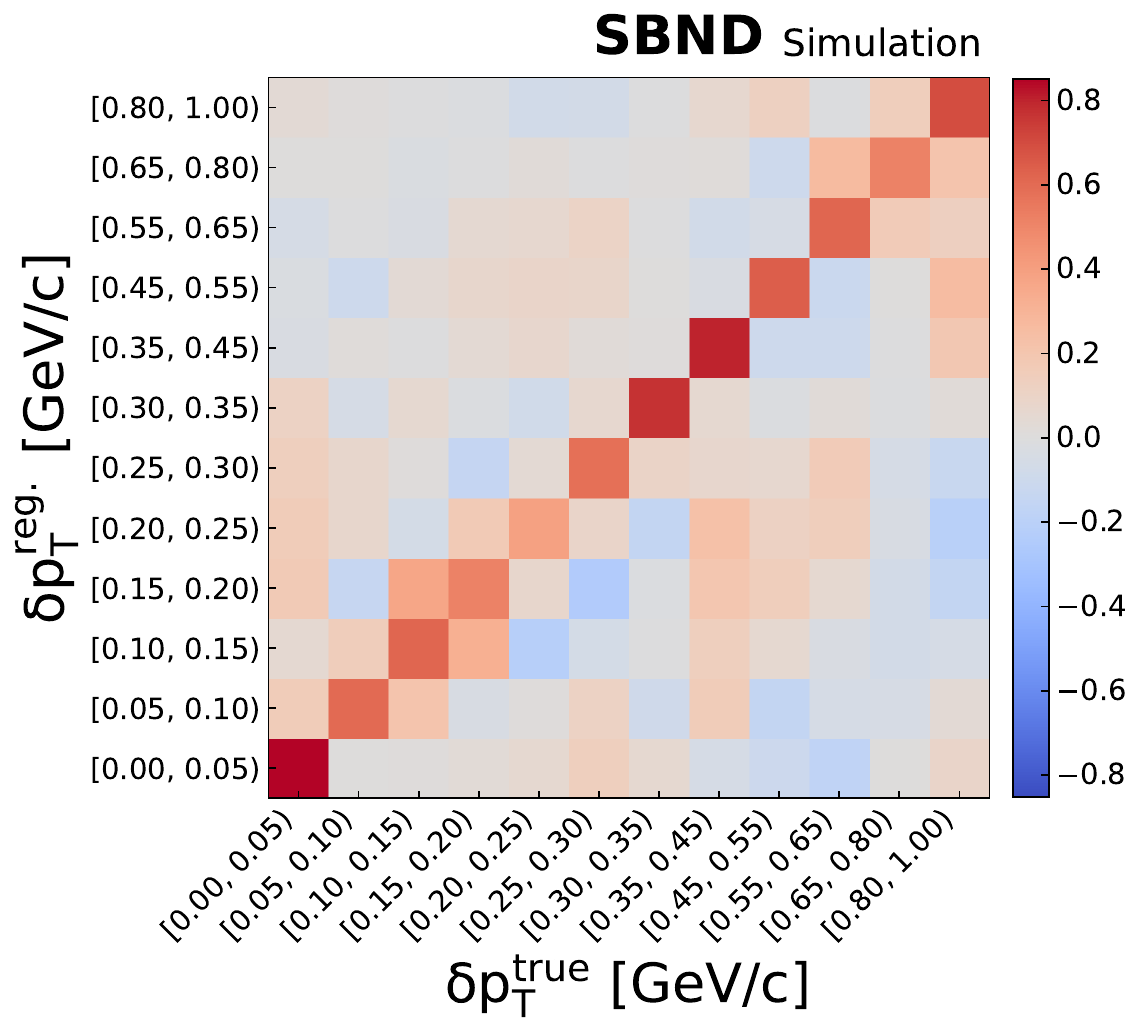} &
\includegraphics[width=0.32\textwidth]{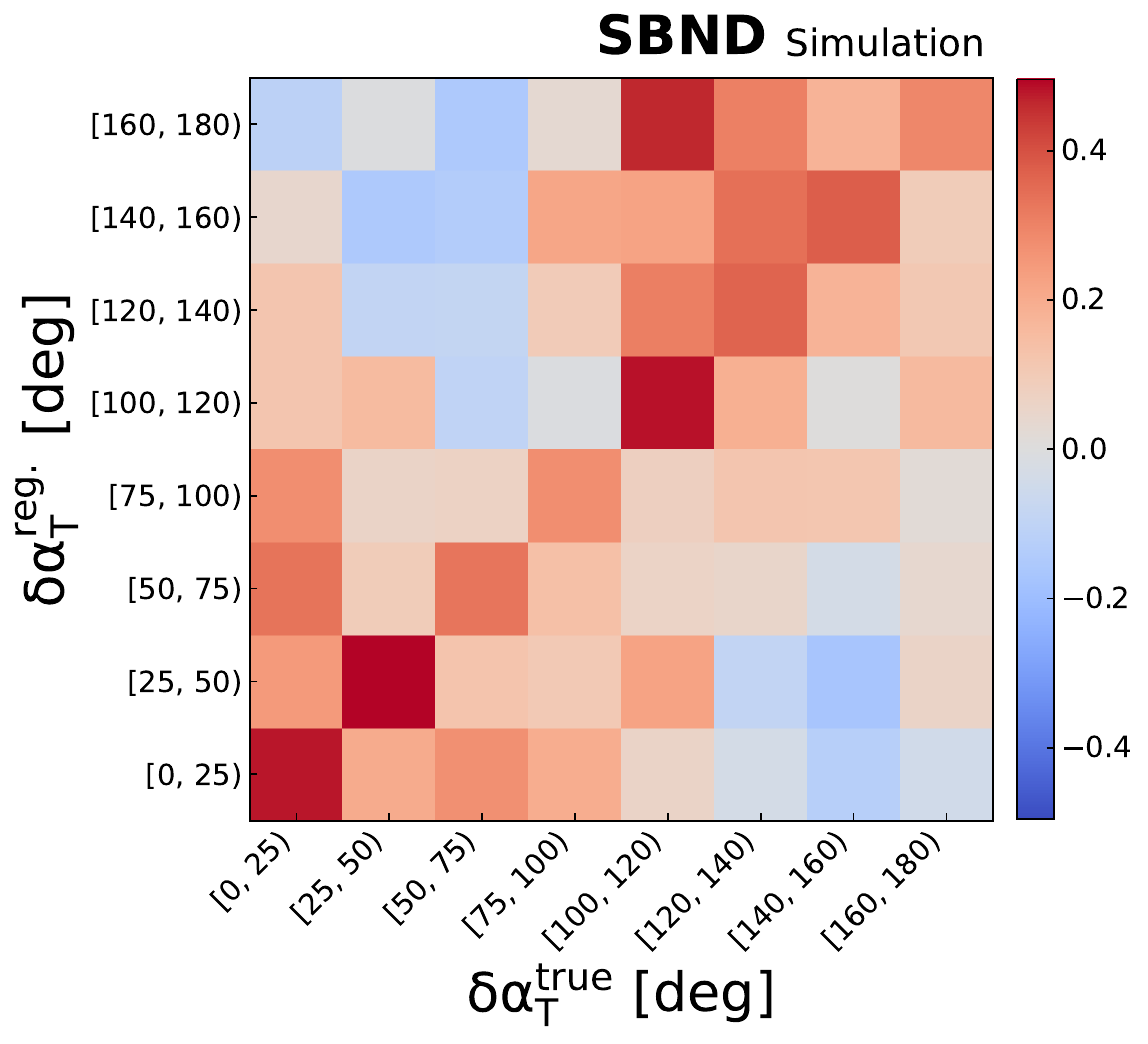} &
\includegraphics[width=0.32\textwidth]{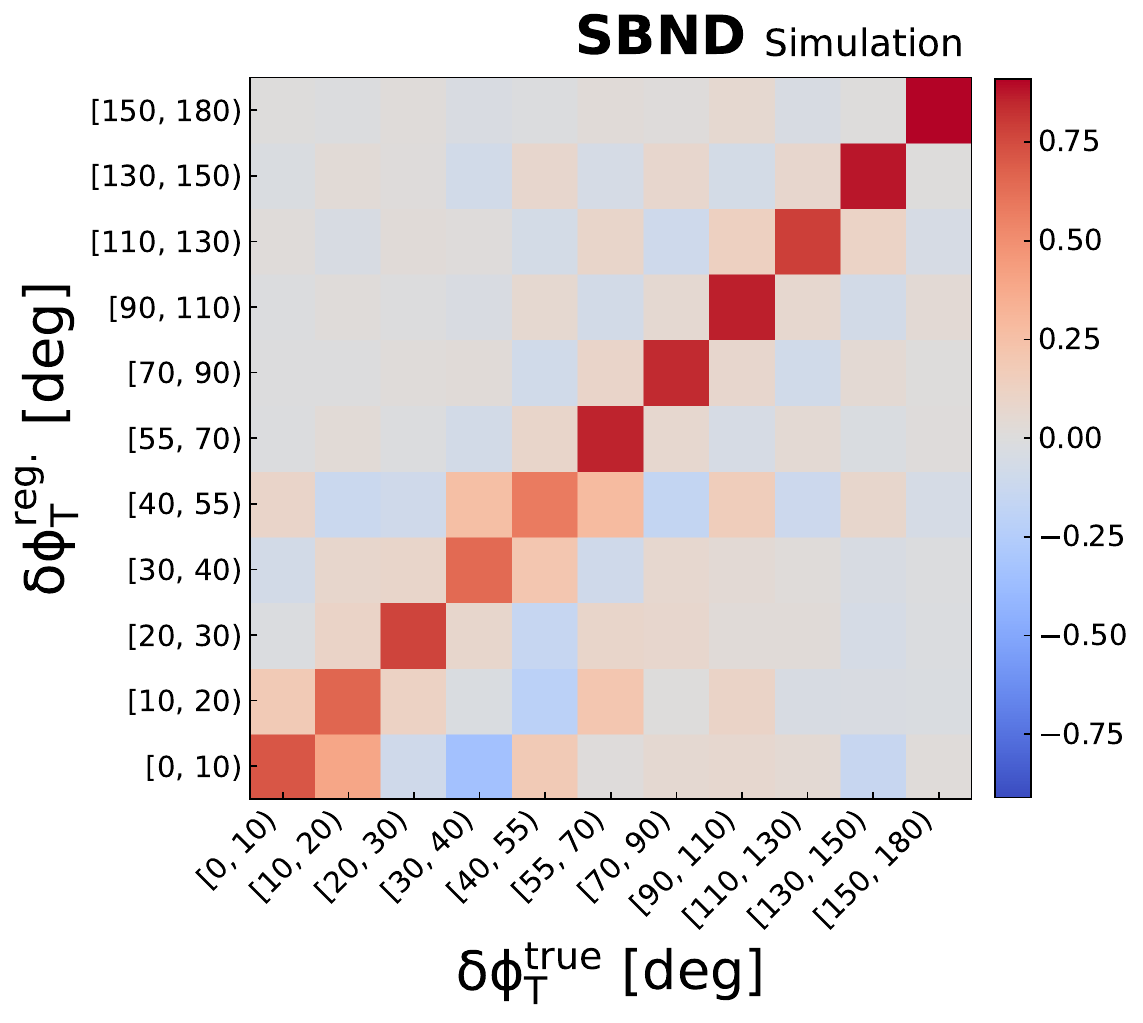} \\
(a) & (b) & (c) \\
\end{tabular}
\caption{Regularization smearing matrices $A_C$ for (a)~$\delta p_T$, (b)~$\delta\alpha_T$, and (c)~$\delta\phi_T$.}
\label{fig:ac-tki}
\end{figure}

In addition to the unfolded spectrum, Wiener-SVD unfolding returns a regularization smearing matrix $A_C$ that maps predictions from true space into the regularized space of the reported measurement, together with the covariance of the unfolded result in that space.
Figures~\ref{fig:ac-muon-proton} and~\ref{fig:ac-tki} show the $A_C$ matrices for muon and proton kinematics and for the TKI variables, respectively. Theory predictions must be multiplied by $A_C$ for comparison with the unfolded data.

\begin{figure}[H]
\centering
\begin{tabular}{@{}cc@{}}
\includegraphics[width=0.32\textwidth]{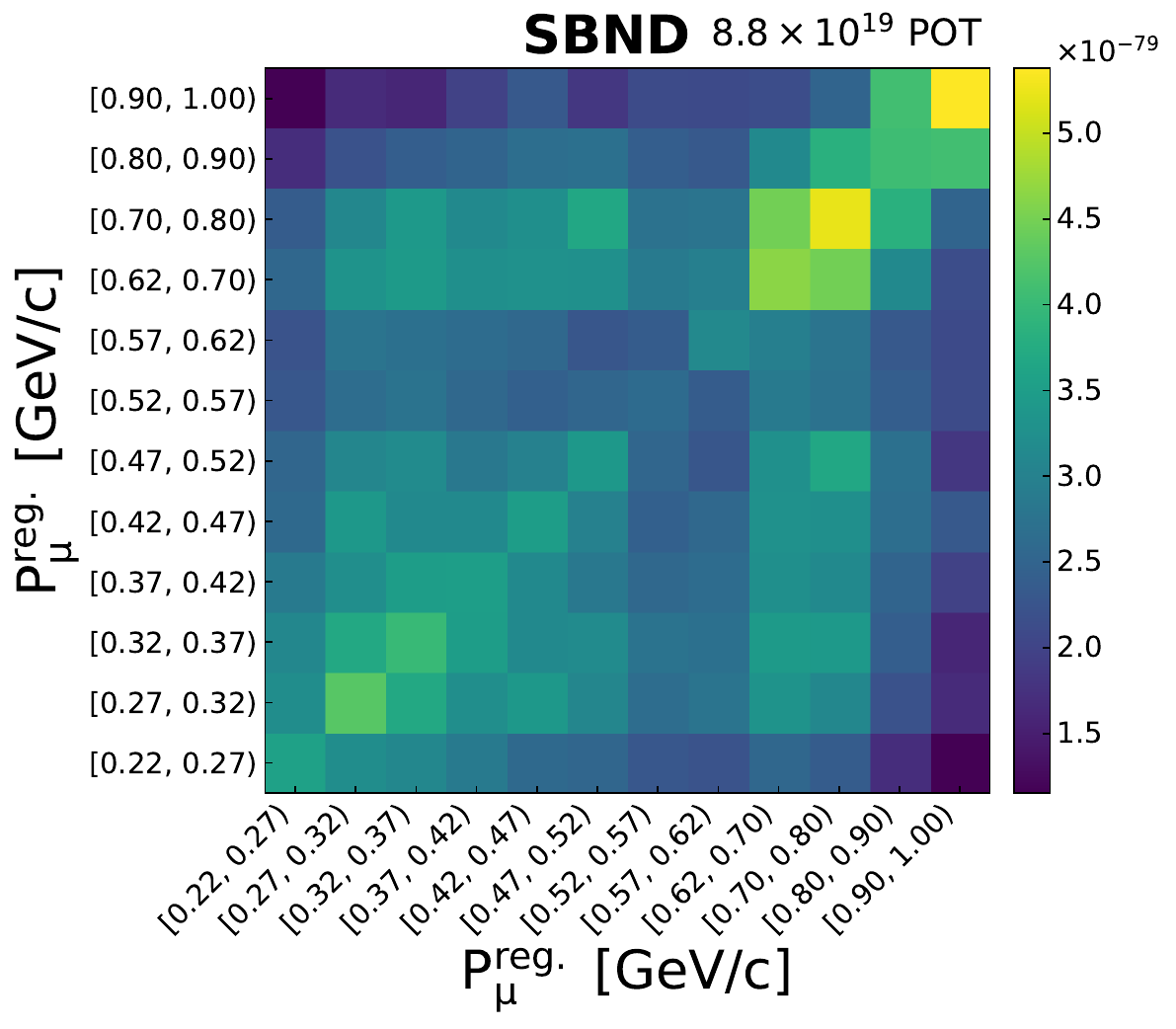} &
\includegraphics[width=0.32\textwidth]{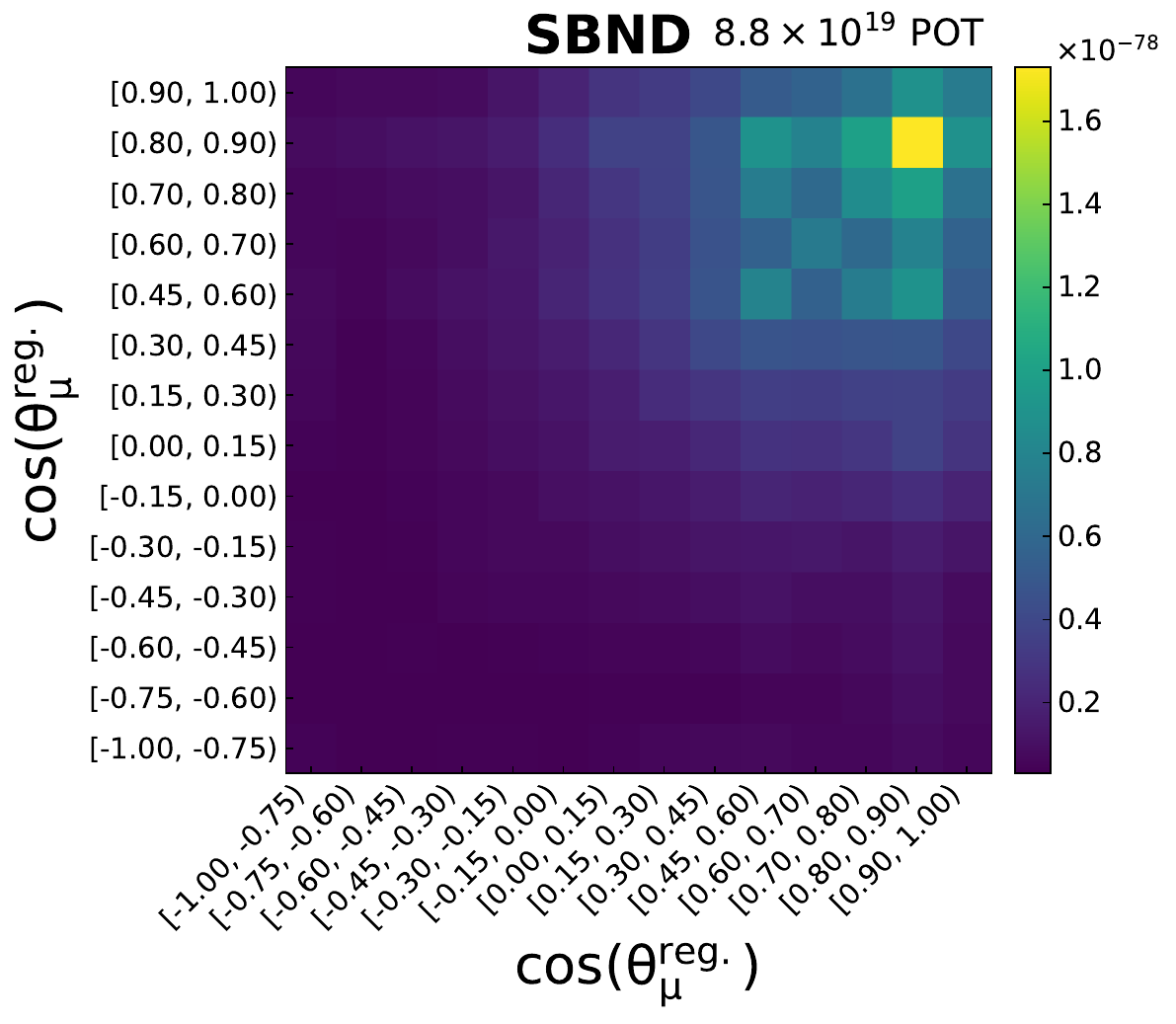}  \\
(a) & (b) \\
\includegraphics[width=0.32\textwidth]{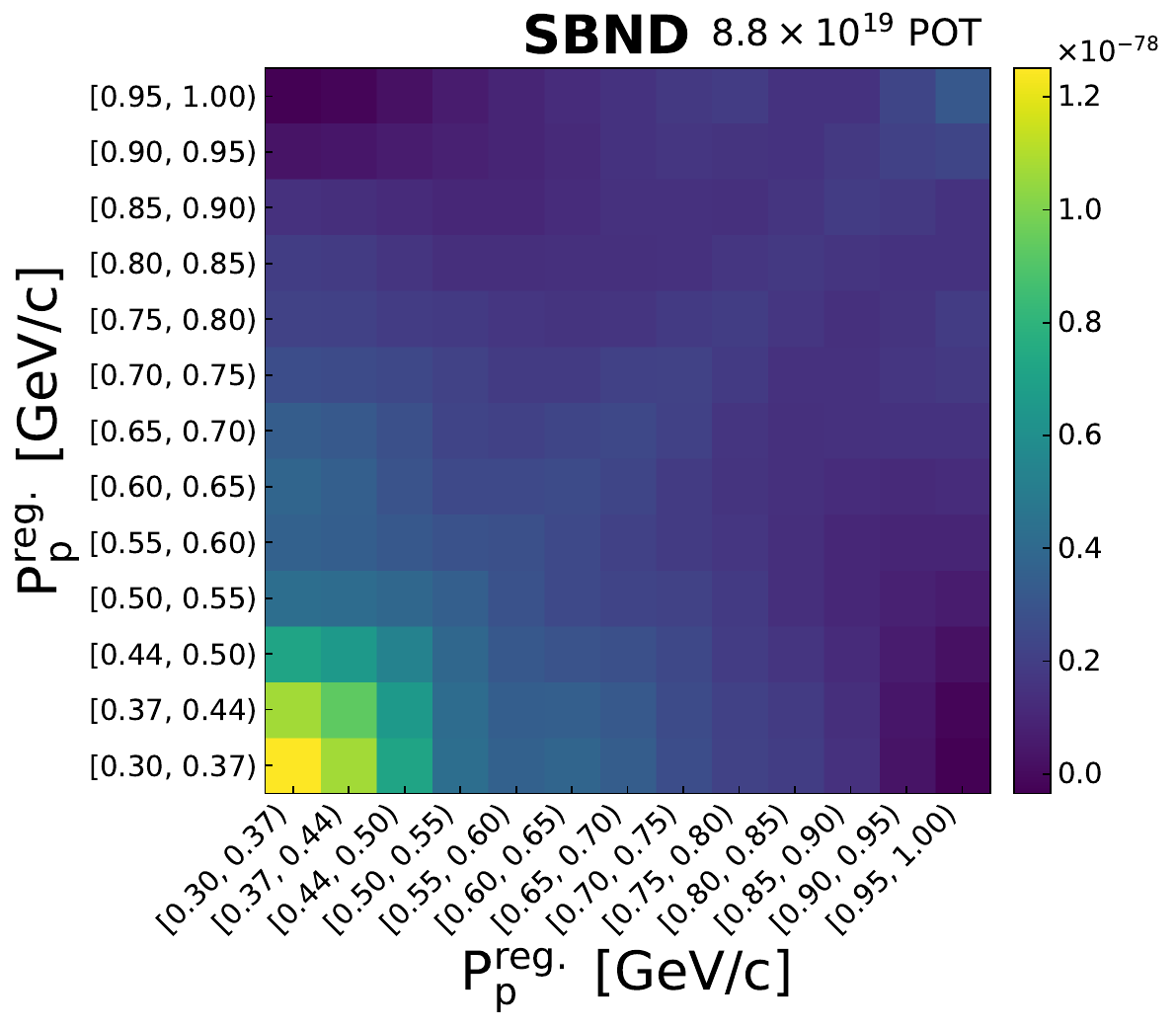} &
\includegraphics[width=0.32\textwidth]{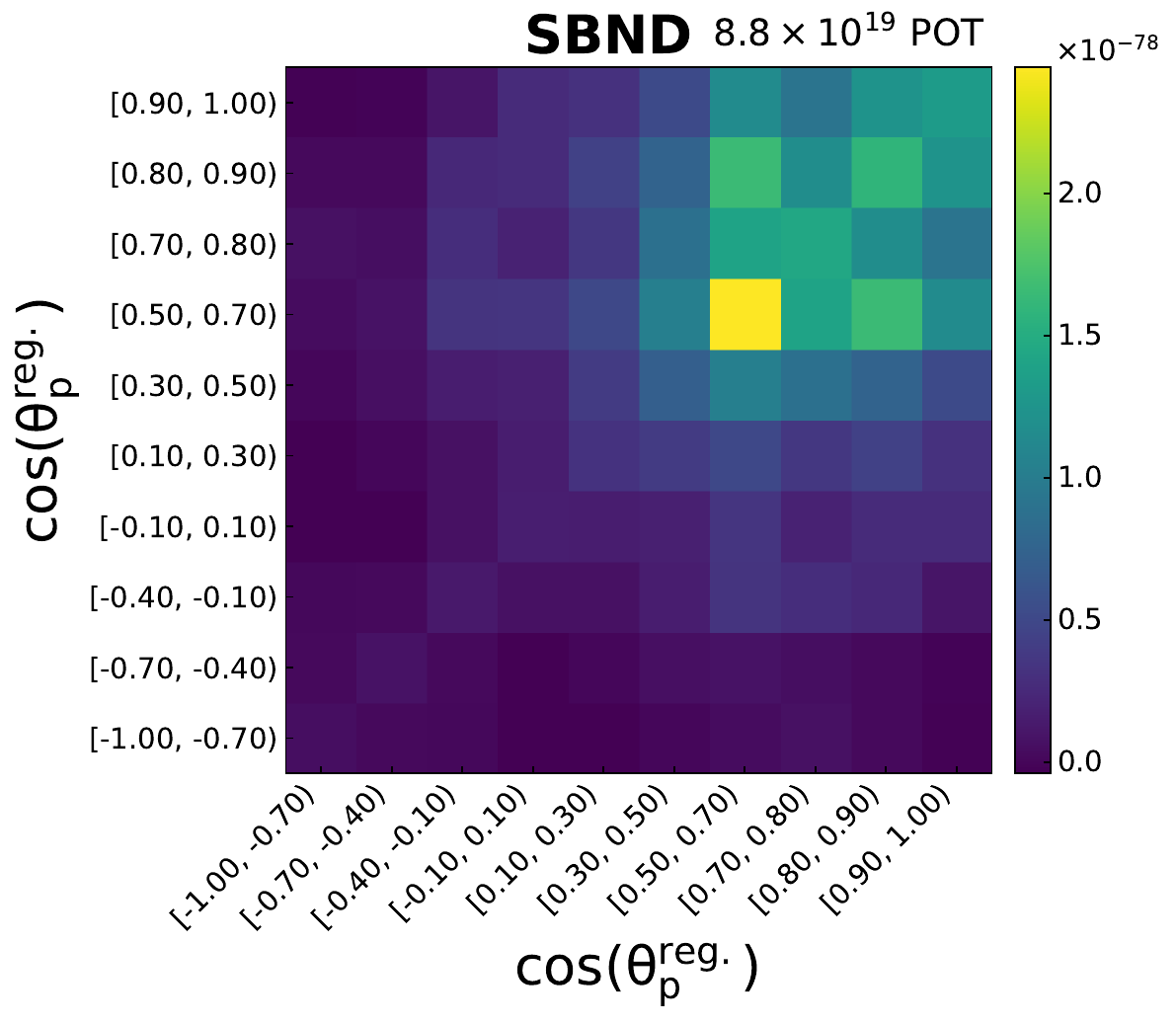} \\
(c) & (d) \\
\end{tabular}
\caption{Unfolded covariance matrices for (a)~muon momentum, (b)~muon direction, (c)~proton momentum, and (d)~proton direction.}
\label{fig:unfold-cov-syst-muon-proton}
\end{figure}

\begin{figure}[H]
\centering
\begin{tabular}{@{}ccc@{}}
\includegraphics[width=0.32\textwidth]{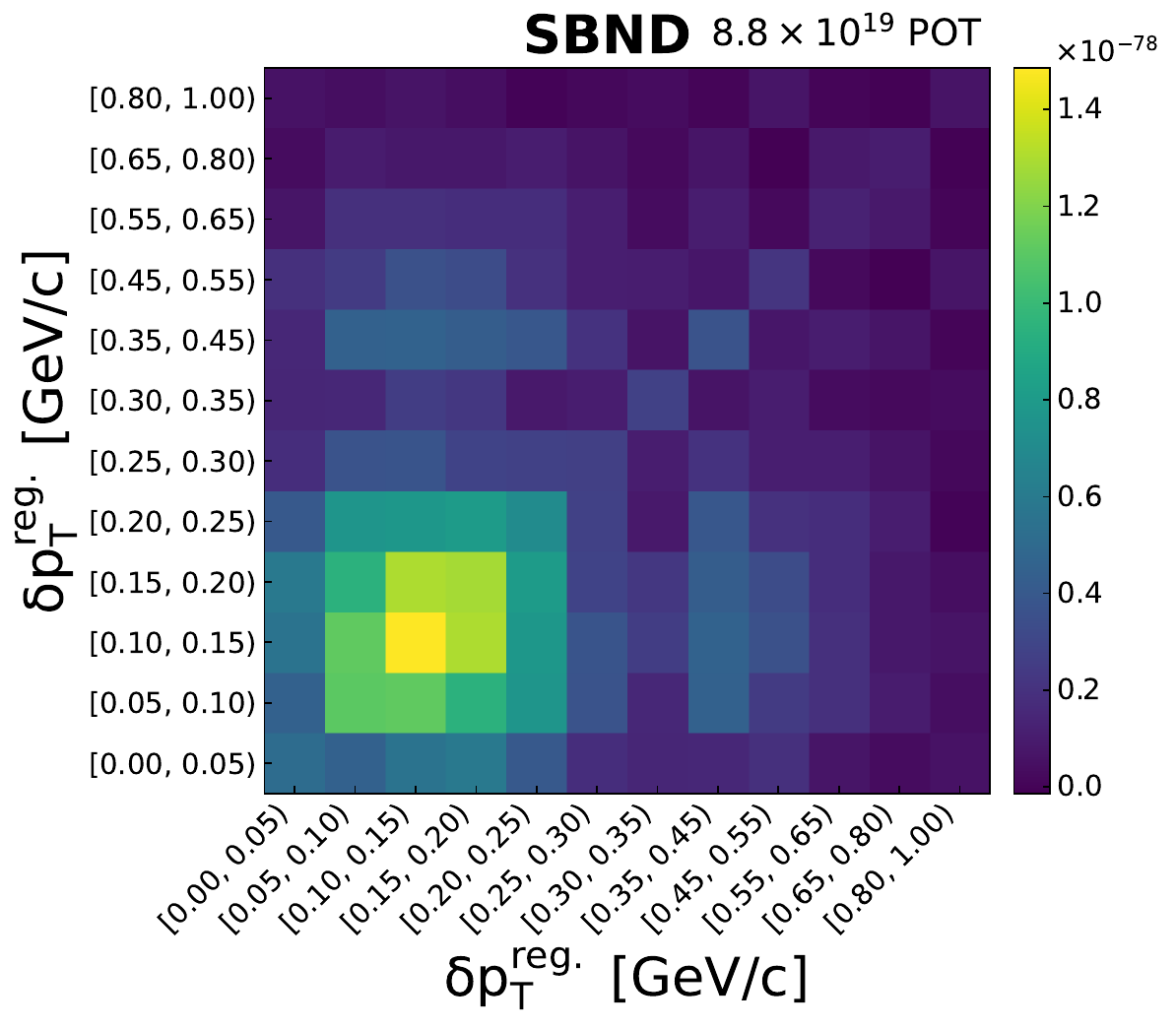} &
\includegraphics[width=0.32\textwidth]{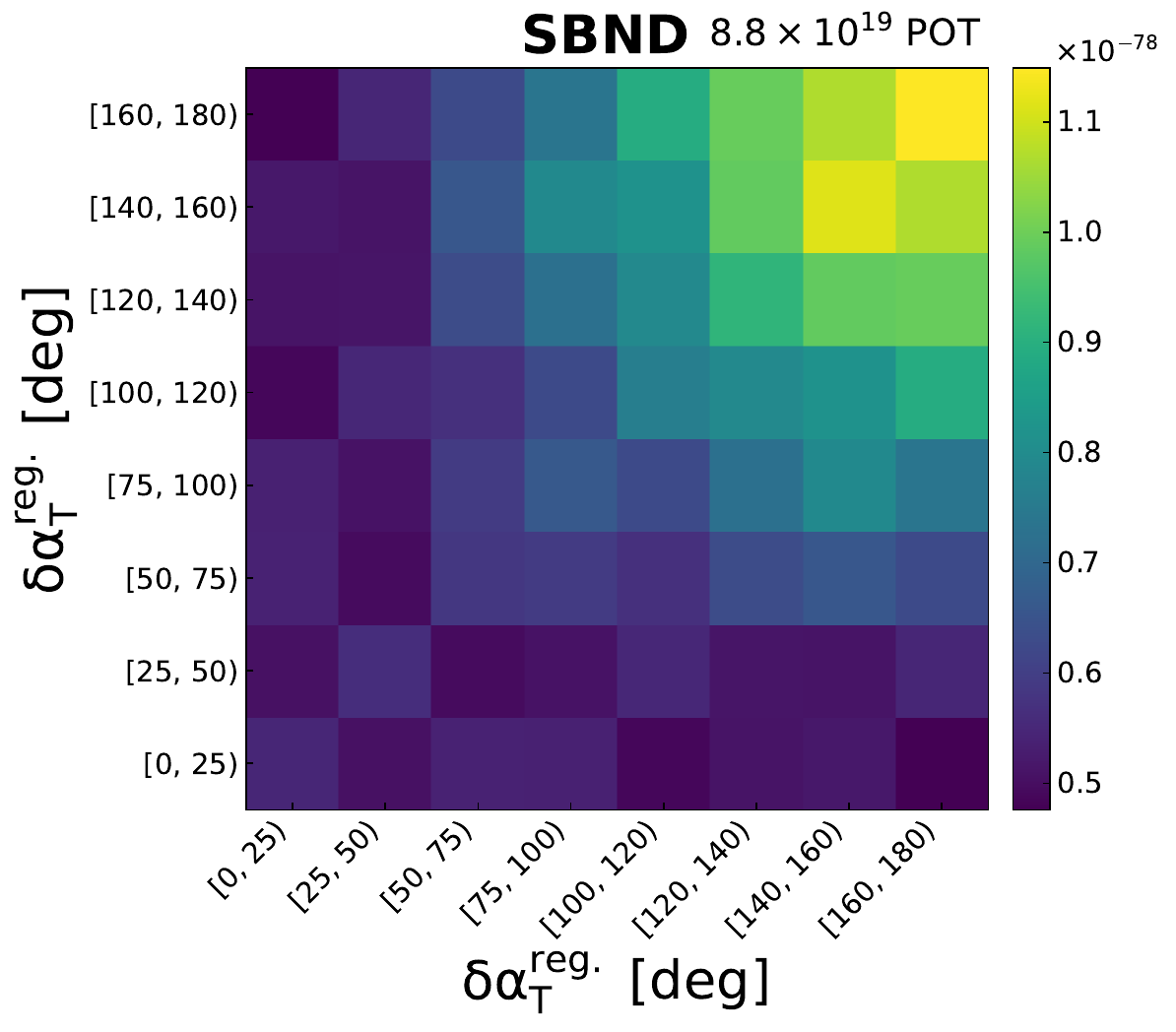} &
\includegraphics[width=0.32\textwidth]{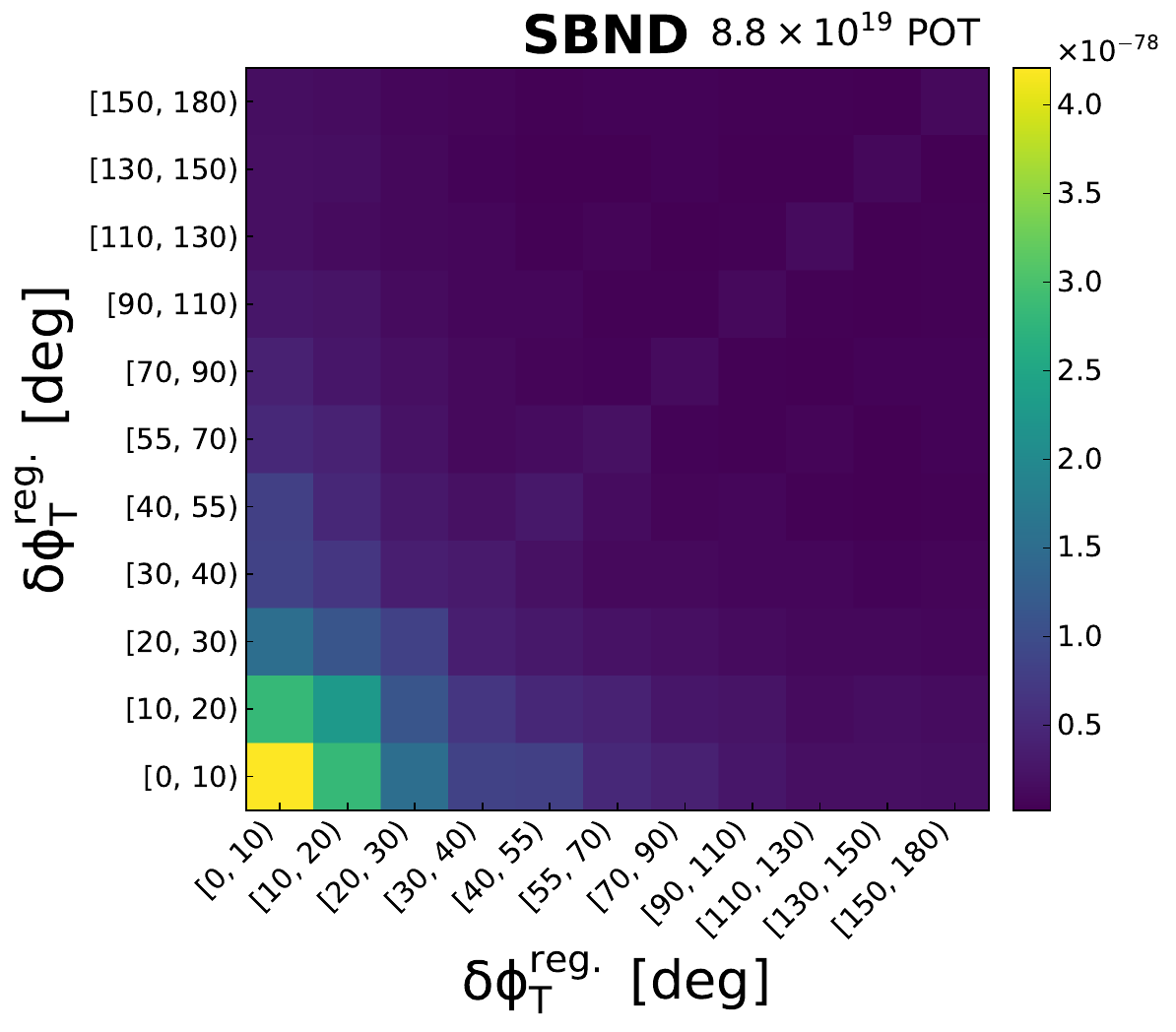} \\
(a) & (b) & (c) \\
\end{tabular}
\caption{Unfolded covariance matrices for (a)~$\delta p_T$, (b)~$\delta\alpha_T$, and (c)~$\delta\phi_T$.}
\label{fig:unfold-cov-syst-tki}
\end{figure}

Figures~\ref{fig:unfold-cov-syst-muon-proton} and~\ref{fig:unfold-cov-syst-tki} show the corresponding unfolded covariance matrices. These matrices encode the covariance of the unfolded cross sections in the regularized space.

\FloatBarrier
\newpage
\section{Model comparisons}
\label{sec:chi2}

Generator predictions are compared to the unfolded measurements as follows.
Each generator is run using the nominal flux described in Section~\ref{sec:flux}.
The generator outputs are processed with the \texttt{NUISANCE} framework~\cite{NUISANCE} to obtain predicted differential cross sections in the same signal definition and kinematic binning as the measurement.
These predictions are then multiplied by the corresponding regularization matrix $A_C$ and compared to the data in the regularized space of the reported results.

In the letter, we compare the unfolded cross sections to predictions from the following generators:
\begin{itemize}
\item \texttt{GENIE AR23\_20i\_00\_000}~\cite{GENIEAR23_0, GENIEAR23_1}: Uses thelocal Fermi gas (LFG) nuclear ground state, Nieves CCQE model~\cite{ValenciaModel}, SuSAv2 MEC (2p2h)~\cite{SuSAv2_MEC}, and hA2018 FSI model~\cite{hA2018}.

\item \texttt{GiBUU 2025}~\cite{Mosel:2019vhx}: Uses a quantum-kinetic transport treatment of initial- and final-state interactions in a common mean-field potential. Uses LFG nuclear ground state~\cite{GiBUU_LFG}, a standard CCQE model~\cite{GiBUU_CCQE}, and an empirical MEC model.

\item \texttt{NEUT v6.1.4}~\cite{NEUT}: Uses the Benhar SF nuclear ground state~\cite{Benhar_SF_1}, Valencia CCQE and MEC models~\cite{NEUT_CCQE,NEUT_MEC}, and an intranuclear-cascade FSI model with in-medium pion corrections.
\end{itemize}

\begin{figure}[H]
\centering
\begin{tabular}{@{}cc@{}}
\includegraphics[width=0.4\textwidth]{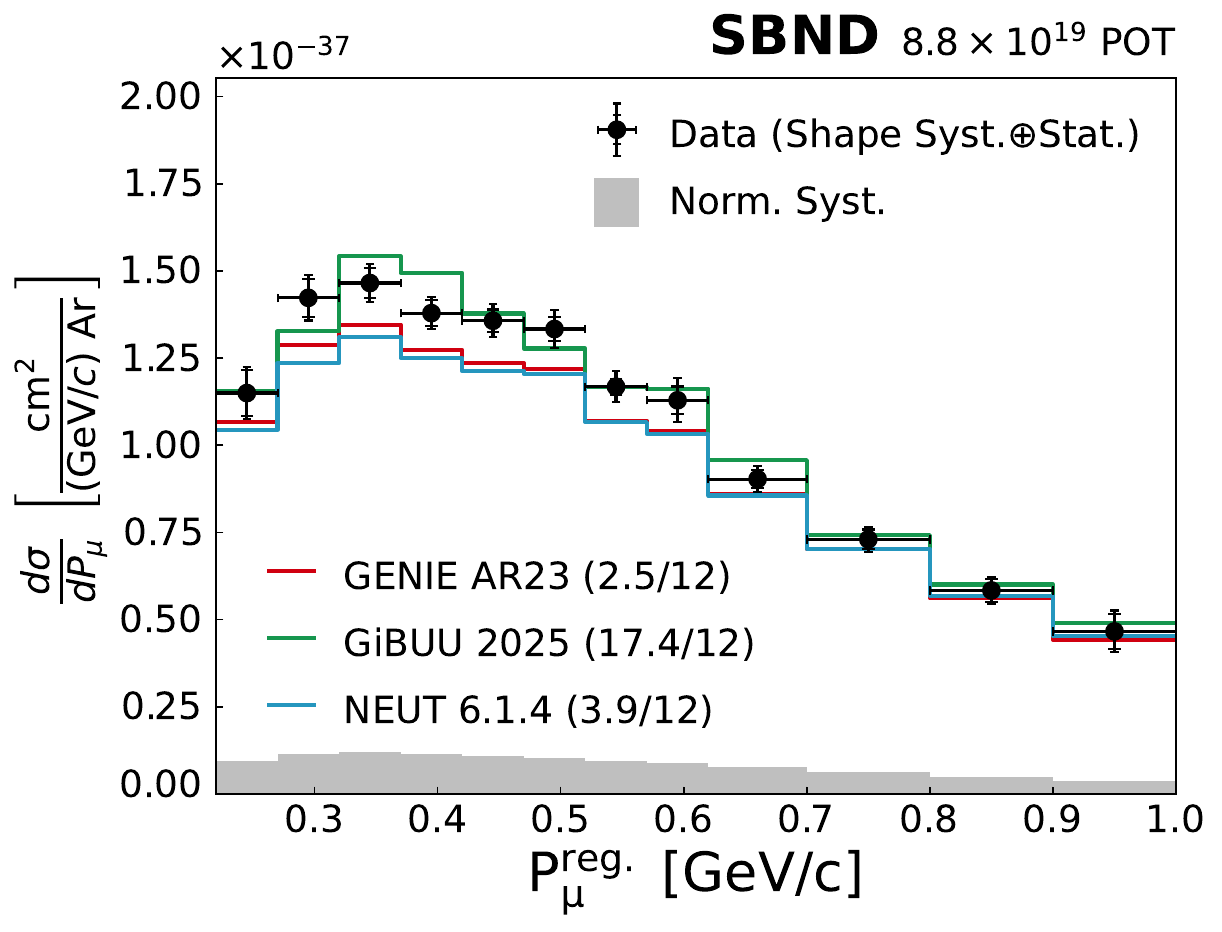} &
\includegraphics[width=0.4\textwidth]{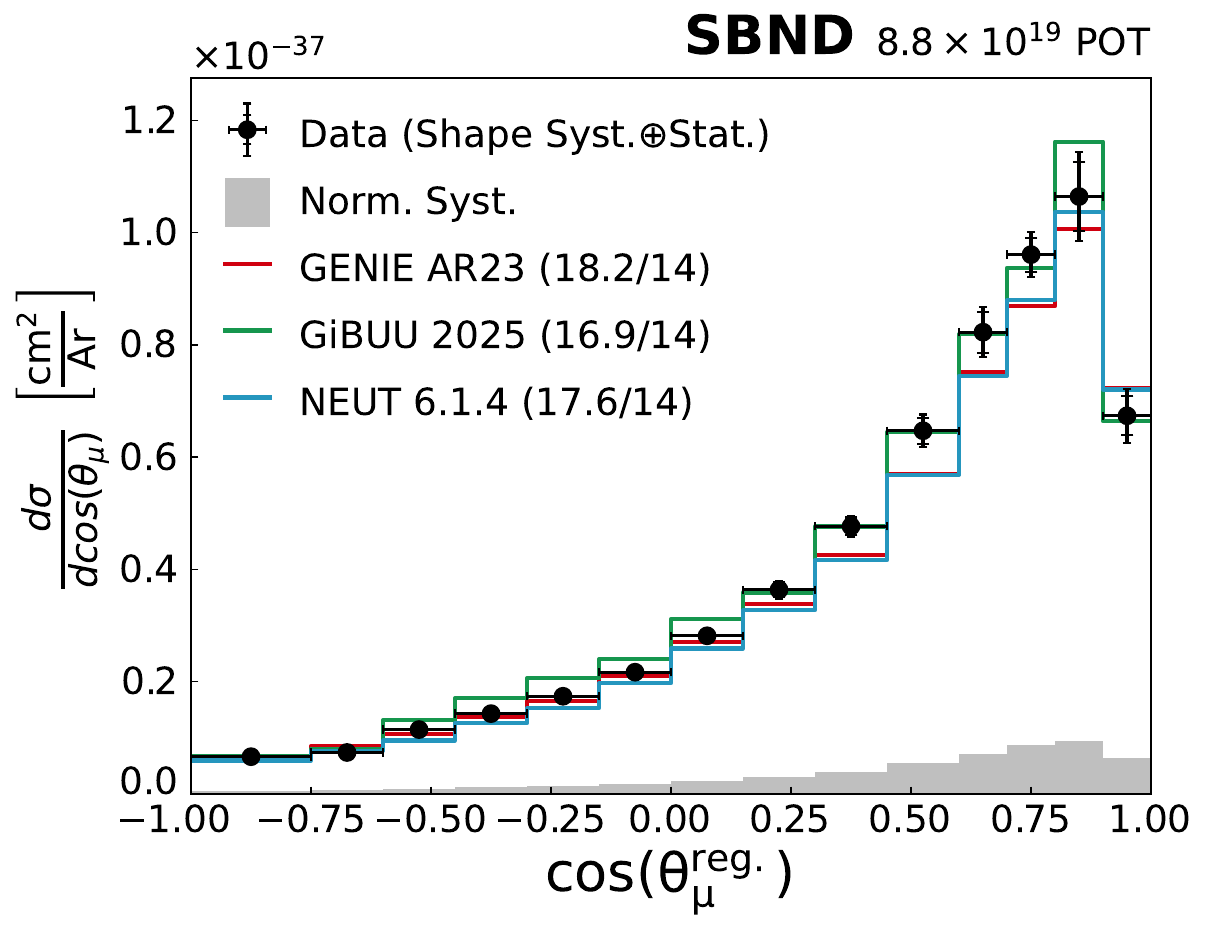}  \\
(a) & (b) \\
\includegraphics[width=0.4\textwidth]{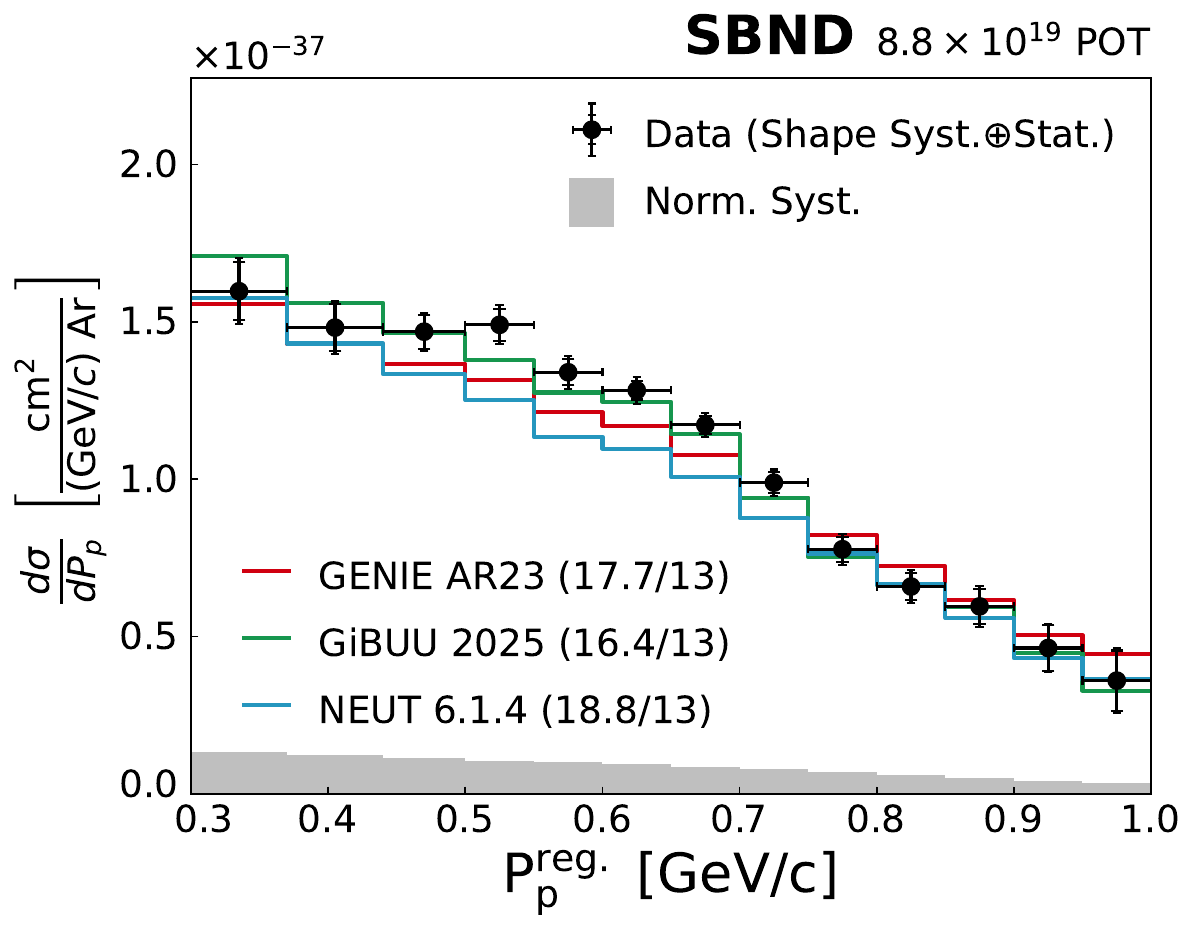} &
\includegraphics[width=0.4\textwidth]{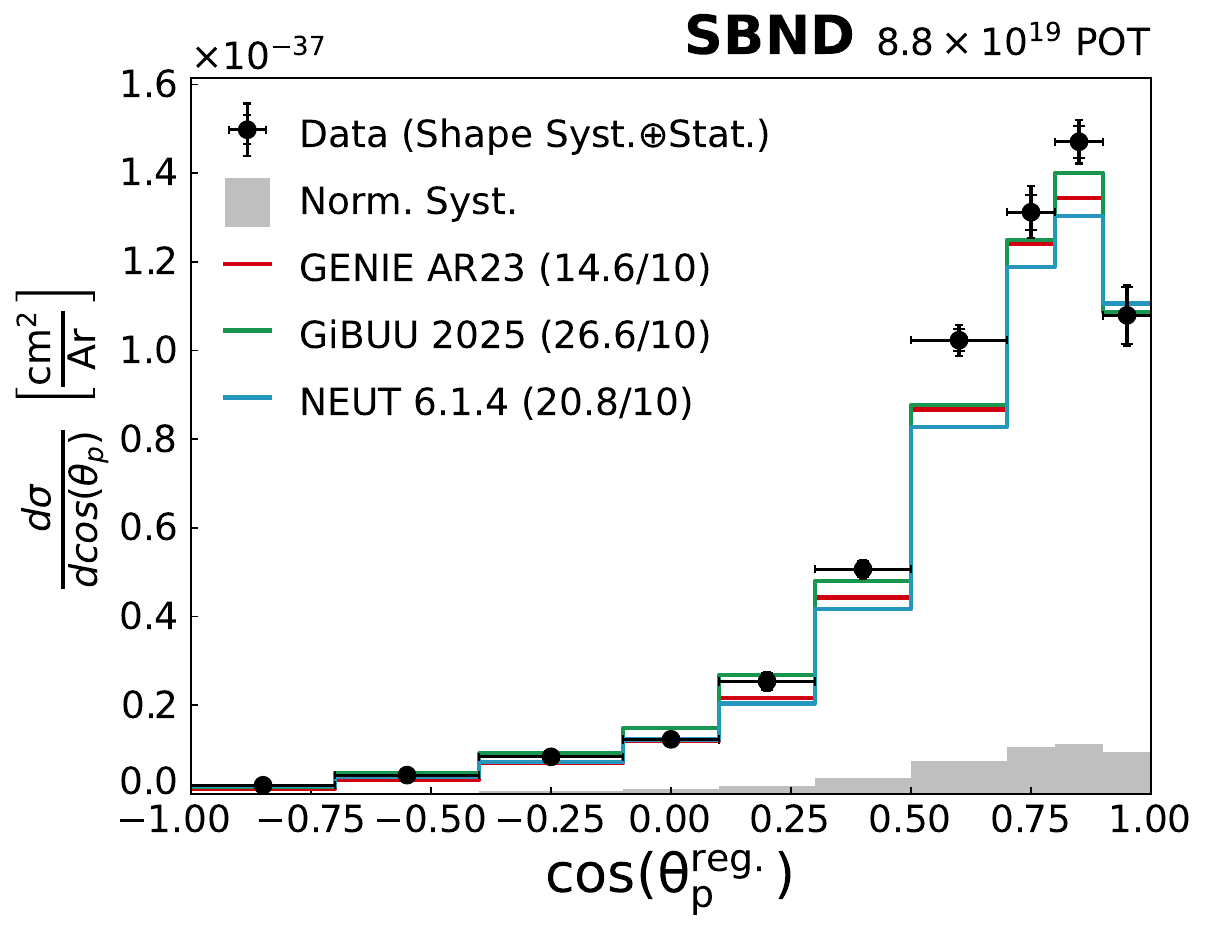} \\
(c) & (d) \\
\end{tabular}
\caption{
Unfolded, flux-averaged $\CCQE$ differential cross sections as functions of (a)~muon momentum, (b)~muon direction, (c)~proton momentum, and (d)~proton direction, compared to \texttt{GENIE AR23\_20i\_00\_000}, \texttt{GiBUU 2025}~\cite{Mosel:2019vhx}, and \texttt{NEUT v6.1.4}~\cite{NEUT}. Inner error bars denote statistical uncertainties, while outer error bars represent the statistical and shape-only systematic uncertainties added in quadrature. The gray bands at the bottom indicate normalization uncertainties. Values of $\chi^2/\mathrm{ndof}$, evaluated using the full unfolded covariance matrix, are given in parentheses.
}
\label{fig:generators-muon-proton}
\end{figure}

The total flux-averaged cross section is $(8.18\pm 0.08_{\text{stat}}\pm 0.66_{\text{syst}}) \times 10^{-38}$ cm$^2$/Ar, at a flux-weighted mean neutrino energy of $\langle E_\nu \rangle = 0.75$ GeV. The unfolded flux-averaged differential cross sections as functions of muon and proton kinematics, compared to the same nominal generators reported in the Letter, is shown in Figure~\ref{fig:generators-muon-proton}.


\begin{figure}[H]
\centering
\begin{tabular}{@{}ccc@{}}
\includegraphics[width=0.328\textwidth]{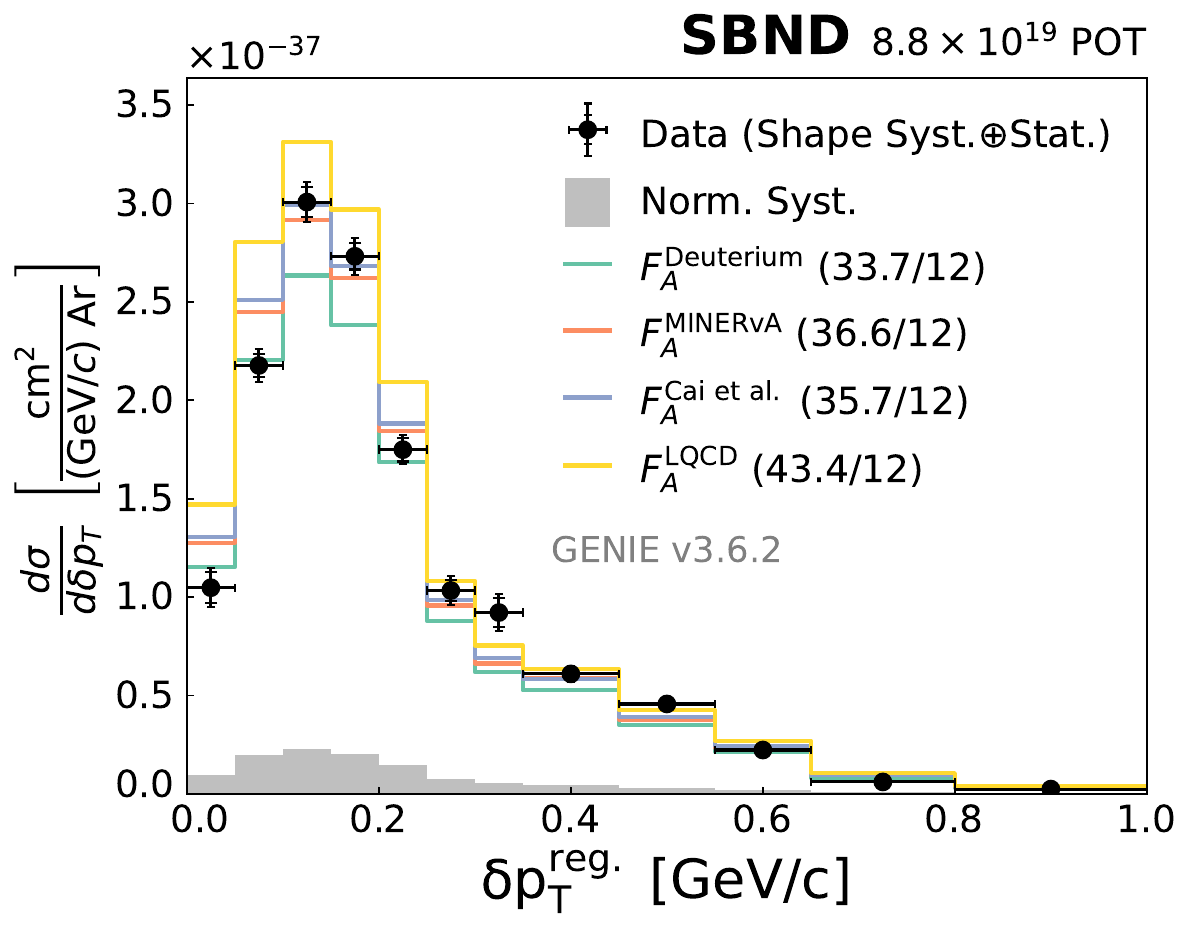} &
\includegraphics[width=0.322\textwidth]{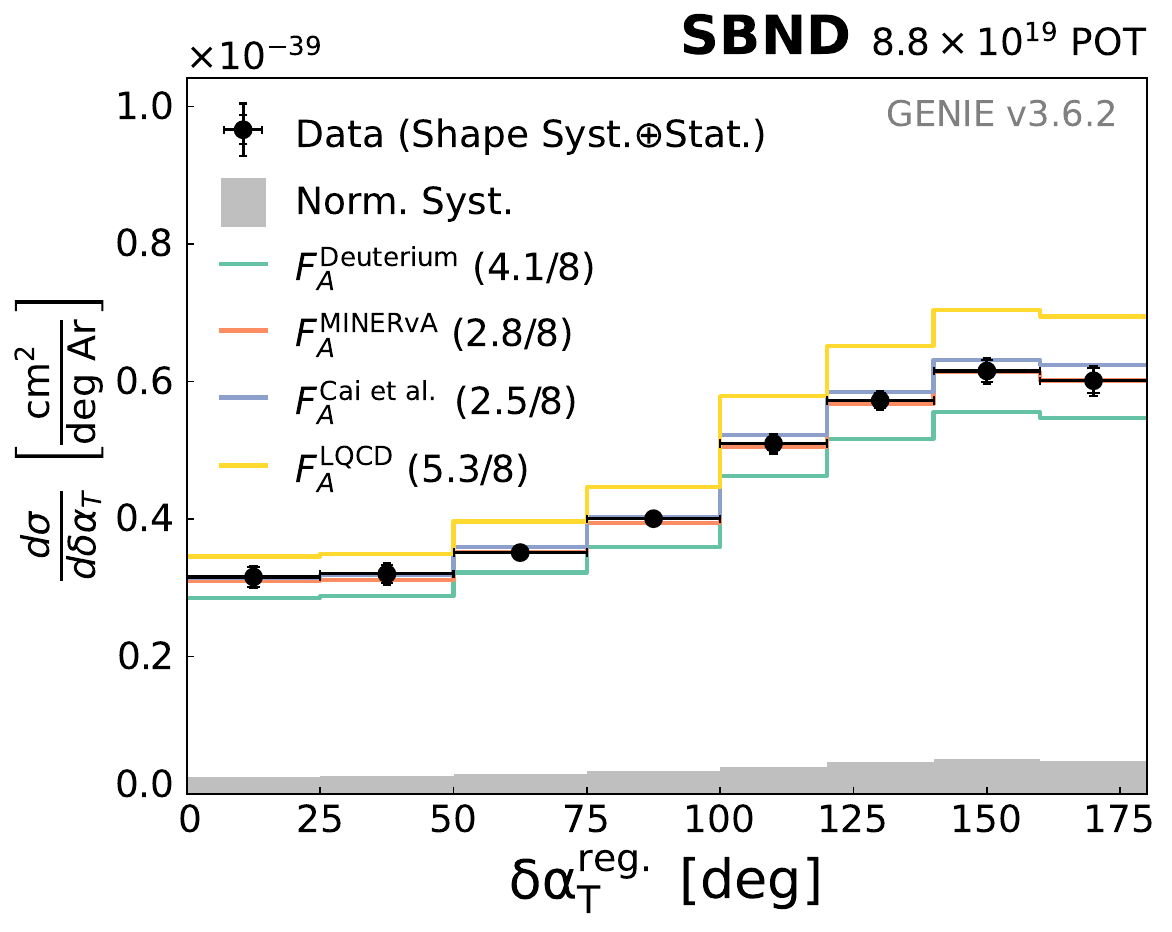} &
\includegraphics[width=0.321\textwidth]{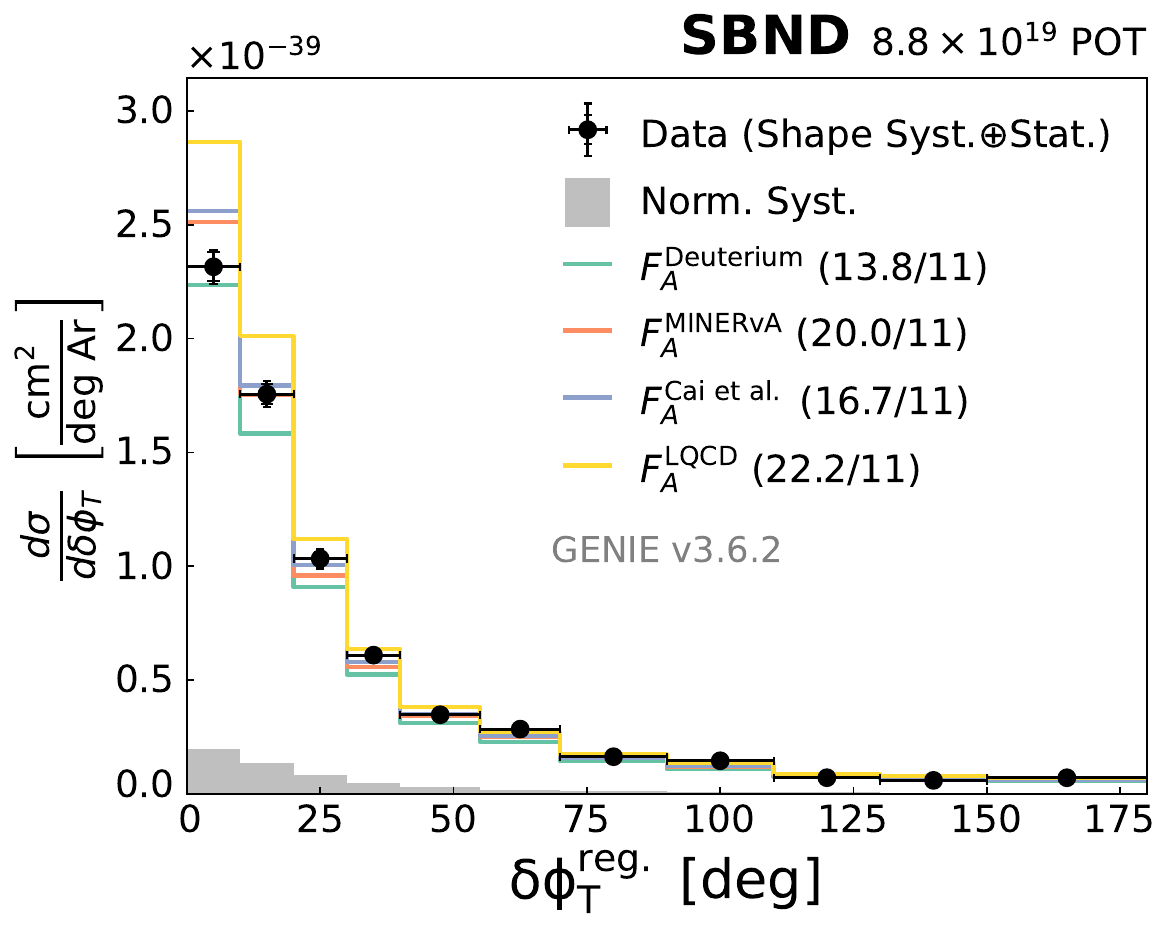} \\
(a) & (b) & (c) \\
\end{tabular}
\caption{
Unfolded, flux-averaged $\CCQE$ differential cross sections as functions of (a)~$\delta p_T$, (b)~$\delta\alpha_T$, and (c)~$\delta\phi_T$, compared with \texttt{GENIE AR23} predictions using alternate axial-vector form factor models. Inner error bars denote statistical uncertainties, while outer error bars represent the statistical and shape-only systematic uncertainties added in quadrature. The gray bands at the bottom indicate normalization uncertainties. Values of $\chi^2/\mathrm{ndof}$, evaluated using the full unfolded covariance matrix, are given in parentheses.}
\label{fig:ar23-tki}
\end{figure}

In addition, we compare the unfolded cross sections to predictions from variations on the \texttt{GENIE} generator. 
We first compare the variations on the axial form factor model, as shown in Fig.~\ref{fig:ar23-tki}. The \texttt{GENIE v3.6.0 AR23\_20i\_00\_000} model with the Nieves QE model bug fix presented in Ref.~\cite{deltp_benchmark} is used as the baseline model. We consider the following form factor variations: MINERvA $z$-expansion with full sum rules, MINERvA $z$-expansion from Ref.~\cite{MINERvA_FormFactor}, and lattice QCD (LQCD) $z$-expansion with full sum rules~\cite{LQCD_FormFactor}.

\begin{figure}[H]
\centering
\begin{tabular}{@{}ccc@{}}
\includegraphics[width=0.333\textwidth]{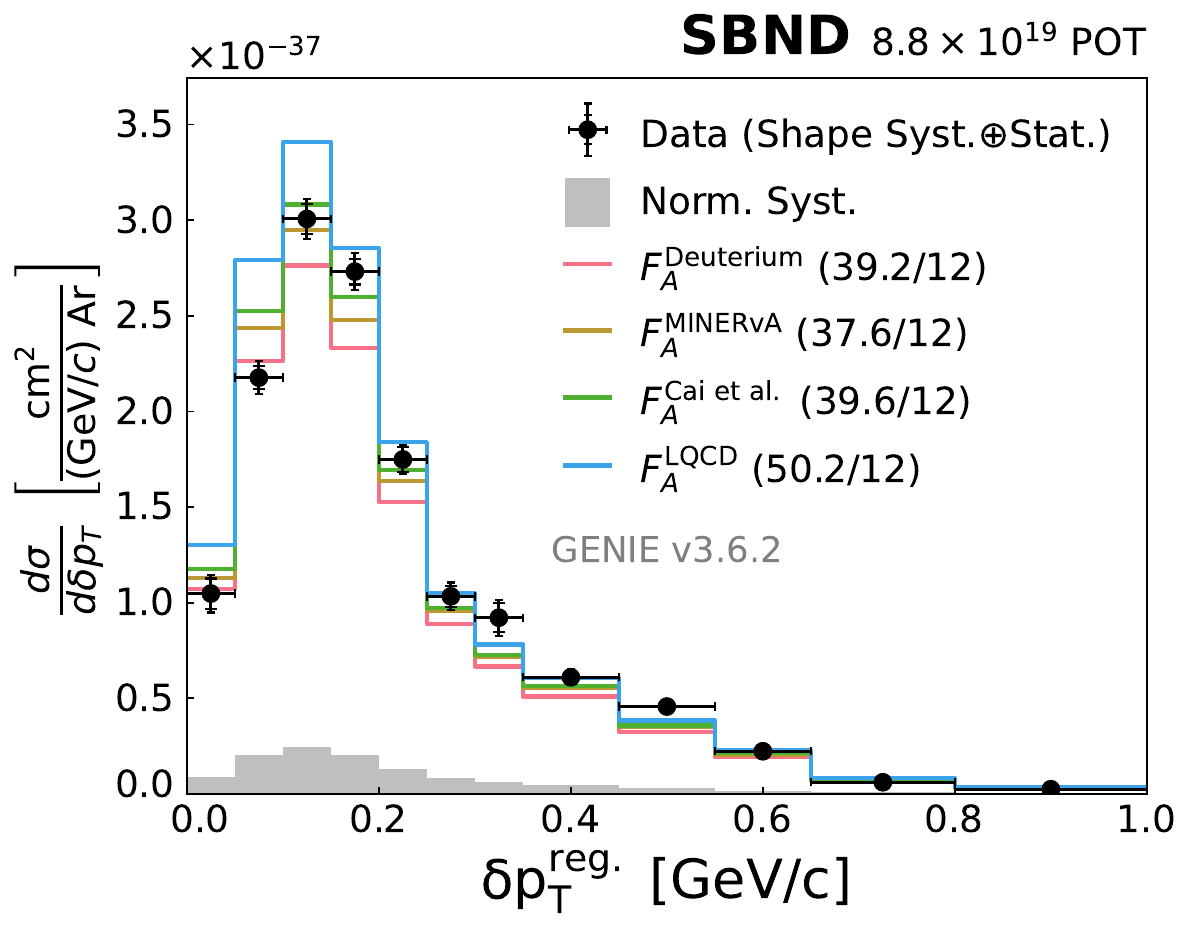} &
\includegraphics[width=0.317\textwidth]{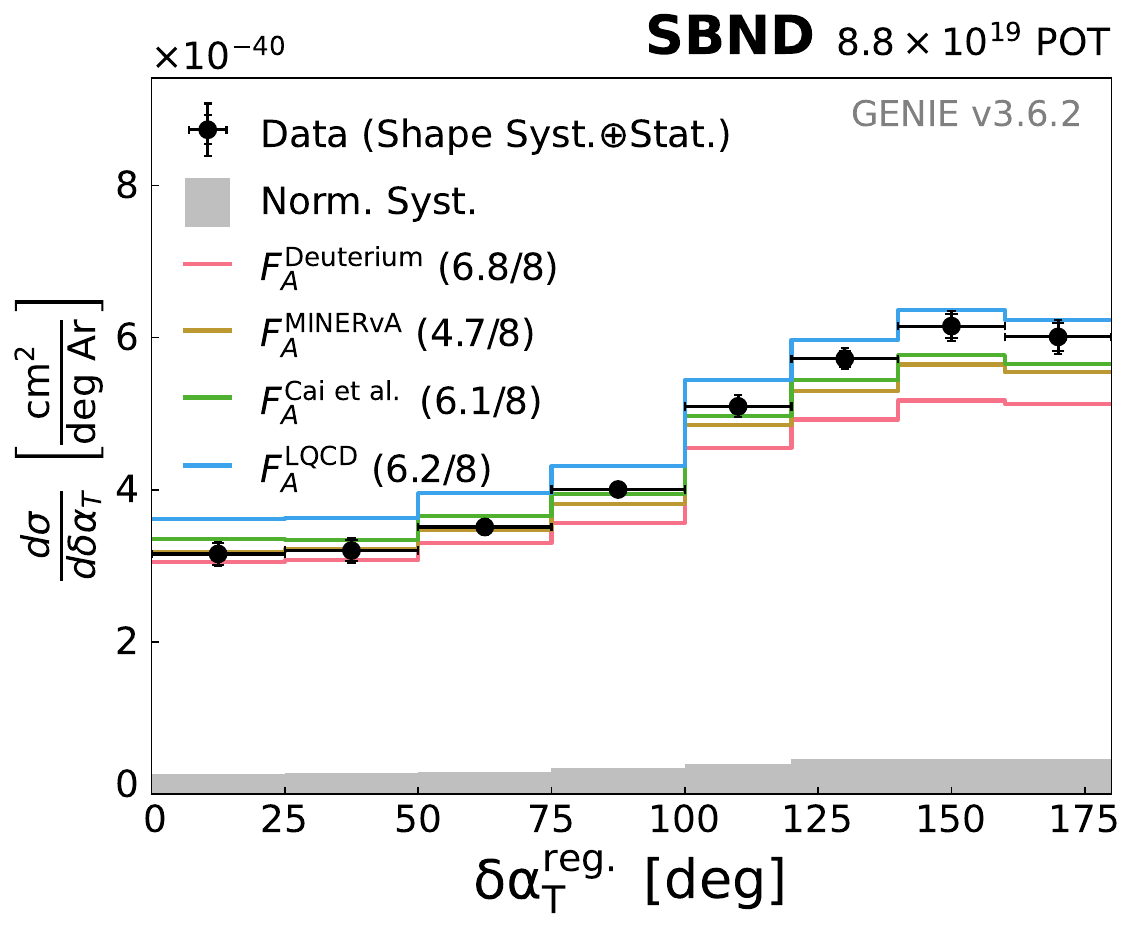} &
\includegraphics[width=0.327\textwidth]{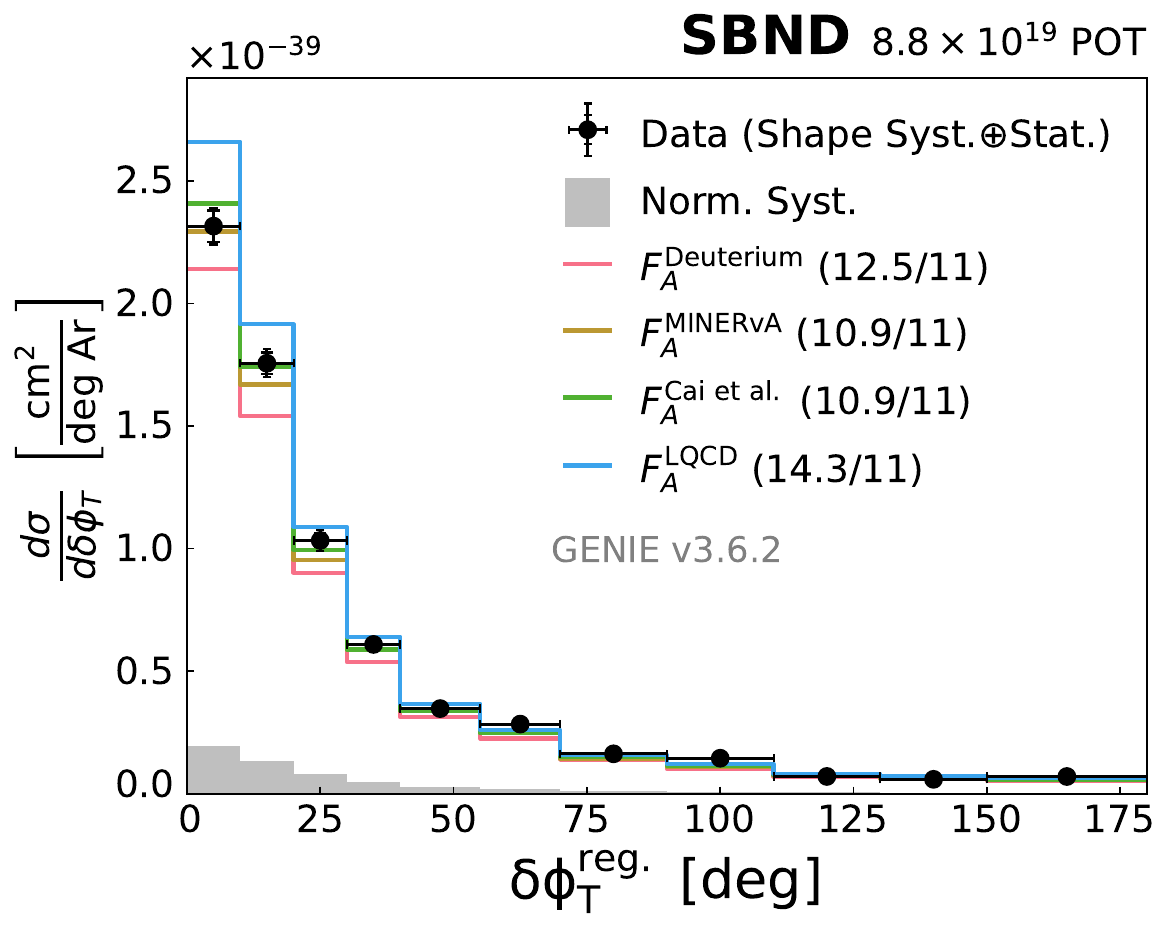} \\
(a) & (b) & (c) \\
\end{tabular}
\caption{
Unfolded, flux-averaged $\CCQE$ differential cross sections as functions of (a)~$\delta p_T$, (b)~$\delta\alpha_T$, and (c)~$\delta\phi_T$, compared with \texttt{GENIE} predictions using a spectral-function nuclear ground state and alternate axial-vector form-factor parametrizations~\cite{MINERvA_FormFactor,LQCD_FormFactor}. Inner error bars denote statistical uncertainties, while outer error bars represent the statistical and shape-only systematic uncertainties added in quadrature. The gray bands at the bottom indicate normalization uncertainties. Values of $\chi^2/\mathrm{ndof}$, evaluated using the full unfolded covariance matrix, are given in parentheses.
}
\label{fig:ar25-tki}
\end{figure}

Next we compare to models using the spectral function (SF) nuclear ground state, with the random-phase approximation (RPA) correction turned off. In addition, we test the above form factor variations with the SF nuclear ground state, as shown in Fig.~\ref{fig:ar25-tki}.

\begin{figure}[H]
\centering
\begin{tabular}{@{}ccc@{}}
\includegraphics[width=0.333\textwidth]{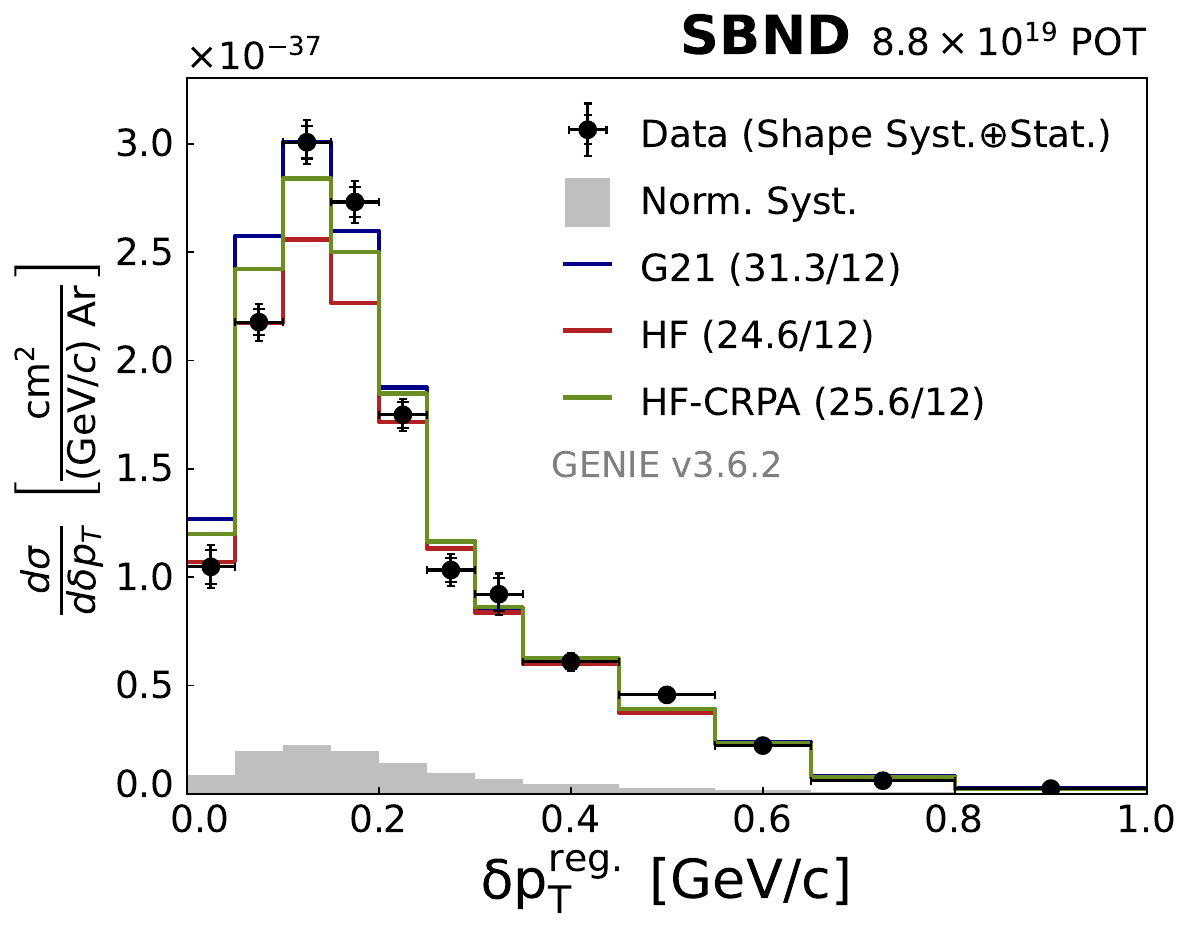} &
\includegraphics[width=0.317\textwidth]{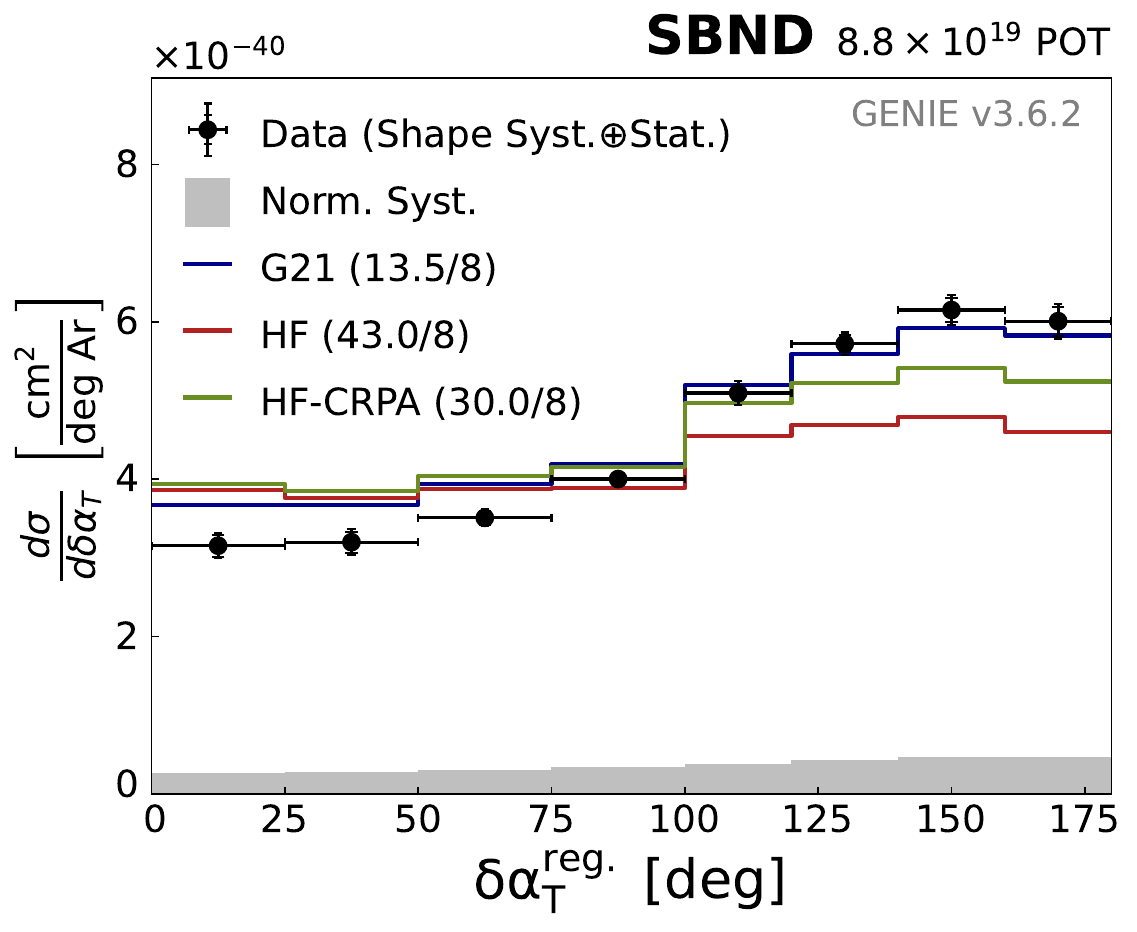} &
\includegraphics[width=0.327\textwidth]{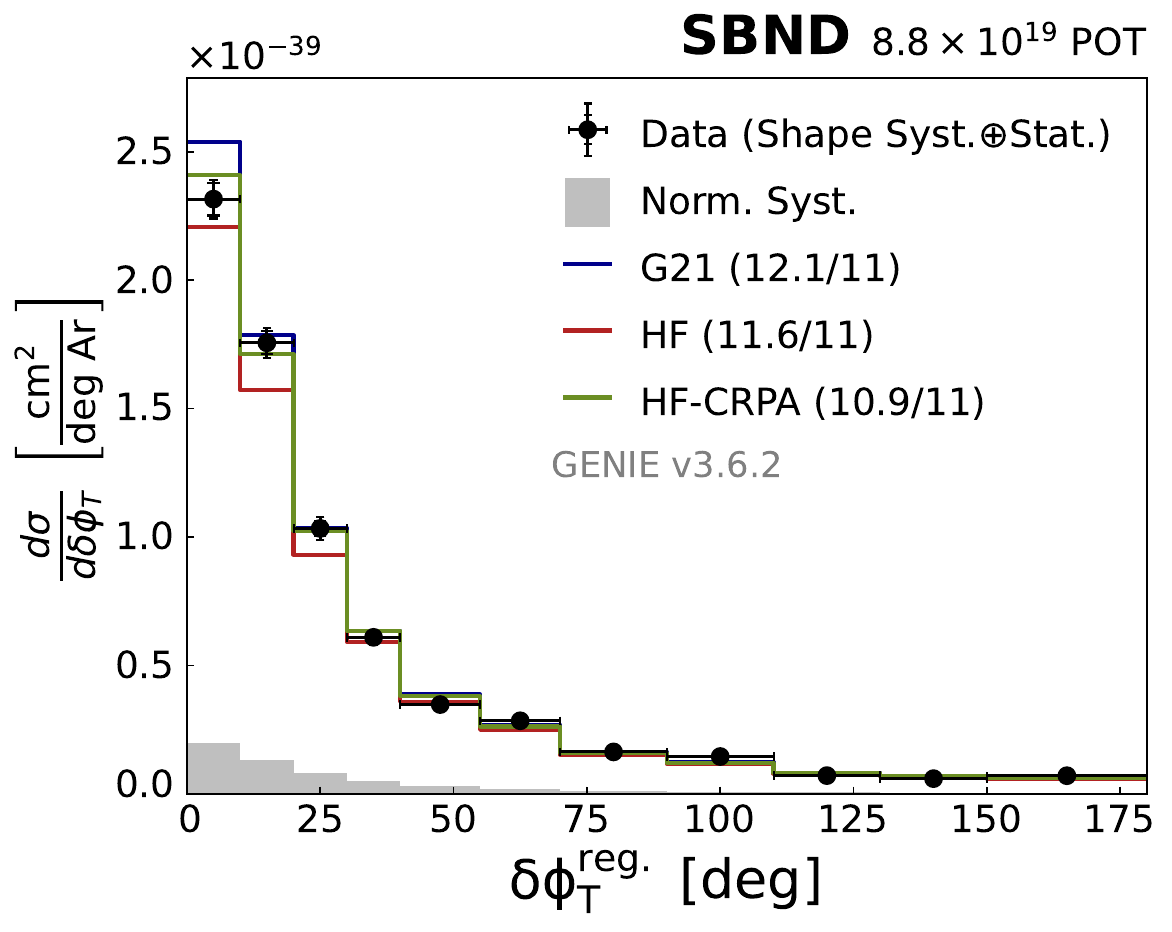}  \\
(a) & (b) & (c) \\
\end{tabular}
\caption{
  Unfolded, flux-averaged $\CCQE$ differential cross sections as functions of (a)~$\delta p_T$, (b)~$\delta\alpha_T$, and (c)~$\delta\phi_T$, compared with \texttt{GENIE G21} Hartree-Fock (HF) and HF-CRPA models~\cite{GENIE_HF}. Inner error bars denote statistical uncertainties, while outer error bars represent the statistical and shape-only systematic uncertainties added in quadrature. The gray bands at the bottom indicate normalization uncertainties. Values of $\chi^2/\mathrm{ndof}$, evaluated using the full unfolded covariance matrix, are given in parentheses.
  }
\label{fig:hf-tki}
\end{figure}

Lastly, as shown in Fig.~\ref{fig:hf-tki}, we compare to models with different QE responses: the \texttt{GENIE G21\_11a\_00\_000} tune, a model that uses the Hartree-Fock (HF) calculation, and the HF model with continuum random phase approximation (CRPA) corrections.

\clearpage

\makeatletter
\gdef\@FMN@list{}%
\makeatother
\bibunitrevtexcontrols
\putbib[supplementalNotes,references]

\end{bibunit}

\end{document}